\documentclass[sigconf,screen,10pt]{acmart}
\renewcommand\footnotetextcopyrightpermission[1]{} 
\usepackage{include/packages}
\usepackage{include/shorthands}
\usepackage{include/macros}

\begin{document}
\title[\nameshort]{\name}

\author{Conor Green, Setsuna Yuki, and Mithuna Thottethodi}

\affiliation{%
\institution{Elmore Family School of Electrical and Computer Engineering, Purdue University}
}
\email{{green456, hou125, mithuna}@purdue.edu}

\renewcommand{\shortauthors}{Green et al.}

\begin{abstract}

Data center network design plays a critical role in AI training by supporting scaling to thousands of accelerators.
An open problem,
designing a near-optimal throughput-oriented network---topology, routing, and collectives---has not been achieved at scale and with broad applicability to physical or implementation constraints. We address this problem with a compelling use-case, Google's~\tpu{} supercomputer where the topology may be reconfigured to achieve higher all-to-all throughput, supporting large, parallelized AI training.
We show that the existing TPU networks leave terabytes per second of throughput on the table and we fill that gap.

This paper presents \name{} (\nameshort{}), an automated network synthesis framework 
that meets the high-throughput demands of modern computing.
\nameshort{} formulates topology synthesis as a linear optimization problem that maximizes a throughput-centric proxy metric, using theory and heuristics to scale to thousands of nodes while producing state-of-the-art network topology performance metrics.
We further introduce a state-of-the-art deadlock-free routing scheme compatible with limited virtual channels and optical switch faults, enabling the synthesized topologies to realize their predicted throughput gains in simulation.
Evaluating uniform random and~\alltoall{} traffic, \nameshort{} networks have a geometric-mean speedups of 2.1$\times$ and 1.6$\times$ over the best~\tpu torus variants.

\end{abstract}

\maketitle

\putsec{intro}{Introduction}

The rapid progress of artificial intelligence (AI) and machine learning (ML) has been driven by scaling model size, dataset size, and training compute~\cite{kaplan2020scaling}. Frontier models now reach billions to trillions of parameters~\cite{meta2025llama,openai2024gpt4technicalreport,mann2020language,thoppilan2022lamda,chowdhery2022palm,brown2020gpt3}, pushing training from single accelerators to distributed execution spanning thousands of devices. At these scales, end-to-end throughput is frequently limited by sustained communication rather than peak compute, and modern training stacks explicitly co-design parallelization and scheduling with the network to achieve acceptable utilization~\cite{narayanan2021megatron,rajbhandari2020zero,xu2021gspmd,zheng2022alpa,barham2022pathways,jiang2024megascale}.

Distributed training uses data, tensor, pipeline, and/or expert parallelism, introducing synchronized collectives into the critical path.
The parallelism techniques require frequent gradient synchronization via \allreduce{} or \reducescatter{} then \allgather{}~\cite{sergeev2018horovod,rajbhandari2020zero,jiang2020byteps} and bandwidth-heavy \allreduce{}/\allgather{} between layers~\cite{shoeybi2020megatron,narayanan2021megatron}.
Although overlap and pipelining can reduce exposed communication~\cite{huang2019gpipe,narayanan2019pipedream}, large-scale studies still report communication bottlenecks and scaling efficiency diminishes as bandwidth per device drops~\cite{fernandez2025diminishing,zheng2022alpa,narayanan2021megatron}. Moreover, mixture-of-expert (MoE) dispatch/combine requires~\alltoall{} communication, dominating step time even with aggressive overlap~\cite{jiang2024lancet}.

Collective libraries continue to improve~\cite{shah2023taccl,cowan2023mscclang,wu2024mccs} but the fundamental throughput bounds of network topologies remain a hard constraint: no schedule can overcome fundamental bottlenecks from insufficient cut capacity, limited path diversity, or load concentration. This motivates a broad line of work showing that irregular and expander-inspired designs can deliver substantially higher throughput than classical structured networks~\cite{singla2012jellyfish,valadarsky2016xpander,basu2024efficientalltoall,zhao2025efficientdirect}.
Motivated by collective-heavy AI/ML training traffic, we focus on direct interconnects because prior works show direct topologies can deliver high aggregate throughput and utilization (at a fixed port budget) and remain robust across traffic patterns~\cite{valadarsky2016xpander,singla2012jellyfish,zhao2025efficientdirect}.

Beyond the topology, the routing must be intelligently designed to account for deadlocks in a lossless fabric and faults in large-scale systems. While previous works have proposed high-performance topologies~\cite{kautz1968kautz,stewart2006kautz,imase1983genkautz,valadarsky2016xpander,singla2012jellyfish,wong1974circulant,guo2009bcube,ahn2009hyperx}, the routing functions are often complicated, require hardware changes, and/or increase virtual channel (VC) resources~\cite{dally1987deadlock}. For this reason, many multihop direct networks maintain simple topologies and use well-established routing like turn model and datelines~\cite{glass1992turnmodel,dally2004textbook}.
Previous works show deadlock freedom in a limited number of VCs~\cite{domke2016nue} but the performance suffers too much to be useful in practice.

Despite clear opportunity, improving these networks faces many obstacles.
\begin{itemize}[leftmargin=0.1in,nosep] %
    \item Objective fidelity: directly optimizing end performance would require modeling routing realizability, deadlock avoidance, and collective scheduling inside the synthesis loop but optimizing weak proxies results in poor performance.
    \item Solution space: the number of possible topology permutations grows super-exponentially while \alltoall{}-style demand induces quadratic number communicating pairs.
    \item Implementability: high-throughput designs are often irregular and inapplicable to many physical constraints.
\end{itemize}

We propose \name{} (\nameshort{}) to address these issues by decomposing the problem of designing high-quality networks into two parts: topology generation and deadlock-free, fault-tolerant routing.
\nameshort{} formulates topology construction as a mixed integer linear program (MILP) targeting the maximum concurrent flow (MCF) which is the strongest proxy metric for \alltoall{} throughput~\cite{basu2024efficientalltoall}.
We combine unique theory, heuristics, and conventional linear programming techniques to enable scalable synthesis without simulation in the loop.
Finally, \nameshort{} achieves high-throughput, low-latency, deadlock-free, and fault-tolerant static routing within a user-provided limited number of virtual channels.
\nameshort redefines the state-of-the-art because each contribution---topology construction and deadlock-free routing---outperforms the best known prior topologies (\autoref{sec:design-topo_optimality}) and routing algorithms (\autoref{sec:design-routing_optimality}).

While \nameshort{} applies broadly, we spend the majority of our effort on a deep dive of a compelling and complex use-case.
To exemplify scalability and compliance to arbitrary constraints, we apply \nameshort{} to Google's tensor processing unit (TPU) AI training supercomputers v4, v5p, and v8t (v4--8t)~\cite{tpuv4,tpuv5specs,tpuv8announcement}. %
Not a niche system, these hyperscale TPU clusters serve and train products for conventional services like YouTube, Gmail, and Google Maps as well as Google's flagship AI model, Gemini~\cite{tpuv5announcement}.
While these systems are critical infrastructure for Google, their inter-chip interconnect (ICI) fabric uses basic torus topologies, which, as we will explore, leave substantial headroom for richer topology choices under the same physical constraints.
The \tpu{} application domain has specific constraints that prohibit many decades of network design from directly applying: routing is implemented via static forwarding tables with a small (two) virtual-channel budget for deadlock freedom, and fault tolerance requires offline routing under failures~\cite{zu2024resiliency}.
We demonstrate \nameshort{}'s systematic approach yields superior performance in topological metrics, uniform random traffic, collective communication, and end-to-end AI training.

This work makes the following contributions.
\vspace{-.4em}
\begin{itemize}[leftmargin=0.1in, itemsep=-2pt]
    \item Novel topology generation linear programming formulation that applies to broad use-cases.
    \item Unique theory-based scaling technique based on triangle inequality properties to reduce formulation size.
    \item Routing algorithm with a new approach to deadlock avoidance with limited VCs by first finding deadlock-free turns/paths and then optimizing path selection.
    \item Theory-based modifications to the routing algorithm for fault tolerance and virtual channel load balance.
    \item Combination of symmetry and relaxation techniques to scale both topology generation and routing to thousands of nodes.

\end{itemize}
\vspace{-.4em}

\putsec{background}{Background}

\subsection{Analytical Models on Performance}
\label{sec:background_metrics}

We distinguish a network's capacity under ideal fractional routing from the throughput achieved by an implementable static routing function~\cite{dally2004textbook,jerger2014nocarchitectures}.
For a given demand matrix, the MCF is the largest scalar $\lambda$ by which all demands can be simultaneously scaled without exceeding link capacities~\cite{shahrokhi1990mcf}. MCF is the exact throughput under fractional multipath routing and therefore upper-bounds any single-path realization. It accounts jointly for cut capacity and link occupancy, whereas individual cut metrics can be loose and can misrepresent actual throughput~\cite{jyoth2016measuringthroughput}. We therefore use MCF as the topology-synthesis objective and as an upper bound for uniform-random and \alltoall{} traffic.
The Moore bound lower-bounds average path length for a given size and radix~\cite{hoffman1960moore}. Combined with total link capacity, it yields the (unrealizable) ceiling on~\alltoall{} throughput (utilized in \autoref{fig:arbitrary_sweep} and
\autoref{fig:a2a_worst_thru_scaling}).
On the other end, topologies can also be characterized by their worst-case throughput, representing throughput under the most adversarial traffic pattern(s)~\cite{towles2002worst,jyoth2016measuringthroughput}.

To evaluate static routing, we use maximum directed-edge load
(\emph{max channel load}). If $L_{\max}$ is the largest aggregate normalized demand assigned to any unit-capacity directed edge, the achievable demand scale is at most $1/L_{\max}$; our routing stage therefore minimizes $L_{\max}$ similar to established approaches~\cite{dally2004textbook,towles2003thruputrouting}.

\subsection{~\tpu{} ICI and OCS}
\label{sec:background_tpu}

Google's \tpu{} ICI uses optical circuit switching (OCS) to scale beyond a single electrically-cabled building block~\cite{tpuv4,tpuv4documentation}.
The system is assembled from 64-chip \emph{cubes} arranged as a $4\times4\times4$ 3D mesh.
TPU pods connect electrically wired cubes using an optically reconfigurable interconnect (Palomar OCS and related designs) to build pod-scale direct networks~\cite{tpuv4,singh2015jupiter,poutievski2022jupiterevolving,urata2022missionapollo,liu2023lightwave,liu2024reconfigurable}.
On its six faces, each cube has optical ICI links that connect to hardwired OCSes to realize a job-level inter-cube topology~\cite{tpuv4}. The OCS is software reconfigurable to directly connect/pair edges via a MEMS-based mirror, effectively creating a directly connected topology~\cite{singh2015jupiter,poutievski2022jupiterevolving,urata2022missionapollo}. 
In production, Google deploys various configurations of 3D prismatic tori (PT) and prismatic doubly twisted tori (PDTT) as baseline job topologies up to 8192 nodes~\cite{tpuv4}.

This setting imposes constraints uncommon in general DCN synthesis: fixed intra-cube wiring, fixed optical port budgets per cube face, and OCS-imposed connection restrictions (circuits connect only within a switch).
Moreover, forwarding is deterministic and configured per job (i.e., static routing tables)~\cite{tpuv4}, so routing choices directly shape max channel load and achievable throughput under \alltoall{} or similar demand (\autoref{sec:design-routing}).
Google utilizes minimal hop XYZ dimension ordered routing (DOR) and datelines for deadlock freedom, performing heuristic or integer linear program (ILP) route optimizations for equivalent distance paths.
Finally, at pod scale, failures are expected.
Google devises ``wild first routing'' (WFR) where XYZ DOR paths that are disallowed due to a fault take a few hops through neighbors following the ``sandwich rule''~\cite{zu2024resiliency}.

The baseline pod topologies (e.g., tori) have vertex/edge symmetry that can reduce synthesis and routing complexity by representing equivalent commodities as a small canonical set~\cite{towles2003thruputrouting}.
From descriptions of the TPU v4 system, a 3D coordinate grid defines a minimal canonical set $C$, a map that returns the transformation taking $u$ to $u_c\in C$, and an induced transform $T_u(v)$ applied to destinations~\cite{zu2024resiliency}.
We write this formally in~\autoref{eq:symmetry}.

{
\setlength{\abovedisplayskip}{3pt}%
\setlength{\belowdisplayskip}{1pt}%
\setlength{\abovedisplayshortskip}{0pt}%
\setlength{\belowdisplayshortskip}{0pt}%
\begin{equation}
\label{eq:symmetry}
\begin{aligned}
(u,v)\in E &\iff (u_c, T_u(v))\in E \quad \text{where } u_c \in S,\\
T_u &: V \to V,\quad T_u(v)=v'.
\end{aligned}
\end{equation}
}

This definition of symmetry and canonical sets will be utilized for the baseline PDTT routing (as described~\cite{zu2024resiliency}) and \nameshort{}'s topology generation and routing.

\subsection{Linear Optimization}
\label{sec:background_optimization}

\nameshort{} casts topology synthesis and routing selection as MILP and linear programs (LP).
MILPs introduce discrete (binary/integer) variables to express combinatorial structure (e.g., selecting edges), at the cost of worst-case NP-hardness~\cite{hartmanis1982miphard}.
A standard scalability tool is relaxation: dropping integrality constraints yields an LP that upper-bounds the MILP objective and often provides a useful guide for synthesis but at the cost of global optimality.

For our class of problems, the interior point method was the fastest and scales by both the number of constraints and variable sparsity.
In practice, model formulations that preserve sparsity can be orders of magnitude faster and less memory-intensive~\cite{gurobi}.
This motivates the symmetry reductions and formulation choices in~\autoref{sec:design-topo} and \autoref{sec:design-routing}.

\section{Design Overview}\label{sec:design_overview}

\begin{figure*}
    \centering
    \includegraphics[width=0.95\linewidth]{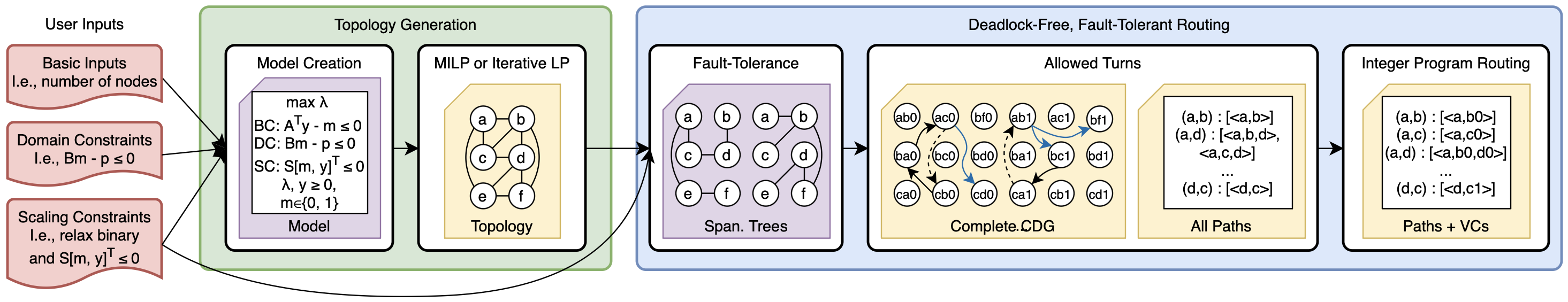}
    \vspace{-.4em}
    \caption{Overall pipeline of \nameshort{} from user inputs (red) through topology generation (green) and routing (blue). Boxes in purple are intermediate data structures while boxes in yellow are files that may be output/ingested by downstream scripts. An example six node (\texttt{a--f}) topology is shown for simplicity. For the complete channel dependency graph, turns required by the spanning trees, allowed, and prohibited are blue, black, and dashed arrows, respectively.}
    \label{fig:total_diagram}
\end{figure*}

The \nameshort{} approach to network design has two major components---generating topologies and a routing function for the topology---that interface by a single, simple file representing the topology. We illustrate the complete pipeline in~\autoref{fig:total_diagram}. In \nameshort{}, the user provides information on the application domain and decides which scaling heuristics to utilize. The final outputs are the topology and routing (paths plus VCs). 
We review works related to direct/multihop topologies and the routing functions used by them.
\autoref{sec:related_topology} evaluates alternative HPC and/or OCS topologies which we later quantitatively evaluate versus \nameshort{} in~\autoref{sec:design-topo}.
\autoref{sec:related_routing} provides a brief discussion of HPC routing with an emphasis on deadlock freedom and VC usage.

\putsubsec{related_topology}{Related Topology Works}

\nameshort{} performs optimization-based network synthesis for optically reconfigurable fabrics.
Rather than selecting from a small family of rule-based templates (e.g., tori), \nameshort{} uses LP/ILP/MILP formulations and relaxations to optimize throughput-oriented proxy objectives and produce irregular, non-regular topologies.
To our knowledge, no prior work directly \emph{generates} topologies using an MCF-based linear optimization formulation. Most either generate candidates heuristically and evaluate them with MCF iteratively~\cite{hu2006communication,singla2012jellyfish,griner2021cerberus,singla2014highthroughput,curtis2012rewire} or optimize different objectives (e.g., power/latency, tail performance, or reconfiguration cost)~\cite{lpbt,netsmith,gangwar2020automatedsynthesis,shukla2025tamingtail,schlinker2015condor,mellette2017rotornet}.
\nameshort{} targets the physical and control-plane constraints of \tpu{} ICI---fixed intra-cube wiring, constrained OCS port groupings, and static forwarding---so we review work on direct/irregular topologies, optimization frameworks, and throughput-centric theory.

\textbf{Direct and irregular datacenter topologies} --
A broad line of work explores direct or server-centric DCNs beyond Clos hierarchies, including DCell~\cite{guo2008dcell}, BCube~\cite{guo2009bcube}, CamCube~\cite{abu2010camcube}, SWDC~\cite{shin2011swdc}, and Scafida~\cite{gyarmati2010scafida}.
While these designs motivate moving beyond regular tori, they do not address \tpu ICI constraints (fixed intra-cube wiring, OCS port groupings) nor the static single-path forwarding model, and thus are not directly portable.

\textbf{Optimization-based network design} --
REWIRE uses an optimization framework for unstructured DCN design but scales only to modest sizes~\cite{curtis2012rewire}.
COUDER optimizes OCS datacenters under traffic uncertainty and evaluates hop-count/throughput improvements under quasi-static reconfiguration~\cite{teh2020couder}, while Perseus jointly optimizes topology and cabling complexity but focuses on physical wire length rather than collective throughput~\cite{mudigonda2011taming}.
In contrast, \nameshort{} targets~\alltoall{} throughput under stringent structural constraints and scales to thousands of nodes while remaining compatible with static forwarding.

\textbf{High-throughput graph constructions and throughput-centric theory} --
Irregular and expander-inspired DCNs such as Jellyfish~\cite{singla2012jellyfish} and Xpander~\cite{valadarsky2016xpander}, and low-diameter HPC networks such as Slim Fly~\cite{besta2014slimfly}, show that high expansion and path diversity can outperform structured designs in throughput.
However, these systems typically assume packet-switched routers with flexible (often multipath) routing, whereas \nameshort{} operates under static single-path forwarding with a small VC budget.
VL2~\cite{greenberg2009vl2} popularized Valiant Load Balancing~\cite{valiant1982scheme} as an oblivious mechanism to support broad traffic matrices, including under failures~\cite{zhang2008faultvlb}, but its assumptions differ from those of the \tpu ICI (i.e., deterministic forwarding).

\putsubsec{related_routing}{Related Routing Works}

We focus our evaluation of routing to those applicable to lossless fabrics where deadlocks may occur and omit exploration of deadlock recovery since it often requires hardware changes.
Deadlocks may be avoided by restricting routing, using VCs, and/or a combination of the two~\cite{dally2004textbook}.
We categorize high-performance routing related works as requiring: an unknown, possibly large, number of VCs or a configurable number of VCs.
\nameshort{}'s routing contribution fits into the latter category.

\textbf{Deadlock Free in Unknown Number of VCs} --
DFSSSP performs iterative routing, online allocating paths to VCs, but has no guarantees on the number of virtual channels required~\cite{domke2011dfsssp}.
LASH makes any routing function deadlock free so it can be as high-performance as the underlying routing function, but again requires possibly many VCs~\cite{lysne2006lash}.

\textbf{Deadlock Freedom with a Fixed VC Budget} --
Dimension-ordered turn models prevent cyclic dependencies by prohibiting
selected turns and often use dateline VCs, but require coordinate-based
topologies such as meshes or tori~\cite{glass1992turnmodel,
dally1987deadlock,dally2004textbook,dally1990vcs}.
Up*/Down* routing instead orients links using a spanning tree and
prohibits down-to-up transitions, favoring hierarchical
topologies~\cite{Schroeder1991updown}.
Nue supports arbitrary topologies by iteratively selecting Dijkstra paths
while rejecting turns that create CDG cycles~\cite{domke2016nue}, but
sacrifices throughput and was only evaluated up to 300 switches.

\section{Topology Design}
\label{sec:design-topo}

\nameshort{} synthesizes a topologies through a linear programming formulation that maximizes the MCF, a representative metric for uniform random and \alltoall{} demand.
The output is an adjacency matrix that is directly implementable in the user-provided domain (e.g., in our use-case, the \tpu ICI OCSes) and serves as the input to the routing and deadlock-avoidance pipeline, described in~\autoref{sec:design-routing}.

We start from an established MCF LP formulation and re-formulate, introducing topology variables and ``base constraints'' (BC) so the program~\emph{generates} a topology rather than only evaluates a fixed graph.
\nameshort{} conform to the \tpu ICI implementation restrictions with ``domain constraints'' (DC).
\nameshort{} scales through three ``scaling constraints'' (SC):
(i) ``one-leg'' reduction that shrinks the triangle-inequality footprint, (ii) symmetry reductions that collapse equivalent commodities/constraints, and (iii) an iterative LP relaxation.
Scaling constraints (ii) and (iii) may reduce the quality of generated topologies and we offer light justification for their inclusion because they unlock scaling.
For users interested in applying \nameshort{} to a different context, the domain and scaling constraints should be tailored.

Furthermore, the formulation can be modified to support user-provided demands (i.e., some specific traffic pattern), non-binary connections (i.e., non-uniform bandwidth topology), and/or different link types (i.e., low- and high-bandwidth tiers) by introducing coefficients to the flow variables, extending the range of the topology variables, and/or introducing multiple topology variables, respectively. However, we omit these extensions for space and to focus on the core ideas presented.

\subsection{Basic Topology Generation}
\label{sec:design_topo-basic}

We use the MCF metric as the guiding proxy and cast synthesis as linear optimization. We start from an existing LP that captures MCF and re-formulate to include topology connections as variables.
As we will see, this novel re-formulation outperforms \emph{all} major rule-based and synthesized topologies from the past few decades in a small-scale sweep.
Later, when we add domain-specific and scaling constraints, we also see this formulation outperforms industry SoTA topologies.

We utilize the dualized MCF formulation as defined by Leighton and Rao (LR)~\cite{leighton1999lrapproxsc}.
The LR formulation is intended to approximate the sparsest cut of a graph by (exactly) finding the MCF.
The formulation takes as input a graph and minimizes a semi-metric to find the MCF.
LR utilizes triangle inequality constraints to define the semi-metric~\cite{leighton1999lrapproxsc}. 
While useful for analyzing existing topologies, LR cannot naively generate topologies. Replacing the formulation's static topology edge values with variables would introduce multiplication between the topology and semi-metric variables.
Furthermore, the objective is minimization so a direct substitution would find the edges and distance values with the minimal MCF.
Therefore, we utilize the theory of duality for linear programs to relate the distance and edge variables by only linear operations and cause a maximization objective sense.

\begin{table*}[htbp]
\centering
\caption{\label{tab:mip_form} \nameshort topology generation formulation, applied to OCS domain and including scaling}
\vspace{-1.0em}

\begin{tabular}{|c|c|c|}
\hline
\textbf{Label} & \textbf{Objective (O), Base, Scaling, \& Domain Constraints (BC, SC, \& DC), Variable (V)} & \textbf{Description}    \\ \hline

\hline
O             & $\underset{m,y}{\text{maximize}} \quad \lambda$     &     MCF maximization                     \\ \hline
\hline
\constrReflex             &  $ m_{i,i}=0 ,\quad y_{\Delta i,i,*} = y_{\Delta i,*,i} = y_{\Delta *,i,i}=0 \quad \forall i \in R$      & Ignore self-adjacency                  \\ \hline
\constrDual            & $ \lambda - \sum_{k\in R} y_{\Delta a,b,k}  + \sum_{j\in R} y_{\Delta a,j,b} + \sum_{i\in R} y_{\Delta i,a,b} - m_{a,b} \leq 0 \quad \forall \text{ distinct } a,b \in R$                 & Dualized LR    \\ \hline
\hline

\constrBiDir             &    $ m_{i,j}=m_{j,i} \quad \forall i,j \in R$    & Bidirectional links                         \\ \hline
\constrInvalid             &  $m_{i,j}=0 \quad \forall (i,j) \notin L_{valid}$                & Prohibited links     \\ \hline
\constrForced             &  $m_{i,j}=1 \quad \forall (i,j) \in L_{known}$                & Known links     \\ \hline
\constrValid             &   $ \sum_{j \in L_{valid,d}} m_{i,j} = 2 \quad \forall i, d \in \{X,Y,Z\}$   & Radix (by dimension)        \\ \hline
\constrFault             & $\lambda \ge \nicefrac{f+1}{32|R|}$                 & Fault-tolerance    \\ \hline
\hline

\constrOneLeg             &  $y_{\Delta i,j,k}=0  \quad \forall (i,k) \notin L_{valid}$                & One-Leg     \\ \hline
\constrSymEdge             &  $\forall j, \quad m_{i,j}=m_{i_c,T_{i}(j)} \ \forall i \text{ s.t. } T_{i}(i)=i_c$                & Canonical Links     \\ \hline
\constrSymDual             &  $\forall j,k, \quad y_{\Delta i,j,k}=y_{\Delta i_c,T_{i}(j), T_i(k)} \ \forall i \text{ s.t. } T_{i}(i)=i_c$                & Canonical Duals     \\ \hline
\hline

V1             & $m_{i,j} \in [0,1]$ or $m_{i,j} \in \{0,1\}$                          &  Adjacency matrix               \\ \hline
V2             & $y_{i,j,k} \in [0,1]$                         &     LR dual variables       \\ \hline

\end{tabular}
\vspace{-.6em}
\end{table*}

\textbf{Dual-Based Formulation} --
The LR formulation dual has a maximization sense and no multiplication between the topology and semi-metric.
By the theory of duality, the objective value of the primal and dual are equal for optimal solutions and for sub-optimal solutions, this dualized objective will be strictly less than the maximally achievable MCF~\cite{schrijver1998duality}.
Dualization creates variables for each constraint of the primal (original) formulation. The conventional notation for dual variables is $y$ and based on their meaning in the primal LR formulation, we call these ``flow'' variables. The first variable in $y$ represents the MCF, $\lambda$, while the rest, $y_{\Delta i,j,k} \forall i,j,k \in R$, represent the triangle inequalities (delta symbol for triangles).
This formulation is now over two variables $y$ and $m$ and seeks to maximize the MCF, $\lambda$, represented by the objective.

This simple formulation is the engine of topology synthesis and the objective (O), base constraints (\constrReflex and \constrDual), and variables (V1 and V2) are provided in~\autoref{tab:mip_form}. 
We provide a complete description of the transformation from LR to \nameshort{} in~\autoref{ap:derivation}.

\putsubsec{design-topo_optimality}{Empirical Validation of Approach}

\begin{figure}[t]
\footnotesize 
\includegraphics[width=0.95\linewidth]{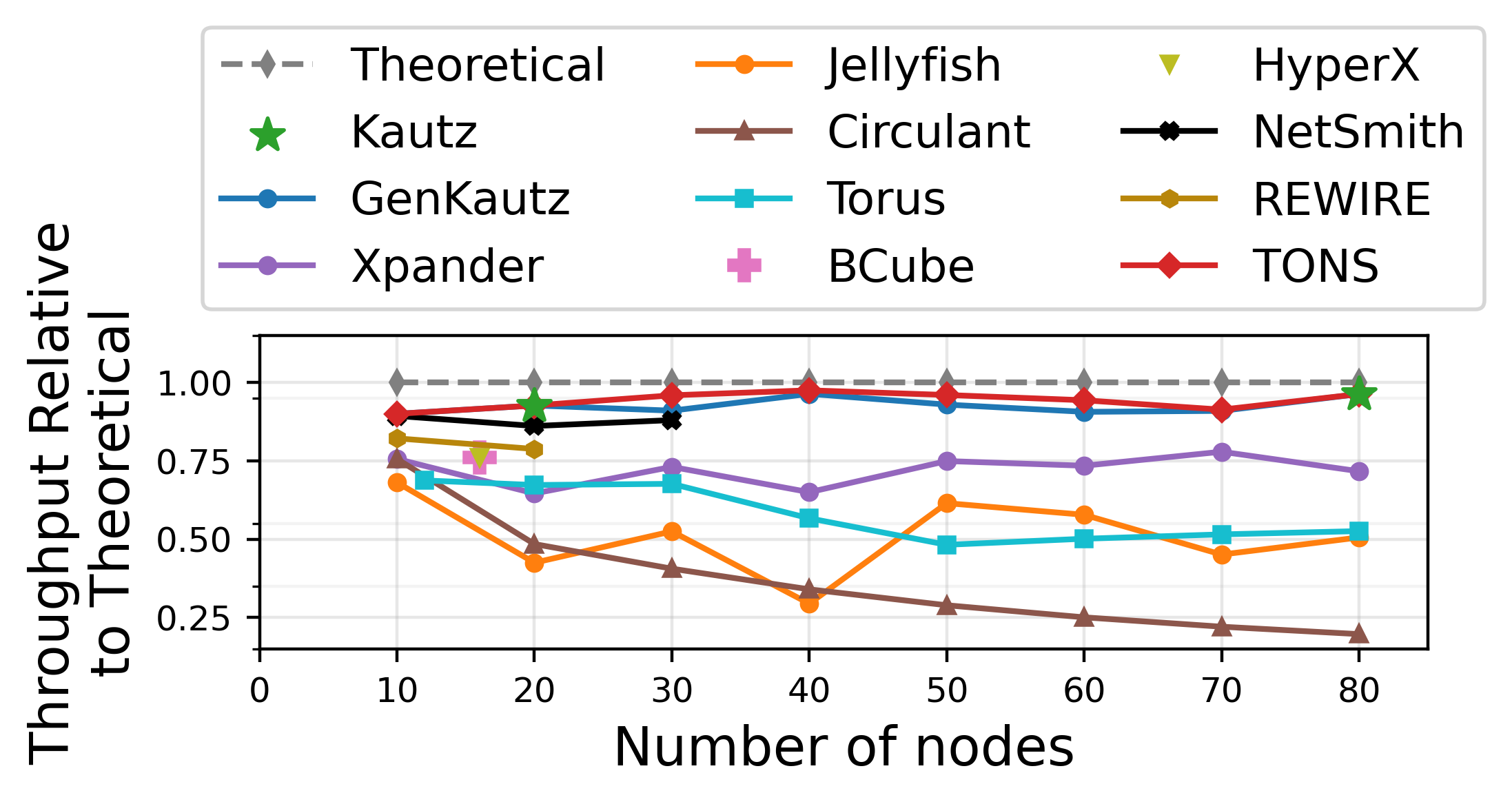}
\captionsetup{skip=2pt}
\caption{Topological bound on throughput of directed, regular four radix topologies from literature (Kautz--HyperX) versus linear programming synthesized topologies (NetSmith--\nameshort{}). Synthesized topologies were allowed one day for generation. For each marker, we plot the topologies MCF value relative to that size's theoretical MCF (higher is better).}
\label{fig:arbitrary_sweep}
\vspace{-1.2em}
\end{figure}

In a preliminary analysis to validate the~\nameshort{} formulation, we compare against topology works form the past decades of research for a broadly applicable domain: four radix, directed links, up to 80 nodes.
We extend the \nameshort{} base formulation to  maximum out and in degrees to less than four.

We evaluated the MCF of commonly cited topology proposals in the data center network (DCN) or AI/ML domain. We evaluate heuristicaly-connected topologies: Kautz~\cite{kautz1968kautz,stewart2006kautz}, GenKautz~\cite{imase1983genkautz}, Xpander~\cite{valadarsky2016xpander}, Jellyfish (random)~\cite{singla2012jellyfish}, Circulant~\cite{wong1974circulant}, 2D torus, BCube~\cite{guo2009bcube}, and HyperX~\cite{ahn2009hyperx}.
Since they are derived from strict rules, not all topologies work for all sizes.
For random, for each size, we created 100 topologies and took the highest value.
We also slightly modified topology synthesis tools NetSmith (average hops objective and no link length limitation)~\cite{netsmith} and REWIRE (no ``cost'' constraints)~\cite{curtis2012rewire}.
All topology synthesis tools (NetSmith, REWIRE, and~\nameshort{}) were given one day and ample compute resources for each generated topology.
The theoretical bound is derived from the Moore bound, as discussed in~\autoref{sec:background_metrics}.

We compare four radix direct topologies for 10 to 80 nodes for the aforementioned approaches to~\nameshort{}-generated topologies in~\autoref{fig:arbitrary_sweep}.
A trend is clear: for all sizes in this sweep,~\nameshort{} topologies are equal to or better by a few percent than all other approaches. For node sizes without a Kautz topology,~\nameshort{} generated novel and strictly superior topologies, and for this example, generated all topologies in less than a day.
The power of \nameshort{} is not only superior performance but broad applicability. Some topologies (Kautz, HyperX, BCube) only work for a few sizes while \nameshort{} applies to any size and radix combination.
These results validate the formulation and give ``proof of opportunity'' to our approach.
We now target the obstacles with LP-based topology generation:~\emph{scaling} up to 100$\times$ the size of these preliminary results and~\emph{implementing} a full network stack for routing, deadlock avoidance, fault-tolerance, and collective communication.

\subsection{Domain Constraints}
\label{sec:design_topo-domain}

Any constraints on topology construction may be added by independently constraining the topology variable, $m$.
Specifically, we constrain the $m$ to valid~\tpu{} ICI topologies that follow: reflexivity, symmetry, and valid optical connections.
Reflexivity and symmetry are provided in~\autoref{tab:mip_form} \constrReflex and \constrBiDir but in practice, handled in code by directly substituting $m_{i,i}$ with $0$ and any $m_{i,j}$ where $i > j$ with $m_{j,i}$.
Similarly, the forced electrical connections (\constrForced in~\autoref{tab:mip_form}) directly set their respective topology variables.
Valid optical connections follow OCS connectivity between $(x,y,z)$ co-ordinates as described in~\autoref{sec:background_tpu} to define the total valid connections, $L_{valid}$ (\constrValid and \constrInvalid in~\autoref{tab:mip_form}).

With these modifications, this formulation accomplishes: a maximization-based exact MCF objective, linear program with edge connection variables, and topology constraints can be added/removed without affecting the MCF maximization.

\putsubsec{design-topo_one-leg}{Formulation Scaling}
\label{sec:design_topo-scaling}

We now look to scale the topology generation to thousands of nodes and present these scaling constraints as \constrOneLeg--\constrSymDual.
We present three modifications to the formulation: a ``one-leg'' reduction, utilizing symmetry (\autoref{sec:background_tpu}), and basic binary variable relaxation. The first does not affect solution quality and introduces sparsity in the flow/dual variables. The second and third do reduce the topology design space and remove global optimality guarantees; however, since they can generate topologies faster, they result in better topologies when limited by time.

\putsubsubsec{design-topo_oneleg}{One-Leg Reduction}

The first obstacle to scalability is the $\Theta(n^3)$ number of triangle inequalities. Let $L_{valid}$ denote the set of possibly connected links; under~\tpu{} ICI constraints, $|L_{valid}|=\frac{n}{64}$. We reduce the variable $y_{\Delta i,j,k}$ footprint to $O(n^2|L_{valid}|)$ by only instantiating $y_{\Delta i,j,k} \text{ s.t. } (i,k) \in L_{valid}$ as seen in \constrOneLeg of~\autoref{tab:mip_form}. In practice, we directly set $y_{\Delta i,j,k}=0$ for known invalid connections $(i,k)$.
This retains the exact objective while substantially reducing memory in practice.

We explain this concept in the primal LR because it is easier to comprehend.
The idea is that for an optimal solution, at least one triangle inequality must be tight because one intermediate $k$ upper bounds the distance $d_{i,j}$ for all pairs. Therefore, we can limit the considered intermediate nodes to a subset $K= \{k | m_{i,k}=1\}$.
This line of thinking is similar to Nguyen and Minoux~\cite{nguyen2021linear}
but has stronger guarantees applied to optimization. Due to space, we present the full proof of this idea for the primal in~\autoref{ap:one_leg} and utilize it in the dual.
We call this variable reduction, ``one-leg'' for the one active leg of the triangle inequality.

\putsubsubsec{design-topo_symmetry}{Vertex Symmetry}

We restrict synthesis to translationally symmetric topologies (\autoref{sec:background_tpu}), using one
64-node cube $C$ as the canonical source set. For each source $i$, let
$i_c \in C$ be its canonical representative and $T_i()$ the corresponding
destination transformation. We instantiate only canonical edge variables
and reconstruct the remainder as
$m_{i,j}=m_{i_c,T_i(j)}$ (\constrSymEdge in
\autoref{tab:mip_form})
.\constrSymDual applies the same reduction to the flow/dual variables.

This reduces the number of variables from $O(n^2|L_{valid}|)$ to $O(n|L_{valid}|)$ and eliminates constraints for non-canonical sources, reducing their count from $\Theta(n^2)$ to $\Theta(n)$. 
Consequently, the formulation size falls from $O(n^4|L_{valid}|)$ to $O(n^2|L_{valid}|)$, enabling large-scale synthesis with little observed loss in topology quality (\autoref{sec:results}).

\putsubsubsec{design-topo_relaxed}{Integrality Relaxation}

Even after the preceding reductions, solving the topology MILP at the largest scales is impractical. We therefore relax the binary edge variables to $[0,1]$ and apply iterative rounding. After each LP solve, we fix $interval$ number of unfixed/unchosen edges with the largest $m_{i,j}$ values and repeat
until the topology is integral (binary connections).
We normally use $interval=1$ to preserve solution quality. \autoref{ap:relax_iter} gives the complete algorithm.

On small instances, iterative rounding reduces MCF by a few percent
relative to a completed MILP solve. Its substantially lower time and
memory requirements, however, produce better topologies under practical
compute limits.

{
\setlength{\abovedisplayskip}{2pt}%
\setlength{\belowdisplayskip}{1pt}%
\setlength{\abovedisplayshortskip}{0pt}%
\setlength{\belowdisplayshortskip}{0pt}%
\begin{figure}[t]
\noindent\hspace*{-\tabcolsep}%
\begin{tabular}{@{}c@{}c@{}} %
\includegraphics[width=.2475\textwidth]{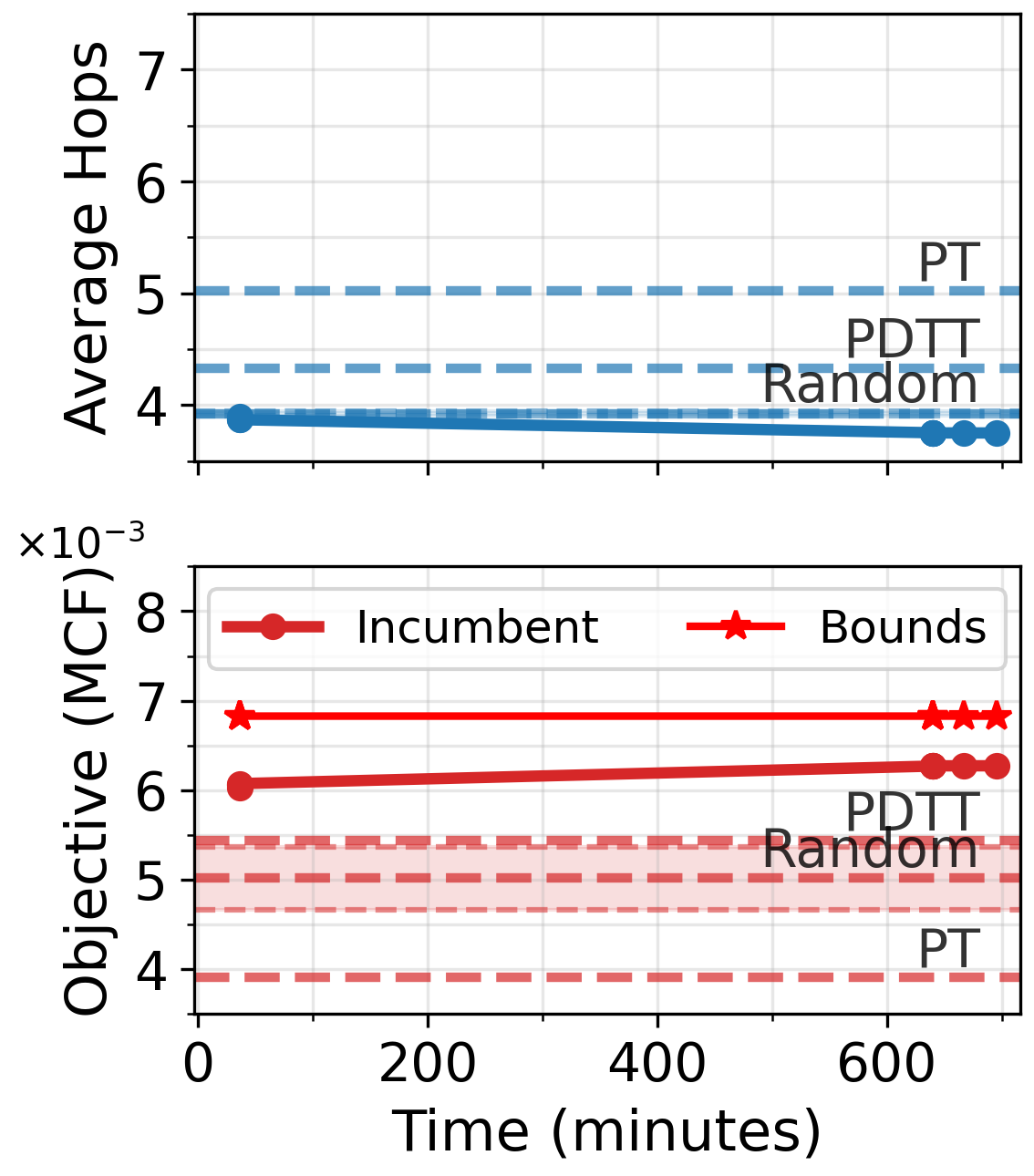} &
\includegraphics[width=.2475\textwidth]{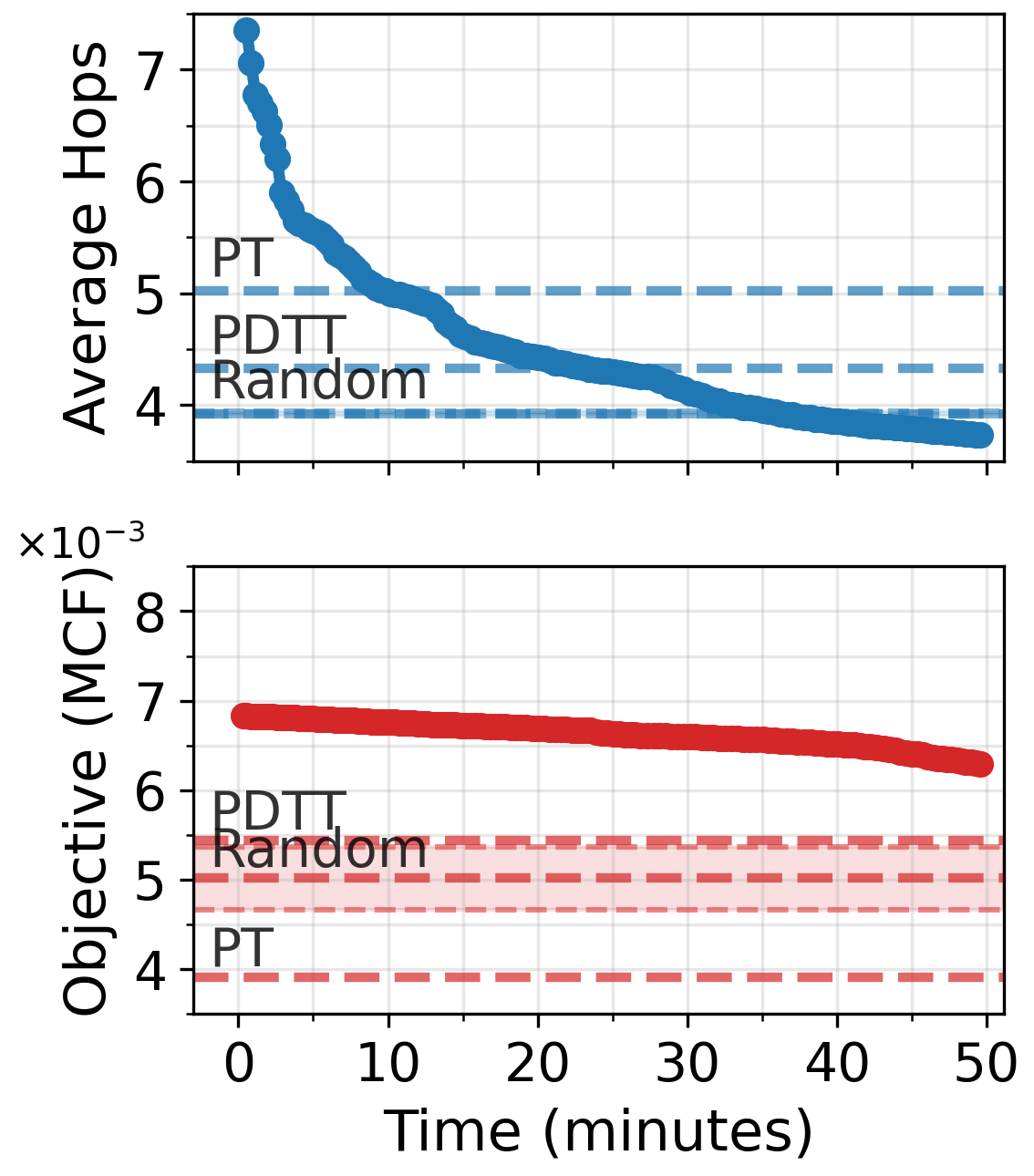} \\
(a) MILP & (b) LP \\
\end{tabular}
\vspace{-.8em}
\caption{Progress of MILP (a) and LP (b) (non-symmetric) variants of~\nameshort{} average hops (blue) and MCF objective (red) over time for a 256-node configuration.
For MILP, each dot is an incumbent solution and each star is the (dual) bound. For LP, each dot is the objective value for each solution (final dot has binary edges). Included is TPU-applicable random with standard deviation shaded.}
\label{fig:progress}
\vspace{-1.2em}
\end{figure}
}

\putsubsec{design-topo_results}{Resultant Topologies}

\begin{figure}[t]
    \centering
    \begin{subfigure}[b]{0.45\linewidth}
        \centering
        \includegraphics[width=\linewidth]{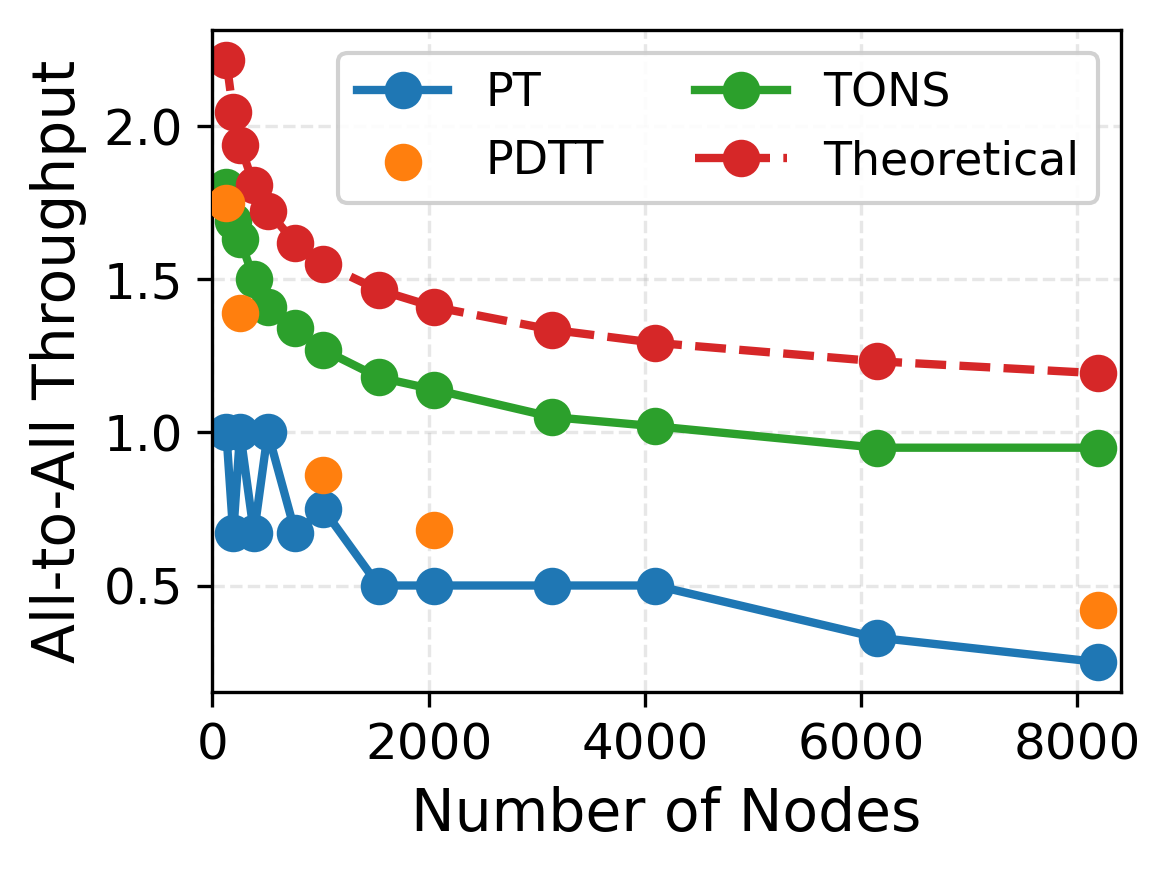}
        \caption{All-to-all throughput}
        \label{fig:a2a_thru}
    \end{subfigure}
    \hfill %
    \begin{subfigure}[b]{0.45\linewidth}
        \centering
        \includegraphics[width=\linewidth]{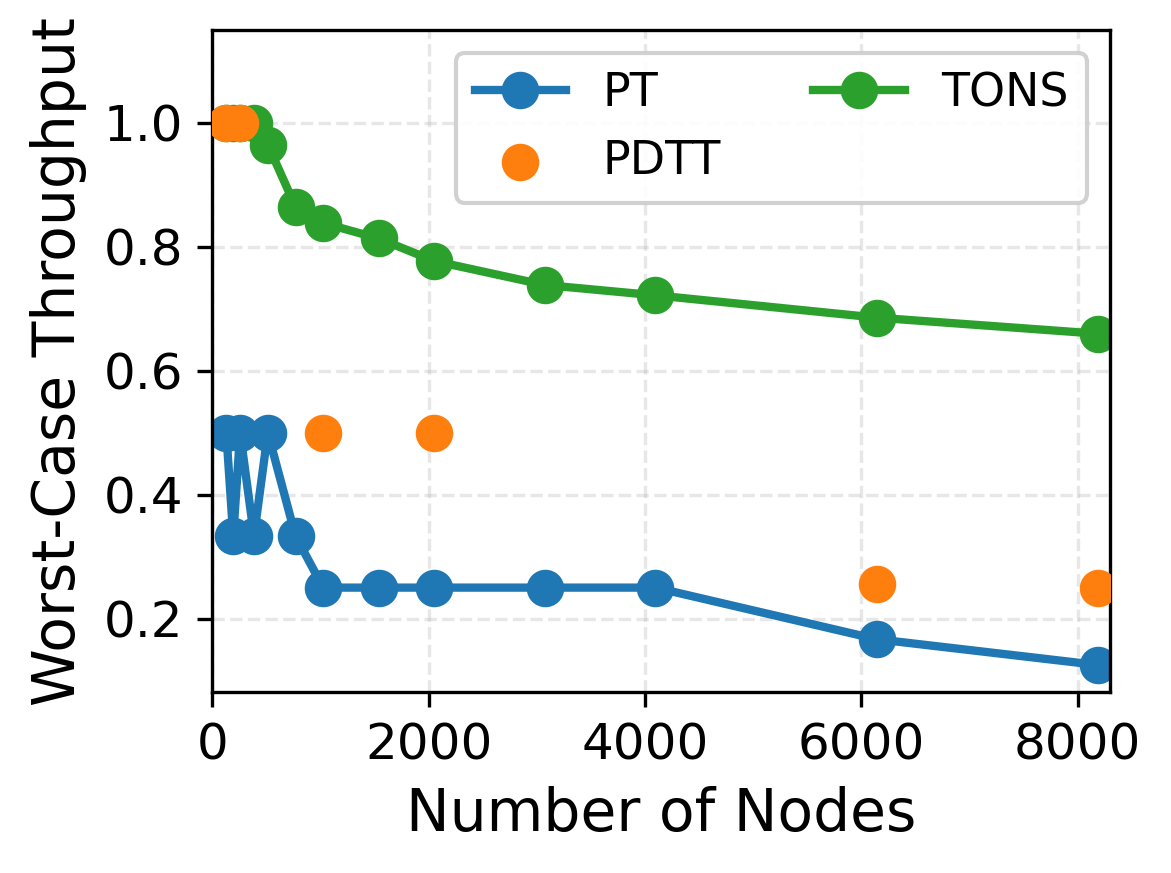}
        \caption{Worst-case throughput}
        \label{fig:worst_case_thru}
    \end{subfigure}
    
    \captionsetup{skip=2pt} 
    \caption{All-to-all (a) and worst-case (b) throughput of the best PT, PDTT, and~\nameshort{} sweeping topology sizes. Theoretical limit (dashed) is based on size and radix but unrealizable for TPU/OCS constraints.}
    \label{fig:a2a_worst_thru_scaling}
    \vspace{-1.2em}
\end{figure}

We generate topologies for all sizes\footnote{While PT and PDTT topologies differ based on the job dimensions due to their algorithmic construction,~\nameshort{} topologies are the same regardless of the configuration.}
using AMD EPYC 7763 ``Milan'' CPUs with up to 256GB of memory and Gurobi version 12.0.0.
An example of the average hops and MCF objective changing over time for a 256-node, non-symmetric synthesis run is plotted in~\autoref{fig:progress}. We see that both the MILP and LP comprehensively beat the heuristic baselines (PT, PDTT) and the standard deviation for a random---randomly selected optical links while adhering to ~\tpu{} ICI constraints---topology. Furthermore, we see the benefit of the iterative LP approach: faster topology synthesis with little objective loss. While these non-symmetric solutions take hours, the symmetric LP variant completes in 3 minutes for 256 nodes. At 8192 nodes, it takes five days.

Only LP with symmetry scales to the complete 8192 nodes and we plot the \alltoall{} throughput (number of nodes times MCF) in~\autoref{fig:a2a_thru}.
Furthermore, we plot the worst-case throughput~\cite{jyoth2016measuringthroughput} in~\autoref{fig:worst_case_thru} with a similar trend: \nameshort{} outperforms at all scales.
Using the proxy metric of the maximum concurrent flow did not sacrifice performance or create adversarial patterns. In fact, even the throughput of the worst possible traffic is drastically improved.
In~\autoref{fig:a2a_thru}, we also include a theoretical bound on the per source injection rate for any six radix, undirected graph as derived by the Moore bound (\autoref{sec:background_metrics}) which does not obey~\tpu{} ICI constraints but is still a valid upper bound to provide context.
\nameshort{} topologies not only have higher analytical throughput than the baselines but scale well, matching the curvature of a (non-realizable) upper bound.
A complete table of analytical metrics is provided in~\autoref{ap:full_analytical} and visualizations of example topologies are in~\autoref{ap:viz}.

As discussed in~\autoref{sec:background_metrics}, the throughput of networks is bounded by either the link occupancy or cut bounds.
PT topologies are limited by cuts which PDTT topologies remedy with the twisted wrap-around links, as utilized in the original proposal~\cite{camara2010twisted}. However, the wrap-around links do not sufficiently reduce hops and the link occupancy bounds are much tighter than cuts. Both PT and PDTT leave one bound with a large slack, which wastes resources.
Instead of maximizing one/two bottlenecks as in twisted tori,~\nameshort{}-generated designs select optical edges to improve multiple cuts and lower average hops simultaneously rather than optimizing a single bisection.
Maximizing the MCF, minimum of \emph{both} throughput bounds, naturally optimizes both the link occupancy and cut bounds.

\section{Deadlock Free and Fault Tolerant Routing}
\label{sec:design-routing}

The topologies synthesized in~\autoref{sec:design-topo} improve analytical throughput metrics but are not directly implementable using torus-specific routing rules.
To realize these gains in a \tpu ICI lossless fabric, we must produce (i) static forwarding tables selecting a single path per source--destination pair, (ii) a VC assignment within a small VC budget, and (iii) fault-tolerant reachability under single-OCS failures.

Making an \emph{arbitrary} routing function deadlock-free within a bounded number of VCs is NP-complete, but it is often possible to construct a \emph{good} deadlock-free routing using a limited VC budget~\cite{domke2016nue}.
Prior work (e.g., Nue) integrates VC allocation and CDG maintenance while committing to routes early, but the resulting routing functions can be overly conservative and throughput-suboptimal~\cite{domke2016nue}.
\nameshort{} reverses the conventional approach of routing then deadlock avoidance by first constructing a deadlock-free turn set and all deadlock-free paths then selecting globally optimal routes from that set.

\subsection{Allowed Turns and All Paths}
\label{sec:design_at}

{
\setlength{\textfloatsep}{6pt plus 2pt minus 4pt}
\setlength{\intextsep}{6pt plus 2pt minus 4pt}
\begin{algorithm}[t]
\small
\caption{Allowed Turns (AT)}
\label{alg:allturns}
\begin{algorithmic}[1]
\State \textbf{Input:}  $G$ \Comment{Topology}
\State \textbf{Initialize:} $\overline{D} \gets$ \func{complete\_cdg}($G$)
\State \textbf{Initialize:} $A \gets \emptyset$ \Comment{Allowed turns}
\If{robust}
    \State $S_{0}$,$S_{1} \gets$ \func{ocs\_disjoint\_spanning\_tree}($G$)
    \State $A$,$\overline{D} \gets$ \func{add\_turns}($A$,$S_{0}$,$\overline{D}$,force\_vc=0)
    \State $A$,$\overline{D} \gets$ \func{add\_turns}($A$,$S_{1}$,$\overline{D}$,force\_vc=1)

\EndIf

\State $S \gets$ \func{spanning\_tree}($G$) \Comment{Enforced routability}
\State $A$,$\overline{D} \gets$ \func{add\_turns}($A$,$S$,$\overline{D}$,force\_vc=0)
\State $T' \gets$ \func{prioritize\_turns}($G$)
\State $A$,$\overline{D} \gets$ \func{add\_turns}($A$,$T'$,$\overline{D}$,single\_turn=True)
\State $A$,$\overline{D} \gets$ \func{add\_turns}($A$,$T'$,$\overline{D}$)

\State P = \func{BFS(G,A)} \Comment{Set of all deadlock-free paths}
\end{algorithmic}
\end{algorithm}
}

Arbitrary topologies complicate deadlock freedom because simple rule-based schemes (e.g., dimension ordering with datelining used for tori) do not generalize: a fixed rule can strand some source--destination pairs without a valid path.
A common approach allocates routes to VCs as layered virtual networks using CDG-based sufficient conditions (acyclicity), but it provides no guarantee on the number of VCs required~\cite{lysne2006lash,domke2011dfsssp}.
Nue demonstrates that one can target a fixed VC budget by routing while maintaining a CDG~\cite{domke2016nue}, but early route commitments reduce path diversity and can degrade throughput.

\nameshort{} instead decouples deadlock freedom from route selection: we first construct a large deadlock-free candidate path set using an \emph{allowed-turn} construction on the CDG, then solve a throughput-maximizing ILP to choose one path per flow.
Starting from the complete CDG $\overline{D}$ over VC-labeled channels, we greedily add turns into a global allowed set $A$ only when the insertion preserves acyclicity in $\overline{D}$. Thus, any routing restricted to $A$ is deadlock-free by construction.
\autoref{alg:allturns} summarizes the approach.
We denote the sorted set of turns as $T'$.
After completely filling the CDG and getting $A$, for each flow, \nameshort{} runs a BFS to find all paths using only allowed turns (\autoref{alg:allturns} line 14). These paths are minimal for the allowed turns but possibly non-minimal on the topology (i.e., if all turns were allowed).

Compared to Nue, AT is lightweight because it operates on turns (not flows) and defers route selection to the routing ILP in~\autoref{sec:design-routing_ilp}.
With symmetry enabled, we add/reject turns in symmetry classes and accept a class only if no member introduces a CDG cycle.

Because turn addition is greedy, the insertion order matters. We evaluate three prioritization heuristics in~\autoref{sec:results_routing}:
\emph{APL} (``all path list'') orders turns by decreasing frequency across the set of all candidate paths; \emph{CPL} (``chosen path list'') orders turns by decreasing frequency in the single-path routing produced by the ILP; and \emph{Random} adds turns in arbitrary order. APL yields the highest path diversity but CPL yields the lowest maximum channel load amongst these variants and is used for topologies under 4096 nodes (faster Random used for larger).

To guarantee routability, we seed $A$ with a spanning-tree turn set, $S$. We build a tree rooted at a central node and add its up/down turns to VC 0/1 (Algorithm~\ref{alg:allturns}, lines 9--10), ensuring at least one path between any pair.
To retain path diversity and reduce bias from the prioritization order, we initially allow at most one VC-labeled turn per base turn (\autoref{alg:allturns}, line 12) before expanding to all admissible VC assignments.

\subsection{Fault Tolerance}
\label{sec:design_fault}

\tpu{} pods must remain operational under optical faults~\cite{zu2024resiliency}.
We adopt a conservative model consistent with reality: an OCS fault disables all links routed through that OCS, the fault is detected, and broadcast before job execution, and routing tables are selected accordingly~\cite{tpuv4}.

We augment all-path discovery (\autoref{alg:allturns}) so that the candidate path set contains deadlock-free backups under any single OCS failure.
A sufficient condition is the existence of $t$ \emph{OCS-disjoint} spanning trees connecting all cubes then at most $t-1$ OCS faults can be tolerated while preserving connectivity via a tree that avoids the failed OCS.
We use Nash-Williams~\cite{nash1961edge} as a sufficient condition for $t$ edge-disjoint spanning trees and extend the argument under \tpu ICI-specific OCS connections.
In our setting, we connect this condition to a lower bound on the number of distinct OCS links crossing any partition as a function of MCF, yielding a conservative requirement of the form $\lambda \ge \frac{t}{32n}$ for $n$ nodes and $t$ OCS-disjoint spanning trees. We derive the complete proof of this property in~\autoref{ap:ocs_fault} but the proof sketch using worst-case analysis is as follows. The MCF, $\lambda$, is tighter than the cut bound and implies the minimum number of edges leaving a cube or groups of cubes and similarly, the lower bound on number of OCS distinct edges. We then perform algebra on Nash-Williams condition to derive a conservative lower bound on $\lambda$ in terms of the number of nodes and possible faults.

During topology generation, we encode fault-tolerance as constraint \constrFault in~\autoref{tab:mip_form} for a user-specified budget for number of tolerable faults, $f$, via $f=t-1$. %
All synthesized~\nameshort{} topologies satisfy this property.
We label this fault-tolerance scheme~\emph{robust}; it is implemented in the AT stage lines 4--8 of~\autoref{alg:allturns}.
For the single-fault model ($f=1 \implies t \ge 2$), we find two OCS-disjoint spanning trees using a concurrent BFS rooted at hop-distance antipodes: the trees are grown simultaneously while marking an OCS as consumed when used by one tree\footnote{We note that constructing many disjoint trees efficiently is nontrivial. We suspect matroid intersection extensions of Edmonds' algorithm are appropriate~\cite{edmonds1965maximum} but leave this for future work.}.
With a~\emph{robust} allowed turns set, $A$, sets of all paths are generated for each possible fault and routed using the ILP (\autoref{sec:design-routing_ilp}) similar to Google's WFR~\cite{zu2024resiliency}.

\subsection{Routing}
\label{sec:design-routing_ilp}

Given the deadlock-free candidate path set produced by AT, we must select exactly one path per source--destination pair (static forwarding).
We choose routes to minimize congestion using \emph{maximum channel load} as the objective, since it directly upper-bounds achievable uniform throughput: if the most-loaded channel carries $L_{\max}$ routes, then under uniform demand the per-flow rate is at most $\nicefrac{1}{L_{\max}}$.
This matches the routing objective used for PDTT in Zu~\etal~\cite{zu2024resiliency} and is standard in throughput-oriented routing formulations~\cite{towles2003thruputrouting,netsmith}.

Our ILP uses one indicator variable per candidate path; selecting a path increments the load variables of every edge on that path.
The solver minimizes the maximum channel load.
As in prior work, the quality of the solution depends on the provided candidate set: shortest-path sets are often sufficient on regular tori, while the AT-generated minimal all paths set is essential for irregular topologies to preserve deadlock-freedom under VC constraints.
Empirically, for both torus baselines and \nameshort{} topologies, the routed bound is typically tight relative to analytical continuous upper bounds, indicating that routing is not the dominant limiter in most configurations (\autoref{sec:results_routing}).

\putsubsec{sec:design-routing_vcalloc}{VC Allocation}
After route selection, the chosen paths and allowed turns list are used to allocate VCs to turns in the path. We consider each path individually and perform a breadth-first search along the complete CDG to find the allowed VC transitions/selections. A na\"{\i}ve approach biases towards VC 0. The BFS starts at VC 0 and traverses along VC 0 until a turn is disallowed when it transitions to VC 1 and continues trying to allocate to VC 0. However, we found this caused significant load imbalance between the VCs so we propose an online load balancing algorithm. As turns are allocated to VCs, the count of hops per VC is maintained. Before beginning VC allocation for the next path, the VC with the lowest hop count is marked ``priority'' and the BFS checks that VC for all turns in that path. We find this achieves near perfect balance (\autoref{sec:results_routing}).

\putsubsec{design-routing_optimality}{Empirical Validation of Approach}

\begin{figure}[t]
\footnotesize 
\includegraphics[width=0.95\linewidth]{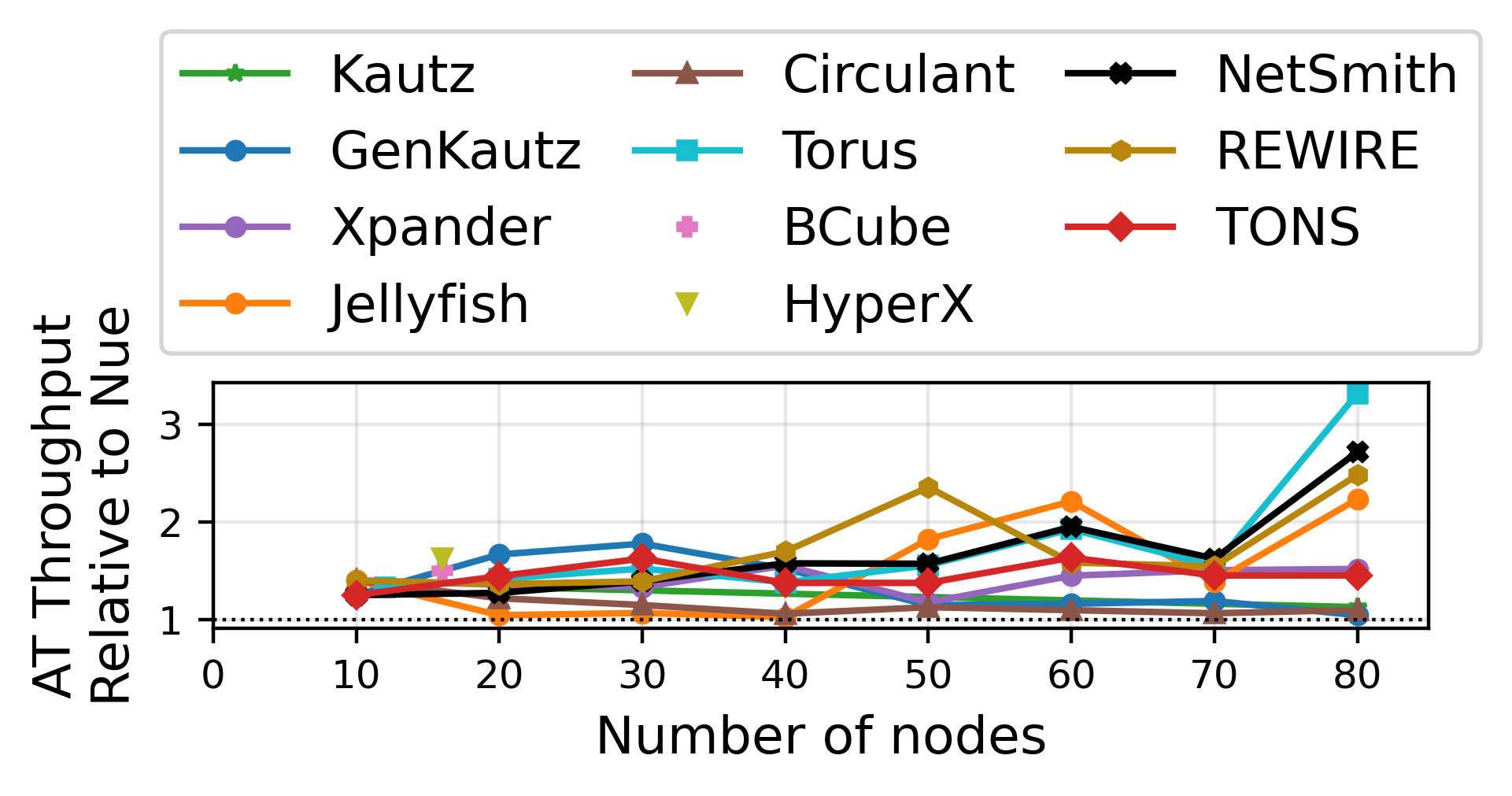}
\captionsetup{skip=2pt}
\caption{Throughput of~\nameshort{} AT relative to Nue static, single path routing using two VCs.}
\label{fig:routing_compare}
\vspace{-1.2em}
\end{figure}

We evaluate the performance of~\nameshort{}'s AT routing approach versus comparable literature\footnote{We omit comparison to Up*/Down* because it severely under-performs}, Nue~\cite{domke2016nue}, which was modified to work for directed graphs.
The topologies evaluated in~\autoref{sec:design-topo_optimality} were routed by~\nameshort{} AT and Nue. As expected, AT outperforms for all data points and seems to improve the gap with scale. AT's performance is unsurprising considering the ILP optimization approach and that Nue routing functions are in the possible solution space of AT.
This provides empirical grounding that~\nameshort{} AT is the globally superior approach to VC-limited routing.

\putsubsec{complexity}{Discussion}

While the \nameshort{} pipeline (\autoref{fig:total_diagram}) is both more complicated than heuristic approaches (e.g., simple mesh or torus) and requires more compute time and resources, the effort is automated and amortized.
The network can be designed per topology size offline and re-used for as long as the target system exists.
Most importantly, if the target system changes, \nameshort{} can be modified and return high-performance networks, without human time and effort.

Admittedly, \nameshort{} does not generate easily comprehensible networks; however, their out-performance imply the utility is worth the algorithmic complexity.
A concrete example is the \tpu use-case. Google, a major hyperscalar, implicitly argues that the performance is worth the complexity.
PT's critical cut-based throughput bounds reduces \alltoall{} throughput and increases latency/diameter too severely.
The PDTT topology is the first step towards complexity because it allows DOR but requires an ILP solver for routing~\cite{zu2024resiliency}. Recently, for low-latency inference serving, Google has introduced an even more complicated topology, ``Boardfly,'' for the explicit purpose of reducing diameter~\cite{tpuv8announcement}.

The presence of link/OCS faults \emph{requires} the routing, deadlock avoidance, collective communication, and workload mapping algorithms support arbitrary topologies.
When a link or OCS fails, the topology becomes arbitrarily connected without link connection invariants.
Even the simplest PT topology requires a complicated fault tolerance scheme, WFR, that tries to graft human intuition onto what should be an algorithmic design space.
Furthermore, while tori may have nice properties for collectives (e.g., multidimensional rings~\cite{de2024swing,juerss2026trivance}) and mapping workloads onto the topology (e.g., the XLA library~\cite{openxla2024xla}), both of these break down during faults, which occur at high probability for large systems~\cite{tpuv4}. Therefore, Google has necessarily already implemented algorithmic routing, deadlock avoidance, collective communication, and workload mapping for a system of this scale.
We simply propose utilizing those degrees of freedom to create better networks.

\putsec{method}{Evaluation Methodology}

We evaluate \nameshort{} using complementary analytical metrics and cycle-level simulation.
Our experiments target the paper’s three claims: (i) synthesized topologies improve throughput-oriented proxies for \tpu ICI constraints, (ii) allowed-turn routing with a small VC budget preserves most of the attainable throughput while remaining deadlock-free, and (iii) these gains translate to~\alltoall{}-style communication without degrading~\allgather{}/\allreduce{}.
Because TPU pods are not externally reconfigurable for arbitrary topologies or routing in practice~\cite{tpuv4documentation}, we rely on analytical evaluation and simulation.

\begin{table}[t]
\caption{Simulation Parameters}
\vspace{-.75em}
\label{tab:params}
\begin{tabular}{|l|l|}
\hline
\textbf{Parameter}  & \textbf{Value}      \\ \hline
Clock frequency     & 1.05\,GHz                            \\ \hline
Link bandwidth      & 128\,GB/s (unidirectional)          \\ \hline
Link latency        & 50/25  cycles electrical/optical    \\ \hline
Router/injection  latency      & 25 cycles                   \\ \hline
Router radix       & 6                  \\ \hline
Flit width          & 128\,B               \\ \hline
Total \& escape VCs           & 4 \& 2              \\ \hline
Buffer slots per VC & 200 flits        \\ \hline
\end{tabular}
\vspace{-1.4em}
\end{table}

\begin{figure*}[ht]
\footnotesize 
\includegraphics[width=0.95\linewidth]{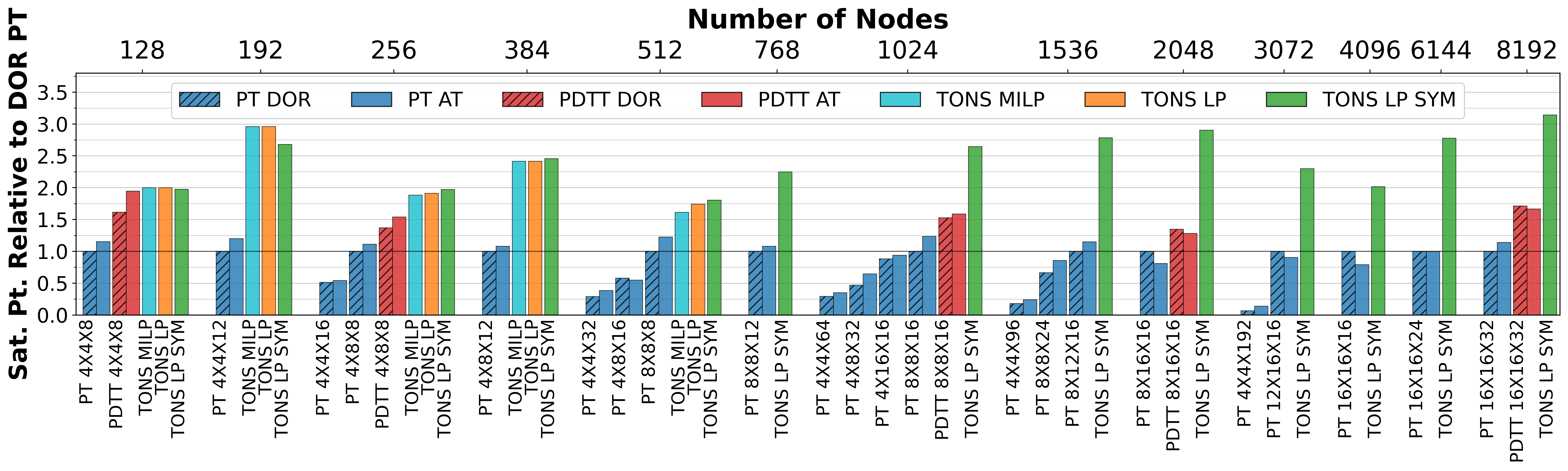}
\vspace{-0.175in}
\caption{Relative saturation points (higher is better) for uniform random traffic simulations normalized to the best PT and DOR (blue hashed). DOR (hashed) and AT (solid) routing were applied to tori while~\nameshort{} uses AT.}
\label{fig:sat_pts}
\vspace{-1.2em}
\end{figure*}

\textbf{Simulation Parameters} --
We tune parameters to match published \tpu{} values where available: TPU v5p clock and link bandwidth rounded to the integer flits per cycle~\cite{tpuv5specs}.
For link latency, we conservatively estimate propagation from published system imagery~\cite{tpuv4} and roughly assume injection latency.
Buffers are sized to sustain the modeled bandwidth delay product with margin.
All simulator parameters follow~\autoref{tab:params}.

\putsubsec{method_simulation}{Synthetic Traffic}

We simulate selected configurations using Chiplet Network Simulator (CNSim)~\cite{feng2024cnsim}, a cycle-accurate, packet-parallel simulator validated against gem5~\cite{gem5} and BookSim2.0~\cite{jiang2013booksim} for synthetic traffic.
We extend CNSim to ingest arbitrary adjacency matrices, static routing tables, and per-flow escape-VC assignments, using four VCs total and reserve two as deadlock-free escape VCs~\cite{duato}.

\textbf{Saturation Point} --
We measure network throughput by simulating uniform random traffic in CNSim~\cite{feng2024cnsim}.
CNSim sweeps injection rates and reports saturation at the first observed timeout (flits sent $>$ flits received).
We use an injection-rate step of 0.01.

\textbf{Fault Tolerance} --
Our fault model follows Google's description for \tpu{} pods~\cite{tpuv4}: a fault disables all links routed through one OCS, at most one OCS faults at a time, and the fault is known before job execution.
For each of the 48 single-OCS fault scenarios, we load the corresponding fault-avoiding routing tables and measure the saturation point under the same traffic and simulator settings.

\putsubsec{method_coll-comm}{Collective Communication}
We evaluate~\allgather{},~\allreduce{}, and~\alltoall{} using existing schedulers.
For~\allgather{} and~\allreduce{}, we use MultiTree~\cite{huang2021multitree} (scales to our largest topologies and achieves near-optimal schedules).
For~\alltoall{}, we use the optimization-based formulations of Basu~\etal~\cite{basu2024efficientalltoall}: decomposed-MCF at sizes where it scales, and pMCF (with symmetry) at larger sizes.
We report schedule quality as link utilization (epochs under cumulative link bandwidth).
For a subset of configurations, we translate link-by-link transfer schedules into traces (MSCCLang-style XML~\cite{cowan2023mscclang} and netrace~\cite{hestness2011netrace}) and simulate them in CNSim~\cite{feng2024cnsim} to validate analytical expectations.

\putsubsec{method_e2e}{End-to-End Simulation}

We use \astrasim{}'s congestion-aware network model to evaluate training time under compute--communication overlap and different parallelization strategies~\cite{won2023astrasim2}, using the parameters in \autoref{tab:params}. \astrasim{} executes Chakra DAGs of compute and communication operations~\cite{sridharan2023chakra}. We model Chakra compute nodes using \astrasim{}'s roofline model with TPU v5p parameters. We expand Chakra collective nodes into topology-aware collective schedules (\autoref{sec:method_coll-comm}) as point-to-point transfers through the collective API library~\cite{yoo2025collectiveapi}.

Because public traces either do not scale to hundreds of TPUs or contain small collective groups~\cite{sridharan2023chakra,hawks2025mlcommons,spcl_atlahs,shen2025atlahs}, we use STAGE to synthesize Chakra traces from model architectures and parallelization strategies~\cite{man2026stage}.
We evaluate three communication profiles: Llama-3.1-8B (dense LLM)~\cite{grattafiori2025llama3}, MLPerf-DCNv2 (DLRM)~\cite{wang2021dcnv2} with Neo column-wise sharding~\cite{mudigere2022dlrm}, and Qwen3-Next-80B-A3B (MoE)~\cite{qwen3technicalreport,qwen2.5-1m}.
For MoE dispatch/combine, we assume ideal token load balancing and use~\alltoall{}, not variable~\alltoall{}, collectives as modeling data-dependent expert selection is outside our scope.

\putsec{results}{Results}

We evaluate \nameshort{} via analytical metrics, cycle-accurate simulation, and event-driven simulation.
We report saturation for uniform random traffic, collective-communication utilization, end-to-end simulation, robustness to OCS faults, and routing ablations that justify our AT turn prioritization and VC load balancing.

\putsubsec{results_saturation}{Saturation Point}

We quantify throughput using uniform random traffic simulation, reported as the saturation point.
\autoref{fig:sat_pts} compares baseline structured topologies (PT and PDTT) against \nameshort{}-generated designs across a wide range of node counts and job shapes.
For each configuration we evaluate tori with both (i) baseline DOR where applicable, and (ii) \nameshort{}’s allowed-turn (AT) routing, isolating the contribution of routing from that of topology.

\nameshort{} topologies deliver large throughput gains across scales.
Depending on configuration,~\nameshort{} improves saturation by roughly $1.6\times$--$3.1\times$ over the best structured torus baseline.
Overall,~\nameshort{} has a geometric mean improvement of 2.07$\times$ over the current baseline.
For sizes/configurations without PDTT,~\nameshort{} has a 2.39$\times$ higher saturation point over the best PT and for sizes with PDTT, a 1.65$\times$ improvement.
These gains not only persist but increase at the largest evaluated sizes (up to 8192 nodes), validating scalability.

Applying AT routing to a baseline torus yields only modest improvements (typically on the order of 1.1--1.2$\times$). The improvements of AT with respect to DOR are due to VC imbalances as explored in~\autoref{sec:results_routing}. The performance loss of AT for PDTT at 2048 and 8192 nodes is likely due to using the random variant of AT and that DOR retains ideal throughput (but poor VC balance). At those large sizes, the random AT loses enough performance that VC balance cannot overcome.
Finally, the~\nameshort{} LP SYM results match or exceed the non-symmetric LP solutions across large node counts, supporting the claim that symmetry can be used as a scalability lever. By scaling better, it improves throughput for generation with equal time and compute resources.

\putsubsec{results_collectives}{Collective Communication}

{
\setlength{\abovedisplayskip}{0pt}%
\setlength{\belowdisplayskip}{0pt}%
\setlength{\abovedisplayshortskip}{0pt}%
\setlength{\belowdisplayshortskip}{0pt}%
\begin{figure}[t]
\footnotesize 
\includegraphics[width=0.99\linewidth]{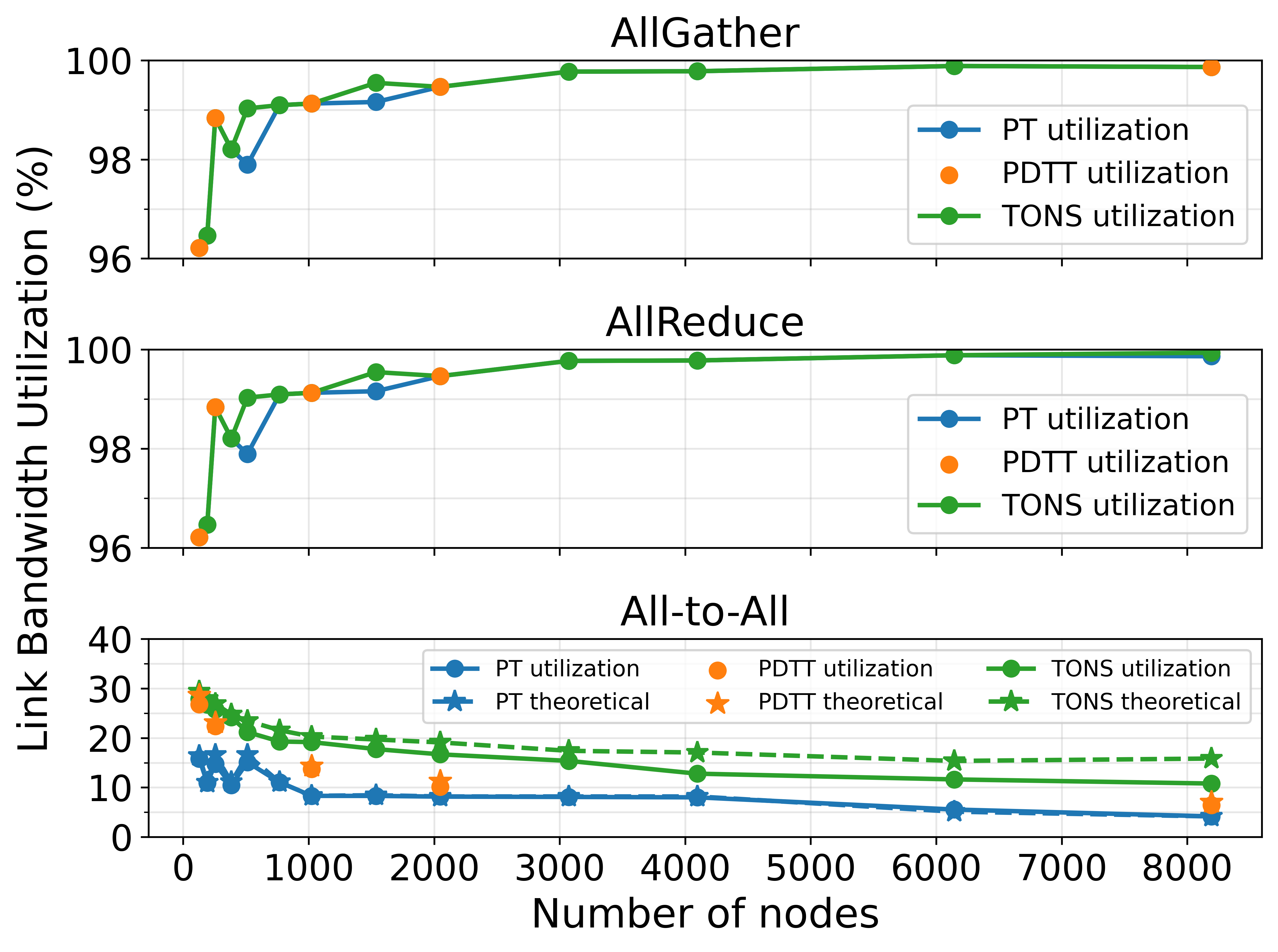}
\vspace{-0.175in}
\caption{Link utilization of generated schedules for~\allgather{} (top),~\allreduce{} (middle), and~\alltoall{} (bottom) for the best PT (blue), PDTT (orange), and~\nameshort{} (green).~\allgather{} and~\allreduce{} have a 100\% theoretical upper bound; for~\alltoall{}, the limit is derived from MCF (dashed). }
\label{fig:analytical_coll_comm}
\vspace{-1.6em}
\end{figure}
}

{
\setlength{\abovedisplayskip}{0pt}%
\setlength{\belowdisplayskip}{0pt}%
\setlength{\abovedisplayshortskip}{0pt}%
\setlength{\belowdisplayshortskip}{0pt}%
\begin{figure}[t]
\footnotesize 
\includegraphics[width=0.98\linewidth]{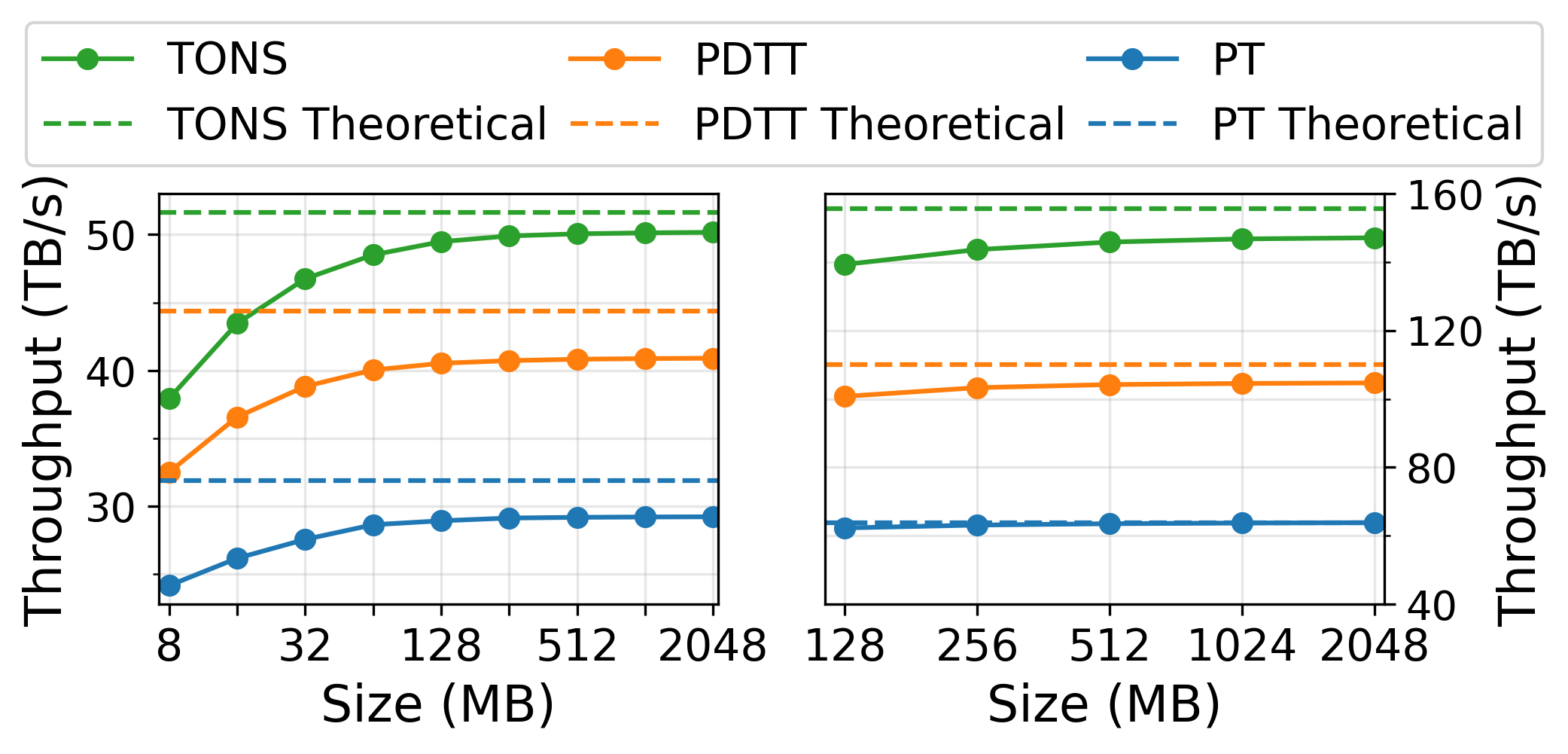}
\vspace{-0.175in}
\caption{Cumulative network throughput under trace-driven simulation (CNSim) for 256-node (left) and 1024-node (right) PT (blue), PDTT (orange), and~\nameshort{} (green), sweeping collective/buffer sizes (log scale) until saturation. MCF-based upper bound (dashed) included.}
\label{fig:coll_comm_simul}
\vspace{-1.0em}
\end{figure}
}

We evaluate~\allgather{},~\allreduce{}, and~\alltoall{} using the scheduling methodology in~\autoref{sec:method_coll-comm}.
For simplicity, we evaluate~\nameshort{} LP SYM and the best PT per size as representative samples.
\autoref{fig:analytical_coll_comm} shows that all designs achieve near-ideal utilization for~\allgather{} and~\allreduce{}, consistent with prior observations that these collectives admit highly efficient schedules on regular low-diameter fabrics~\cite{tpuv4,huang2021multitree}.
Importantly, \nameshort{} does not sacrifice performance\footnote{
We suspect the few data points of higher utilization are due to diameter and average hop differences affecting the greedy MultiTree~\cite{huang2021multitree} algorithm.}

For~\alltoall{},~\nameshort{} consistently achieves higher utilization than PT/PDTT, tracking the higher topological (MCF-based) limit, indicating improved concurrent throughput under the most communication-intensive pattern targeted by our synthesis objective.
\autoref{fig:coll_comm_simul} substantiates the schedule-based analysis with trace-driven simulation for two representative topology sizes, 256 and 1024 nodes, where the smallest possible\footnote{128\,B flits imply minimums.} collectives are not immediately in saturation.
The networks soon reach saturation where~\nameshort{} networks show 9\,TB/s and 47\,TB/s higher cumulative throughput for 256 and 1024 nodes, respectively, exactly matching analytical expectations.

\putsubsec{results_e2e}{End-to-End}

\begin{figure}[t]
    \centering
    \begin{subfigure}[b]{0.5\linewidth}
        \centering
        \includegraphics[width=\linewidth]{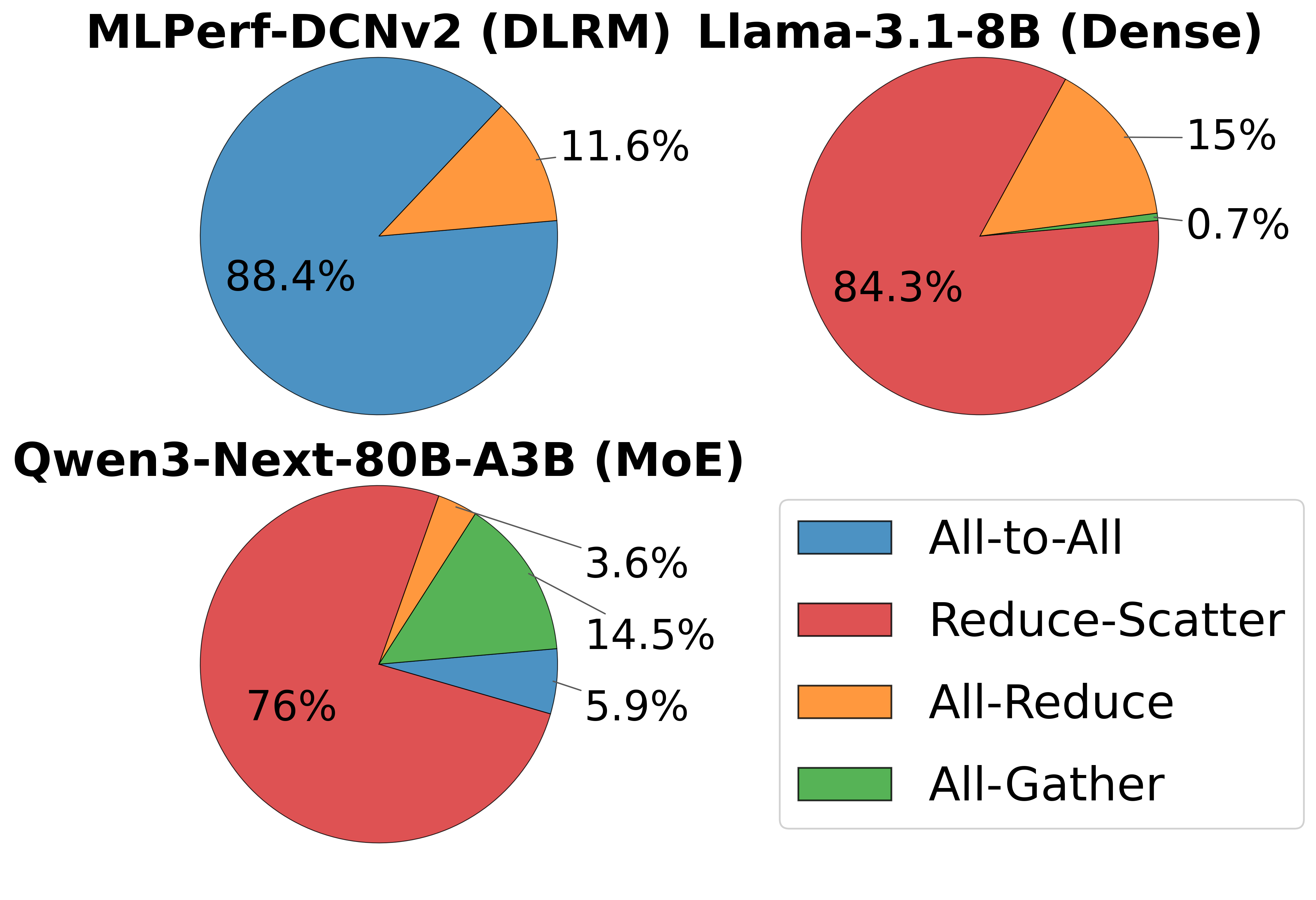}
        \caption{Breakdown of communication types weighted by bytes transferred.}
        \label{fig:e2e_bytes}
    \end{subfigure}%
    \hfill
    \begin{subfigure}[b]{0.47\linewidth}
        \centering
        \includegraphics[width=\linewidth]{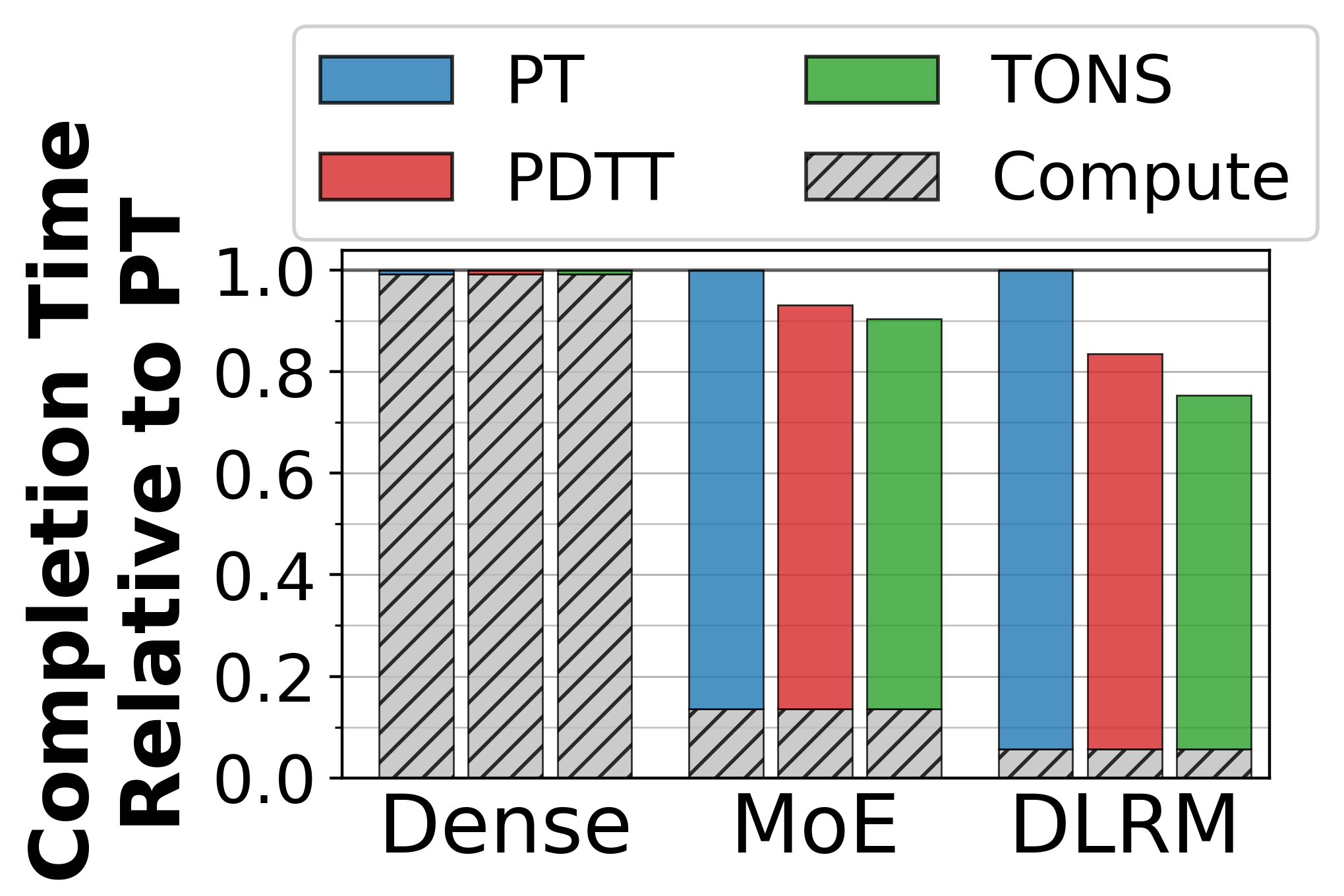}
        \caption{End-to-end completion time (lower is better) relative to PT. Compute ratio is unaffected by topology.}
        \label{fig:e2e_time}
    \end{subfigure}
    \captionsetup{skip=2pt}
    \caption{Communication workload (a) and completion time (b) for dense LLM, DLRM, and MoE training traces on a 256-node topologies.}
    \label{fig:e2e}
    \vspace{-1.0em}
\end{figure}

As described in~\autoref{sec:method_e2e}, we generate synthetic training traces for three diverse AI/ML models: Llama3.1-8B, DCNv2, and Qwen3-Next and measure their distributed training performance on 256-node PT, PDTT, and~\nameshort{} topologies. \autoref{fig:e2e_bytes} plots the distribution of collective communication.
For DCNv2, we use column-wise sharding which results in mostly~\alltoall{} traffic.
For Llama3, we use 256 way fully sharded data parallelism which incurs~\reducescatter{} on the backwards pass~\cite{zhao2023fsdp}.
Because the choice of optimal parallelism is an open research question~\cite{isaev2023calculon}, we do a rough sweep of expert (EP) and tensor parallelism (TP) dimensions and select the best performing configuration, which, for Qwen3-Next, was 32 way EP and 8 way TP with collective distribution in~\autoref{fig:e2e_bytes}.

We plot the end-to-end completion time for the training in~\autoref{fig:e2e_time}. As demonstrated by~\autoref{fig:analytical_coll_comm}, we expect~\nameshort{} to only improve~\alltoall{} communication. Therefore, we see speedups limited by Amdahl's law based on exposed,~\alltoall{} communication. Llama3's high compute and lack of~\alltoall{} results in no speedup. Qwen3-Next is a middleground where we see~\nameshort{} improve end-to-end performance by 10\% over PT. Finally,~\nameshort{} speeds up DLRMs---the original motivating use-case from Google~\cite{tpuv4}--- by 30\% and 7\% over PT and PDTT state-of-the-art topologies.
These results validate expectations derived from individual collective analysis. Previous results (\autoref{fig:sat_pts}--\autoref{fig:coll_comm_simul}) imply that speedups
would continue to grow at larger scales.

\putsubsec{results_fault}{Fault Tolerance}

{
\setlength{\abovedisplayskip}{0pt}%
\setlength{\belowdisplayskip}{0pt}%
\setlength{\abovedisplayshortskip}{0pt}%
\setlength{\belowdisplayshortskip}{0pt}%
\begin{figure}[t]
\centering
\vspace{-0.4em}
\footnotesize 
\includegraphics[width=0.95\linewidth]{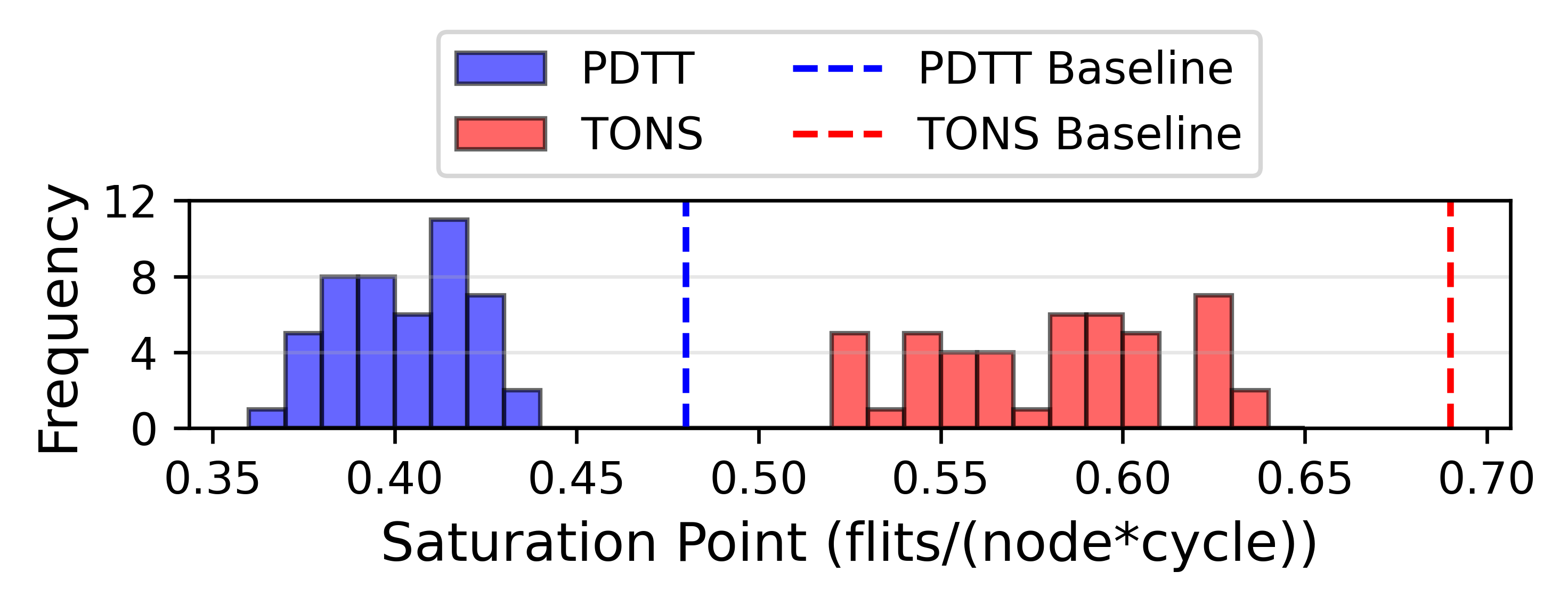}
\captionsetup{skip=1pt} %
\caption{Saturation points for all possible OCS faults for 256 node baseline PDTT WFR (blue) and \nameshort{}~\emph{robust} AT (red). The baseline, no OCS fault saturation point for each is shown by the dotted line.}
\label{fig:ocs_fault}
\vspace{-1.6em}
\end{figure}
}

The performance of the routing technique for jobs with OCS faults was measured by their saturation points for each of the 48 possible single OCS faults.
We compare a 256 node PDTT using WFR (\autoref{sec:background_tpu}) against~\nameshort{} using~\emph{robust} routing and include each design’s no-fault saturation point as a reference.
The histograms of the saturation points are plotted in~\autoref{fig:ocs_fault}.
\autoref{fig:ocs_fault} shows two clear outcomes.
First,~\nameshort{} sustains substantially higher absolute throughput than PDTT under every fault scenario, indicating that the~\emph{robust} AT routing is not sacrificing significant performance.
Second, while both designs experience degradation under faults,~\nameshort{}’s distribution remains centered at a much higher saturation point; qualitatively,~\nameshort{} shifts the entire fault-throughput, leaving significant headroom even in the degraded regime.

\putsubsec{results_routing}{AT Routing}

We isolate \nameshort{} routing choices: turn-prioritization heuristics within the AT framework and the greedy VC load-balancing mechanism relative to DOR on torus baselines.

\begin{figure}[t]
\centering
\footnotesize 
\includegraphics[width=0.99\linewidth]{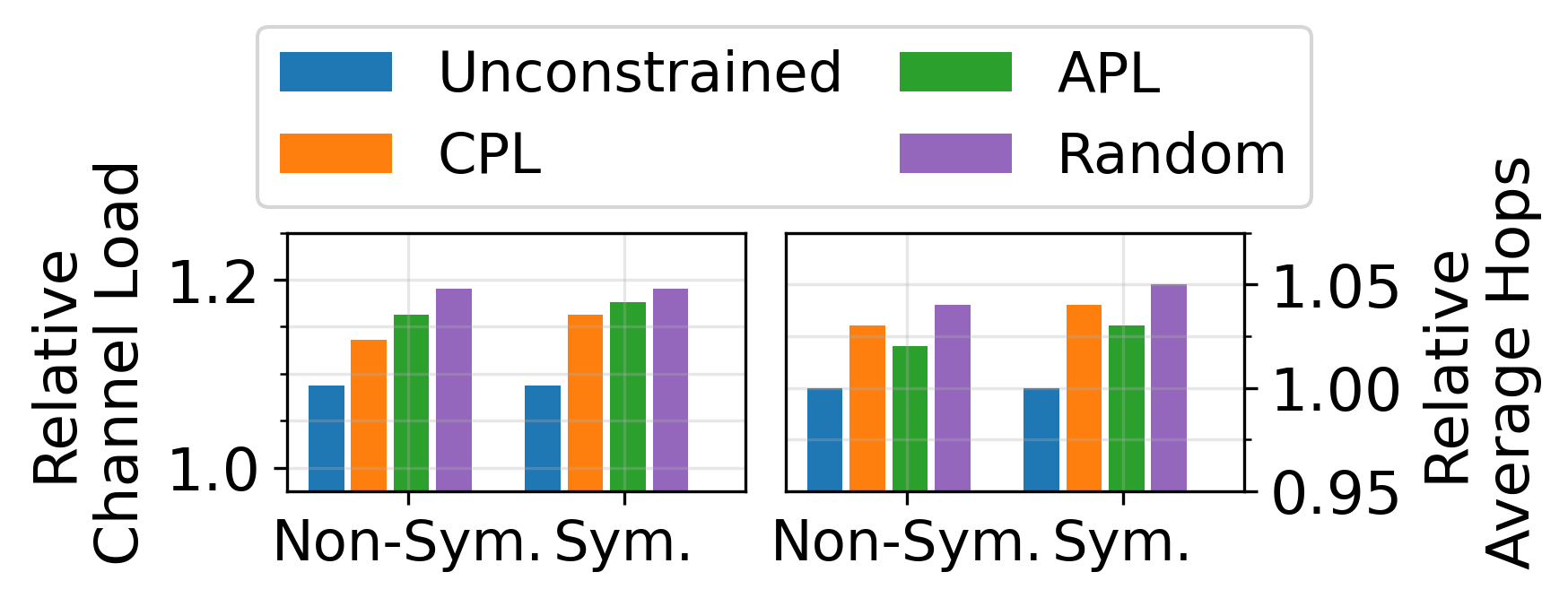}
\captionsetup{skip=2pt}
\caption{Isolation of AT turns prioritization measuring (left) maximum channel load and (right) average hops (both, lower is better) relative to topology bounds.
}
\label{fig:at_priority_iso}
\vspace{-1.8em}
\end{figure}

\textbf{Turn prioritization and symmetry} --
\autoref{fig:at_priority_iso} reports maximum channel load (left) and average hops (right), both normalized to topological bounds, for AT variants and a (likely deadlocky) unconstrained routing function.
Two takeaways matter for the paper’s claims.
First, symmetry-reduced routing preserves throughput: both variants achieve similar normalized performance across all prioritization schemes.
Second, enforcing deadlock freedom via AT incurs low loss relative to unconstrained routing and the choice of prioritization affects that loss.
Our preferred turn priority scheme, CPL, only incurs 5\% higher concentrated channel load relative to an unconstrained variant.
By contrast, weaker prioritization (e.g., random) both reduces throughput and increases hops.

\begin{figure}[t]
\centering
\footnotesize 
\includegraphics[width=0.9\linewidth]{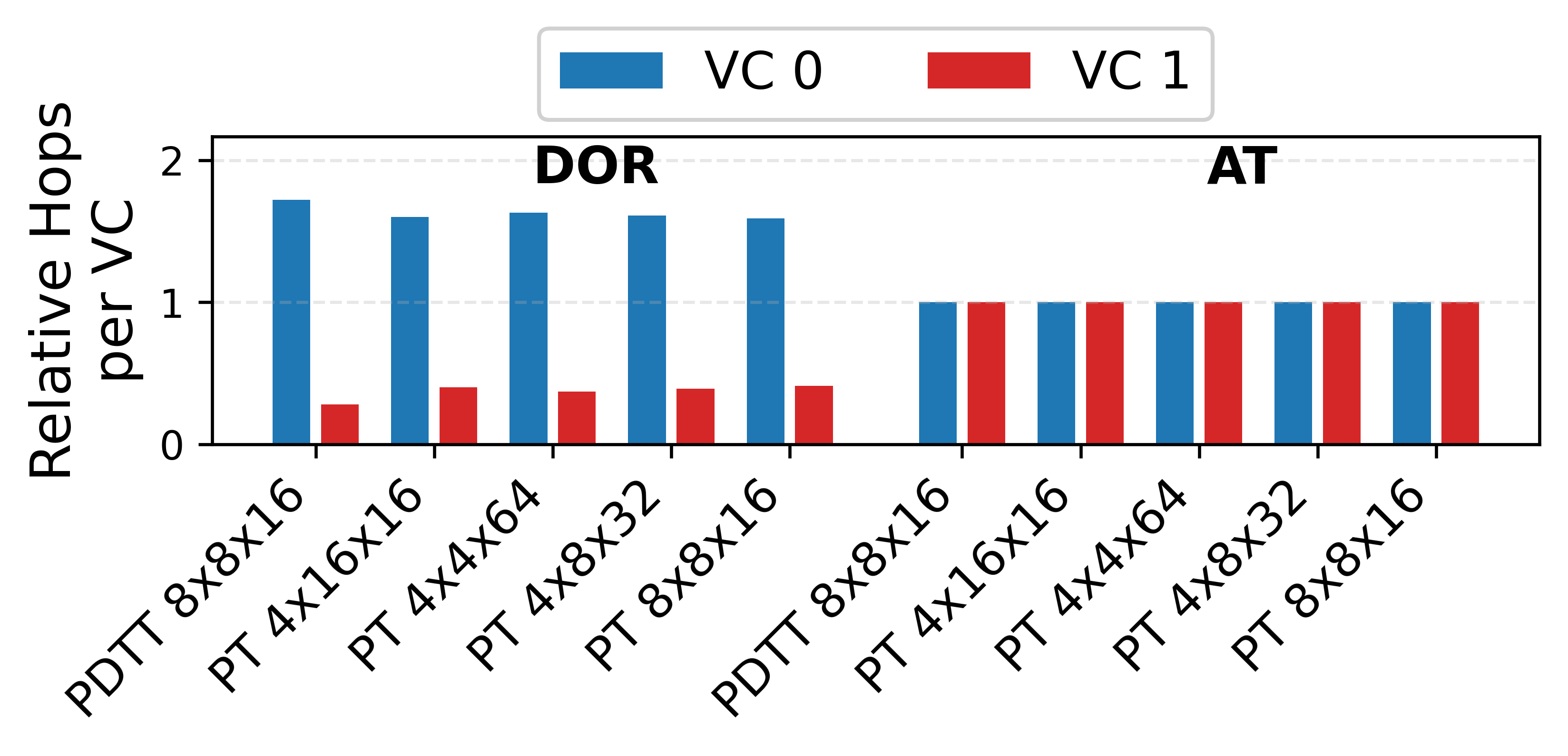}
\captionsetup{skip=1pt} %
\caption{Analytical evaluation of the number of hops per VC for load balanced AT versus DOR.%
}
\label{fig:dor_at_vc_lb}
\vspace{-1.2em}
\end{figure}

\textbf{VC load balancing} --
We compare VC utilization between traditional DOR and AT on torus baselines.
While DOR and AT can have similar max channel load on regular tori, their VC occupancy differs substantially because DOR’s datelining structure often concentrates traffic into a subset of VCs.
\autoref{fig:dor_at_vc_lb} shows that DOR skews hops heavily toward VC 0, leaving VC 1 underutilized, whereas AT achieves substantially more balanced VC usage by construction, indicating that the online algorithm effectively spreads paths/flows across the available VC budget.

\putsec{conclusion}{Conclusion}

We demonstrate that optimization-driven network synthesis can produce large-scale networks that outperform established structured designs under throughput-oriented demand.
\nameshort{} generates direct topologies via linear programming guided by a flow proxy and routes with deadlock-free, static routing under with limited VCs. Across analytical metrics and cycle-level simulation, \nameshort{} delivers higher sustained throughput and remains robust to the targeted faults.
Broadly, these results suggest that \nameshort{} can improve current accelerator fabrics and remains adaptable to new constraints in next generation computing.

\newpage
\clearpage 

\appendix
\newpage

\section{Symmetry Definitions}
\label{ap:symmetry}

\noindent\textbf{Transformation.}
Let $G=(V,E)$ be a (di)graph whose vertices lie on a 3D grid with dimensions
$d=(d_0,d_1,d_2)$, i.e., each vertex $v\in V$ has coordinates
$x(v)\in \mathbb{Z}_{d_0}\times\mathbb{Z}_{d_1}\times\mathbb{Z}_{d_2}$.
We model topology symmetries using axis reflections (enabled per-dimension by a mask $r\in\{0,1\}^3$)
and translations by a displacement vector $\delta$.
The combined transform $g_{\delta,r}$ applies a reflection followed by a translation:

\begin{equation}
\label{eq:transforms}
\begin{aligned}
R_{r}(x)_a &=
\begin{cases}
-x_a \bmod d_a, & r_a = 1,\\
\phantom{-}x_a, & r_a = 0,
\end{cases}\\
T_{\delta}(x) &= x+\delta \pmod{d}, \\
g_{\delta,r}(x) &\triangleq T_{\delta}(R_{r}(x)).
\end{aligned}
\end{equation}

\noindent\textbf{Canonical set.}
We choose a small set of \emph{canonical} vertices $C\subseteq V$ (e.g., one representative cube/region),
and a deterministic tie-break rule (e.g., smallest linearized coordinate).
For any vertex $v$, we define its canonical representative as the image of $v$ that lands in $C$:
\[
\mathrm{can}(v) \triangleq g_v(v), \qquad
g_v \in \arg\min_{\substack{g_{\delta,r}:\\ g_{\delta,r}(v)\in C}} \mathrm{key}\!\left(g_{\delta,r}(v)\right).
\]
In words: $g_v$ is the specific reflection+translation that maps $v$ into the canonical set, and
$\mathrm{can}(v)$ is the resulting canonical vertex.

\noindent\textbf{Edge symmetry.}
We call $G$ \emph{edge-symmetric} (under these transforms) if applying the same transform to both
endpoints preserves edges:
\[
(u,v)\in E \;\Rightarrow\; (g(u),g(v))\in E \quad \text{for all transforms } g=g_{\delta,r}.
\]
Equivalently, for any edge $(v,w)\in E$, mapping $v$ to its canonical representative also maps the edge:
\[
(v,w)\in E \;\Rightarrow\; \big(\mathrm{can}(v),\, g_v(w)\big)\in E.
\]

\section{WFR Deadlock Freedom Counterexample}
\label{app:wfr_counterexample}

This appendix presents a small two-dimensional counter-example illustrating that
\emph{wild-first routing} (WFR), if implemented as ``DOR + extra wild turns'' without
a carefully enforced virtual-channel (VC) discipline, can reintroduce cyclic channel
dependencies and therefore admits wormhole deadlock.

\begin{figure}[ht]
    \centering
    \includegraphics[width=0.95\linewidth]{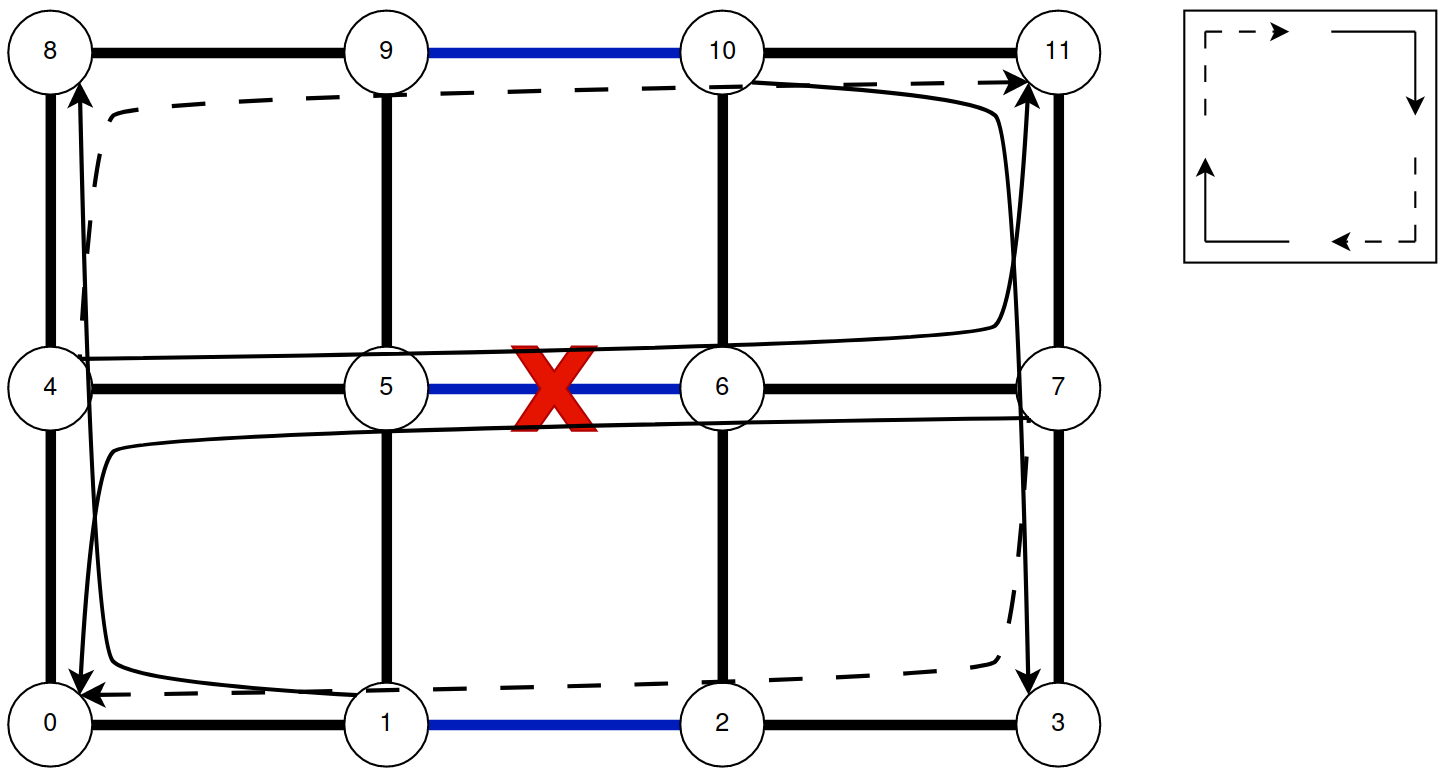}
    \caption{Counter-example topology and routes. Black: electrical links. Blue: optical links.
    Red $\times$: failed optical link. Solid directed arcs: nominal 2D DOR (XY) routes.
    Dashed arcs: candidate WFR detours. Inset: turn-dependency sketch; solid segments depict
    the baseline (DOR) turn restrictions, dashed segments depict additional turn dependencies
    introduced by allowing the WFR detours.}
    \label{fig:wfr_deadlock_counterexample}
\end{figure}

\paragraph{Topology and link types.}
Figure~\ref{fig:wfr_deadlock_counterexample} shows a $3\times 4$ 2D slice.
Thick black edges are the baseline electrical mesh links. Blue edges represent
optical links (conceptually, cross-cube/OCS connectivity). The red $\times$ marks
a single unavailable optical link.
Nodes are labeled for reference only.

\paragraph{Baseline DOR routing.}
Under standard 2D dimension-order routing (DOR) with order $X \rightarrow Y$,
packets traverse $X$ first and then $Y$ (modulo the usual \emph{sign} choices for
shortest-path wrap-around in a torus, or the boundary choices in a mesh).
The important property for deadlock freedom in wormhole routing is that the set
of \emph{allowed turns} induces an acyclic channel-dependency graph (CDG): the
routing function never permits a sequence of channel acquisitions that forms a
directed cycle. Intuitively, this is what ``ordering dimensions'' achieves: once a
packet has transitioned from $X$-channels to $Y$-channels, it never transitions back.

\paragraph{Fault handling via WFR and the new turns it introduces.}
When the failed link lies on the unique (or chosen) minimal DOR path for some
source--destination pairs, WFR introduces detours by allowing a packet to take
a bounded number of \emph{wild} hops in another dimension and then revert to DOR.
In 2D, this commonly manifests as a ``one hop in $Y$ followed by $XY$ DOR'' pattern
(e.g., ``yXY'' in the paper's shorthand). In Fig.~\ref{fig:wfr_deadlock_counterexample},
the solid arcs illustrate nominal DOR paths that would traverse the failed optical
edge, while the dashed arcs show representative WFR alternatives that bypass the
fault by stepping into the other row and then continuing.

Crucially, enabling these WFR detours is not a benign local change: it adds new
\emph{turns} at the routers where a packet transitions from the wild hop back into
the DOR phase. In the 2D case, this typically introduces a $Y \rightarrow X$ turn
at the ``re-entry'' router, even though baseline $X \rightarrow Y$ DOR already
permits the complementary $X \rightarrow Y$ turn elsewhere (when a packet finishes
its $X$ progress and begins its $Y$ progress). The inset of
Fig.~\ref{fig:wfr_deadlock_counterexample} sketches this phenomenon: solid segments
represent the baseline allowed-turn structure, while dashed segments represent
additional turn dependencies required to realize the WFR detours.

\paragraph{Deadlock mechanism.}
In wormhole routing, a blocked packet holds input buffers (and often the upstream
channel/credits) while requesting its next output channel; a \emph{directed cycle}
of such holds-and-requests implies deadlock. The turn-model viewpoint is that if
the permitted turn set completes an abstract cycle (e.g., a four-turn cycle in a
2D plane), then the CDG can contain a directed cycle, and wormhole deadlock becomes
possible.

Our counter-example exploits exactly this: with the failed optical edge, multiple
flows are forced onto the WFR detours (dashed routes). Consider four packets whose
routes occupy the four sides of the perimeter ``ring'' induced by these detours
in Fig.~\ref{fig:wfr_deadlock_counterexample}. Each packet holds one channel class
(e.g., an east-going segment) and requests the next channel class at the corner
router (e.g., a north-going segment) according to the newly enabled WFR turn.
Because WFR reintroduces the ``forbidden'' direction transition (conceptually,
$Y \rightarrow X$) while baseline DOR already allows $X \rightarrow Y$, the four
corner transitions collectively form a directed dependency cycle:
\[
    (X\text{-class}) \rightarrow (Y\text{-class}) \rightarrow (X\text{-class}) \rightarrow (Y\text{-class}) \rightarrow \cdots
\]
In steady state, it is therefore possible to realize a circular wait in which
each packet is waiting on buffer/credit availability held by the next packet in
the cycle, and no packet can advance.

\paragraph{Interpreting WFR ``fault tolerance'' claims.}
The above does \emph{not} claim that the TPU v4 ICI necessarily deadlocks in production.
Rather, it shows that a natural interpretation of WFR as ``permit a bounded wild
hop and otherwise keep DOR'' is \emph{insufficient by itself} to guarantee deadlock
freedom under wormhole flow control. To be safe, WFR must be coupled with an
explicit acyclicity argument, typically via:
(i) a VC-based phase ordering (e.g., separating ``wild'' vs.\ ``DOR'' phases into
different VC classes), and/or
(ii) additional turn prohibitions so the augmented turn set remains acyclic.

Indeed, the paper itself notes that WFR requires two VCs and an additional
restriction on the ordering of wild hops relative to the DOR order to be
deadlock-free. Our counter-example can be read as a cautionary illustration:
if those VC/ordering constraints are relaxed, misapplied, or violated by the
specific set of detours chosen around a fault, the augmented allowed-turn set can
admit a CDG cycle and therefore has the \emph{potential} for wormhole deadlock.

\section{Derivation of \nameshort{} Formulation}
\label{ap:derivation}

\subsection{Inputs}

For the original Leighton-Rao (LR) formulation, the user provides the adjacency matrix, $M$, or equivalently some adjacency list.
In this case, the routable element IDs, $R$, are the row IDs of $M$.
For the \nameshort{} formulation, the user directly provides inputs for the routable element IDs, $R$. This is often just a list of integers $0, 1, ..., n-1$ for some topology of size $n$.
Let the number of nodes, $n$, be $n=|R|$.
The (LR) formulation requires binary topology values but when modified for \nameshort{} this is not necessary so for simplicity we assume basic, non-negative real values.

\subsection{Primal}

The LR formulation introduces a semi-metric, $d$, defined for all pairs of nodes, $i,j \in R$. Colloquially, this semi-metric is the ``distance.''
The semi-metric is defined by a constraint setting the sum over all pairs to be greater than one and a triangle inequality for all unique tuples $(i,j,k)$.

The formulation as presented originally defines the objective via $(i,j)$ tuple indexing on the semi-metric variable to add all distances over topology edges. This objective is the exact maximum concurrent flow (MCF) which we label $\lambda$.
Assume the adjacency matrix and semi-metrics are flattened as vectors (dimensions $M,d \in \mathbb{R}^{n^2}$) then this objective expression is equivalently a dot product (written via transpose) between the topology adjacency matrix and the semi-metric.
The primal formulation is~\textit{(LR)}.
{
\setlength{\abovedisplayskip}{3pt}%
\setlength{\belowdisplayskip}{2pt}%
\setlength{\abovedisplayshortskip}{0pt}%
\setlength{\belowdisplayshortskip}{0pt}%
\begin{lpformulation}[(LR)]
    \lpobj*{min}{\lambda = \sum_{(i,j) \in E} d_{i,j} = M^T d}
    \lpeq*{\sum_{i \in V} \sum_{j \in V, j \geq i} d_{i,j} \ge 1}{}
    \lpeq*{ d_{i,j} - d_{i,k} - d_{k,j} \leq 0 }{ }
    \lpeq*{ \quad }{ \text{ distinct } i,j,k \in V}
    \lpeq*{ d_{i,j} \geq 0}{i,j}
    \lplabel{lp:lr}
\end{lpformulation}
}

For discussion later, let the first constraint be called ``demand satisfaction'' and the second constraint be the ``triangle equality.''

We define the constraint matrix, $A$, and the right hand size (RHS), $b$, such that when written in matrix-vector format, the formulation is as follows. Reminder of meaning of objective added for clarity.
{%
\setlength{\abovedisplayskip}{0pt}%
\setlength{\belowdisplayskip}{1pt}%
\setlength{\abovedisplayshortskip}{0pt}%
\setlength{\belowdisplayshortskip}{0pt}%
\begin{align}
    \mathop{\text{maximize }}_{d} \lambda = M^T d \notag \\
    \text{s.t. } Ad \geq b\notag \\
    d \geq 0\notag
\end{align}
}

\subsection{Basic Dual}

We apply duality to~\textit{(LR)}. This introduces a dual variable, $y$, for each constraint (demand satisfaction and triangle inequality), re-arranges the objective coefficients and RHS, and changes constraint sense.
In the primal, the rows of $A$ correspond to variables $d \in \mathbb{R}^{n^2}$ and the columns of $A$ correspond to the constraints (first column for sum of distances and other $n^3$ for triangle inequality).
In the dual, $A^T$ has the same meaning but swapping rows and columns.
By the theory of duality, the objective value ($\lambda$) of the primal and dual are equal for optimal solutions and for sub-optimal solutions, this dualized objective will be strictly less than the maximally achievable MCF~\cite{schrijver1998duality}.
Remark, that the RHS, $b$, was equal to zero except the single demand satisfaction constraint. Therefore, the first element of $y$ is equal to our objective, $\lambda$. For clarity, we relabel that element $\lambda$ and write it explicitly as a variable.

The dualized formulation in matrix-vector form is as follows.
{%
\setlength{\abovedisplayskip}{0pt}%
\setlength{\belowdisplayskip}{1pt}%
\setlength{\abovedisplayshortskip}{0pt}%
\setlength{\belowdisplayshortskip}{0pt}%
\begin{align}
    \mathop{\text{maximize }}_{y} \lambda = b^T y = [1,0,0,...]y \notag \\
    \text{s.t. } A^{T} y \leq M\notag \\
    \lambda, y \geq 0\notag
\end{align}
}

\subsection{Base \nameshort{}}

We now consider the edge set as \emph{variables}, not user-provided inputs.
We label $m_{i,j}$ for each edge $(i,j)$ to distinguish from the previous user-input $M$.
This formulation is now over two variables $y$ and $m$ and seeks to maximize the MCF, $\lambda$, represented by the objective.

{
\setlength{\abovedisplayskip}{1pt}%
\setlength{\belowdisplayskip}{1pt}%
\setlength{\abovedisplayshortskip}{0pt}%
\setlength{\belowdisplayshortskip}{0pt}%
\begin{lpformulation}[(\nameshort{})]
    \lpobj*{max}{\lambda}
    \lpeq*{A^{T} y - m \leq 0}{}
    \lpeq*{\lambda, y \geq 0, \quad m \in \{0,1\} }{ }
    \lplabel{lp:base_dual}
\end{lpformulation}
}

\subsection{Full \nameshort{}}

Intuitively, the basic, dualized formulation would connect every single node, i.e. $m_{i,j}=1 \ \forall \text{ distinct } i,j \in V$.
Therefore, we add domain-specific constraints using a constraint matrix, $B$, and RHS, $p$. Obvious examples are maximum radix or link symmetry.

For scaling, we may need to introduce constraints on the topology ($m$) and/or the dual variables ($y$). We add these a constraint matrix, $S$, and a RHS. However, for our purposese, the RHS was a zero vector so we omit it.

With the base, domain, and scaling sections of the formulations, we define the \emph{\nameshort{}} formulation:

{
\setlength{\abovedisplayskip}{1pt}%
\setlength{\belowdisplayskip}{1pt}%
\setlength{\abovedisplayshortskip}{0pt}%
\setlength{\belowdisplayshortskip}{0pt}%
\begin{lpformulation}[(\nameshort{})]
    \lpobj*{max}{\lambda}
    \lpeq*{A^{T} y - m \leq 0}{}
    \lpeq*{ Bm \leq p }{ }
     \lpeq*{ S[m, y]^T \leq 0 }{ }
    \lpeq*{\lambda, y \geq 0, \quad m \in \{0,1\} }{ }
    \lplabel{lp:asc_milp}
\end{lpformulation}
}

\section{Proof of Optimal MCF using ``One-Leg'' Triangle Inequality}
\label{ap:one_leg}

Let $G = (V,E)$ be a connected undirected graph.
Consider the two linear programs:

\paragraph{(LR) Full metric formulation.}
Variables $d_{ij}$ for $i,j \in V$:
\begin{align*}
\min_{d} \quad & \sum_{(i,j)\in E} d_{ij} \\
\text{s.t.}\quad
& \sum_{i\in V}\sum_{j\in V} d_{ij} \;\ge 1, \\
& d_{ij} \le d_{ik} + d_{kj} \quad \forall\, i,j,k\in V \text{ distinct}, \\
& d_{ij} = d_{ji} \ge 0 \quad \forall\, i,j\in V, \\
& d_{ii} = 0 \quad \forall\, i\in V.
\end{align*}
Let $d^*$ be an optimal solution and denote $z_{d^*} := \sum_{(i,j)\in E} d^*_{ij}$.

\paragraph{(LR\_OL) One-leg formulation.}
Variables $b_{ij}$ for $i,j \in V$:
\begin{align*}
\min_{b} \quad & \sum_{(i,j)\in E} b_{ij} \\
\text{s.t.}\quad
& \sum_{i\in V}\sum_{j\in V} b_{ij} \;\ge 1, \\
& b_{ij} \le b_{ik} + b_{kj} \quad \forall\, i,j,k\in V \text{ distinct with } (i,k)\in E, \\
& b_{ij} = b_{ji} \ge 0 \quad \forall\, i,j\in V, \\
& b_{ii} = 0 \quad \forall\, i\in V.
\end{align*}
Let $b^*$ be an optimal solution and denote $z_{b^*} := \sum_{(i,j)\in E} b^*_{ij}$.

Then
\[
z_{b^*} = z_{d^*}.
\]
Moreover, from any optimal solution $b^*$ of \textnormal{(LR\_OL)} one can construct a feasible solution $d'$ of \textnormal{(LR)} with $d'_{ij} \ge b^*_{ij}$ for all $i,j$ and
$d'_{ij} = b^*_{ij}$ for all $(i,j)\in E$, so $d'$ is optimal for \textnormal{(LR)}.

We split the argument into two inequalities:
\[
z_{b^*} \le z_{d^*}
\quad\text{and}\quad
z_{d^*} \le z_{b^*}.
\]

\medskip
\noindent\textbf{1. Feasible-set inclusion: $z_{b^*} \le z_{d^*}$.}

By definition, $d^*$ satisfies
\[
d^*_{ij} \le d^*_{ik} + d^*_{kj} \quad \forall\, i,j,k\in V \text{ distinct}.
\]
In particular, this holds for all triples with $(i,k)\in E$. Thus $d^*$ satisfies all the triangle inequalities required by \textnormal{(LR\_OL)}, and it clearly satisfies the normalization and symmetry/nonnegativity conditions as well. Hence
\[
d^* \text{ is feasible for \textnormal{(LR\_OL)}}.
\]
Since \textnormal{(LR\_OL)} minimizes the same objective over a (weakly) larger feasible region, we obtain
\[
z_{b^*}
= \min_{b \text{ feasible for (LR\_OL)}} \sum_{(i,j)\in E} b_{ij}
\;\le\; \sum_{(i,j)\in E} d^*_{ij}
= z_{d^*}.
\]

\medskip
\noindent\textbf{2. Shortest-path closure: $z_{d^*} \le z_{b^*}$.}

Let $b^*$ be an optimal solution of \textnormal{(LR\_OL)}. Define edge weights on $G$ by
\[
w_{uv} := b^*_{uv} \quad \text{for each } (u,v)\in E.
\]
For any $i,j\in V$, define $d'_{ij}$ to be the shortest-path distance from $i$ to $j$ in $G$ with edge weights $w$:
\[
d'_{ij} := \min_{P:i\rightsquigarrow j} \sum_{(u,v)\in P} w_{uv},
\]
where the minimum is over all paths $P$ from $i$ to $j$ in $G$.
Since $G$ is connected, every pair has at least one such path, so $d'_{ij}$ is finite.

\medskip
\noindent\emph{2.1. Metric property and basic conditions.}

By construction, $d'$ is a shortest-path metric on $V$ with nonnegative symmetric weights. In particular,
\[
d'_{ij} = d'_{ji}, \quad d'_{ii}=0, \quad d'_{ij} \ge 0 \quad \forall\, i,j\in V,
\]
and for all $i,j,k\in V$,
\[
d'_{ij} \le d'_{ik} + d'_{kj}.
\]
Thus $d'$ satisfies the full triangle inequalities required by \textnormal{(LR)}.

\medskip
\noindent\emph{2.2. One-leg inequalities imply $b^*_{ij} \le d'_{ij}$.}

We first show that for any feasible $b$ of \textnormal{(LR\_OL)}, and for any $i,j\in V$ and any simple path
\[
P: i = v_0, v_1, \ldots, v_m = j
\quad\text{with}\quad (v_r,v_{r+1})\in E \text{ for all } r,
\]
we have
\begin{equation}
\label{eq:path-ineq}
b_{ij} \;\le\; \sum_{r=0}^{m-1} b_{v_r v_{r+1}}.
\end{equation}

We prove \eqref{eq:path-ineq} by induction on the path length $m$.

\emph{Base case $m=1$.}
Then $P$ consists of a single edge $(i,j)$, and \eqref{eq:path-ineq} reads
\[
b_{ij} \le b_{ij},
\]
which is trivially true.

\emph{Inductive step.}
Assume \eqref{eq:path-ineq} holds for all paths of length $m-1$.
Now consider a path of length $m\ge 2$:
\[
i = v_0, v_1, \ldots, v_m = j.
\]
Let $k := v_1$. Since $(i,k) = (v_0,v_1)\in E$, the one-leg triangle inequality for \textnormal{(LR\_OL)} gives
\[
b_{ij} \le b_{ik} + b_{kj}.
\]
The suffix $k=v_1, v_2,\dots, v_m = j$ is a path of length $m-1$, so by the induction hypothesis,
\[
b_{kj} \le \sum_{r=1}^{m-1} b_{v_r v_{r+1}}.
\]
Combining these inequalities yields
\[
b_{ij} \le b_{ik} + b_{kj}
\le b_{v_0 v_1} + \sum_{r=1}^{m-1} b_{v_r v_{r+1}}
= \sum_{r=0}^{m-1} b_{v_r v_{r+1}},
\]
completing the induction.

Applying this to $b = b^*$, and then taking the minimum over all paths from $i$ to $j$, we obtain
\[
b^*_{ij} \le \min_{P:i\rightsquigarrow j} \sum_{(u,v)\in P} b^*_{uv}
= d'_{ij} \quad \forall\, i,j\in V.
\]
Thus
\begin{equation}
\label{eq:b-less-eq-dprime}
b^*_{ij} \le d'_{ij} \quad \forall\, i,j\in V.
\end{equation}

\medskip
\noindent\emph{2.3. Normalization for $d'$.}

Since $b^*$ is feasible for \textnormal{(LR\_OL)}, we have
\[
\sum_{i\in V}\sum_{j\in V} b^*_{ij} \ge 1.
\]
Using \eqref{eq:b-less-eq-dprime}, we get entrywise $d'_{ij} \ge b^*_{ij}$, hence
\[
\sum_{i\in V}\sum_{j\in V} d'_{ij}
\;\ge\; \sum_{i\in V}\sum_{j\in V} b^*_{ij}
\;\ge\; 1.
\]
Together with the metric property and nonnegativity/symmetry established in 2.1, this shows that $d'$ is feasible for \textnormal{(LR)}.

\medskip
\noindent\emph{2.4. Objective equality on edges.}

We now compare $d'$ and $b^*$ on edges.
Fix any edge $(i,j)\in E$.

By definition of $d'$, there exists a path $P$ from $i$ to $j$ realizing the minimum:
\[
d'_{ij} = \sum_{(u,v)\in P} b^*_{uv}.
\]
In particular, the single-edge path $P = \{(i,j)\}$ is one candidate, so
\[
d'_{ij} \le b^*_{ij}.
\]
On the other hand, applying \eqref{eq:b-less-eq-dprime} with $(i,j)$ we have
\[
b^*_{ij} \le d'_{ij}.
\]
Hence, for all $(i,j)\in E$,
\[
d'_{ij} = b^*_{ij}.
\]

The objective of both \textnormal{(LR)} and \textnormal{(LR\_OL)} is
\[
f(x) = \sum_{(i,j)\in E} x_{ij}.
\]
Therefore,
\[
f(d') = \sum_{(i,j)\in E} d'_{ij}
= \sum_{(i,j)\in E} b^*_{ij}
= f(b^*)
= z_{b^*}.
\]

Since $d'$ is feasible for \textnormal{(LR)}, optimality of $d^*$ implies
\[
z_{d^*}
= \min_{d \text{ feasible for (LR)}} f(d)
\le f(d')
= z_{b^*}.
\]

\medskip
\noindent\textbf{3. Combining the inequalities.}

From Part~1 we have $z_{b^*} \le z_{d^*}$, and from Part~2 we have $z_{d^*} \le z_{b^*}$. Thus
\[
z_{b^*} = z_{d^*},
\]
and $d'$ constructed from $b^*$ is an optimal solution of \textnormal{(LR)}.

\section{Integrality Relaxation Iterative Algorithm}
\label{ap:relax_iter}

\begin{algorithm}[H]
\caption{Relaxed, Iterative LP}
\label{alg:relaxed_lp}
\begin{algorithmic}[1]
\State \textbf{Input:} $D=(x,y,z,c)$ \Comment{Given system dimensions}
\State \textbf{Input:} $interval$ \Comment{Recalculation interval}
\State \textbf{Output:} $M$ \Comment{Complete (binary) topology}
\State Initialize $M \gets$ \func{electrical\_connections}($D$)
\State Initialize $V \gets$ \func{valid\_connections}($M$)
\While{$|M| < 6xyz$}
    \State $\hat{M} \gets$ \func{LP\_\nameshort{}}($M$,$V$,$D$) \Comment{Continuous map $\hat{M}$}
    \For{$i < interval$}
        \State $e \gets $ \func{max\_optical\_connection}($\hat{M}$,$V$)
        \State $M  \gets M  \cup e$
    \EndFor

\EndWhile
\end{algorithmic}
\end{algorithm}

\section{Complete Analytical Metrics}
\label{ap:full_analytical}

\begin{table}[H]
\begin{tabular}{|l|l|l|l|l|}
\hline
\textbf{Size}          & \textbf{Topology}                   & \textbf{Diameter}           & \shortstack{\textbf{Avg.}\\\textbf{Hops}}          & \textbf{MCF}                    \\ \hline
                       & PT 4x4x8                            & 8                           & 4.032                          & 0.00781                         \\ \cline{2-5} 
                       & \cellcolor[HTML]{DAE8FC}PDTT 4x4x8  & \cellcolor[HTML]{DAE8FC}6   & \cellcolor[HTML]{DAE8FC}3.465  & \cellcolor[HTML]{DAE8FC}0.01364 \\ \cline{2-5} 
                       & \nameshort{} MILP                           & 6                           & 3.373                          & 0.01401                         \\ \cline{2-5} 
                       & \cellcolor[HTML]{DAE8FC}\nameshort{} LP     & \cellcolor[HTML]{DAE8FC}6   & \cellcolor[HTML]{DAE8FC}3.359  & \cellcolor[HTML]{DAE8FC}0.01406 \\ \cline{2-5} 
\multirow{-5}{*}{128}  & \nameshort{} LP SYM                         & 6                           & 3.368                          & 0.01403                         \\ \hline
                       & \cellcolor[HTML]{DAE8FC}PT 4x4x12   & \cellcolor[HTML]{DAE8FC}10  & \cellcolor[HTML]{DAE8FC}5.026  & \cellcolor[HTML]{DAE8FC}0.00347 \\ \cline{2-5} 
                       & \nameshort{} MILP                           & 6                           & 3.623                          & 0.00867                         \\ \cline{2-5} 
                       & \cellcolor[HTML]{DAE8FC}\nameshort{} LP     & \cellcolor[HTML]{DAE8FC}6   & \cellcolor[HTML]{DAE8FC}3.560  & \cellcolor[HTML]{DAE8FC}0.00882 \\ \cline{2-5} 
\multirow{-4}{*}{192}  & \nameshort{} LP SYM                         & 6                           & 3.560                          & 0.00883                         \\ \hline
                       & \cellcolor[HTML]{DAE8FC}PT 4x8x8    & \cellcolor[HTML]{DAE8FC}10  & \cellcolor[HTML]{DAE8FC}5.020  & \cellcolor[HTML]{DAE8FC}0.00391 \\ \cline{2-5} 
                       & PT 4x4x16                           & 12                          & 6.024                          & 0.00195                         \\ \cline{2-5} 
                       & \cellcolor[HTML]{DAE8FC}PDTT 4x8x8  & \cellcolor[HTML]{DAE8FC}6   & \cellcolor[HTML]{DAE8FC}4.329  & \cellcolor[HTML]{DAE8FC}0.00544 \\ \cline{2-5} 
                       & \nameshort{} MILP                           & 7                           & 3.833                          & 0.00606                         \\ \cline{2-5} 
                       & \cellcolor[HTML]{DAE8FC}\nameshort{} LP     & \cellcolor[HTML]{DAE8FC}6   & \cellcolor[HTML]{DAE8FC}3.750  & \cellcolor[HTML]{DAE8FC}0.00627 \\ \cline{2-5} 
\multirow{-6}{*}{256}  & \nameshort{} LP SYM                         & 6                           & 3.739                          & 0.00636                         \\ \hline
                       & \cellcolor[HTML]{DAE8FC}PT 4x8x12   & \cellcolor[HTML]{DAE8FC}12  & \cellcolor[HTML]{DAE8FC}6.016  & \cellcolor[HTML]{DAE8FC}0.00174 \\ \cline{2-5} 
                       & \nameshort{} MILP                           & 7                           & 4.127                          & 0.00380                         \\ \cline{2-5} 
                       & \cellcolor[HTML]{DAE8FC}\nameshort{} LP     & \cellcolor[HTML]{DAE8FC}7   & \cellcolor[HTML]{DAE8FC}4.029  & \cellcolor[HTML]{DAE8FC}0.00389 \\ \cline{2-5} 
\multirow{-4}{*}{384}  & \nameshort{} LP SYM                         & 6                           & 4.055                          & 0.00392                         \\ \hline
                       & \cellcolor[HTML]{DAE8FC}PT 8x8x8    & \cellcolor[HTML]{DAE8FC}12  & \cellcolor[HTML]{DAE8FC}6.012  & \cellcolor[HTML]{DAE8FC}0.00195 \\ \cline{2-5} 
                       & PT 4x8x16                           & 14                          & 7.014                          & 0.00098                         \\ \cline{2-5} 
                       & \cellcolor[HTML]{DAE8FC}PT 4x4x32   & \cellcolor[HTML]{DAE8FC}20  & \cellcolor[HTML]{DAE8FC}10.020 & \cellcolor[HTML]{DAE8FC}0.00049 \\ \cline{2-5} 
                       & \nameshort{} MILP                           & 8                           & 4.384                          & 0.00260                         \\ \cline{2-5} 
                       & \cellcolor[HTML]{DAE8FC}\nameshort{} LP     & \cellcolor[HTML]{DAE8FC}7   & \cellcolor[HTML]{DAE8FC}4.258  & \cellcolor[HTML]{DAE8FC}0.00276 \\ \cline{2-5} 
\multirow{-6}{*}{512}  & \nameshort{} LP SYM                         & 7                           & 4.259                          & 0.00276                         \\ \hline
                       & \cellcolor[HTML]{DAE8FC}PT 8x8x12   & \cellcolor[HTML]{DAE8FC}14  & \cellcolor[HTML]{DAE8FC}7.009  & \cellcolor[HTML]{DAE8FC}0.00087 \\ \cline{2-5} 
\multirow{-2}{*}{768}  & \nameshort{} LP SYM                         & 7                           & 4.642                          & 0.00169                         \\ \hline
                       & \cellcolor[HTML]{DAE8FC}PT 8x8x16   & \cellcolor[HTML]{DAE8FC}16  & \cellcolor[HTML]{DAE8FC}8.008  & \cellcolor[HTML]{DAE8FC}0.00049 \\ \cline{2-5} 
                       & PT 4x16x16                          & 18                          & 9.009                          & 0.00049                         \\ \cline{2-5} 
                       & \cellcolor[HTML]{DAE8FC}PT 4x8x32   & \cellcolor[HTML]{DAE8FC}22  & \cellcolor[HTML]{DAE8FC}11.011 & \cellcolor[HTML]{DAE8FC}0.00024 \\ \cline{2-5} 
                       & PT 4x4x64                           & 36                          & 18.018                         & 0.00012                         \\ \cline{2-5} 
                       & \cellcolor[HTML]{DAE8FC}PDTT 8x8x16 & \cellcolor[HTML]{DAE8FC}12  & \cellcolor[HTML]{DAE8FC}6.976  & \cellcolor[HTML]{DAE8FC}0.00084 \\ \cline{2-5} 
\multirow{-6}{*}{1024} & \nameshort{} LP SYM                         & 8                           & 4.852                          & 0.00119                         \\ \hline
                       & \cellcolor[HTML]{DAE8FC}PT 8x12x16  & \cellcolor[HTML]{DAE8FC}18  & \cellcolor[HTML]{DAE8FC}9.006  & \cellcolor[HTML]{DAE8FC}0.00033 \\ \cline{2-5} 
                       & PT 8x8x24                           & 20                          & 10.007                         & 0.00022                         \\ \cline{2-5} 
                       & \cellcolor[HTML]{DAE8FC}PT 4x4x96   & \cellcolor[HTML]{DAE8FC}52  & \cellcolor[HTML]{DAE8FC}26.017 & \cellcolor[HTML]{DAE8FC}0.00005 \\ \cline{2-5} 
\multirow{-4}{*}{1536} & \nameshort{} LP SYM                         & 8                           & 5.116                          & 0.00077                         \\ \hline
                       & \cellcolor[HTML]{DAE8FC}PT 8x16x16  & \cellcolor[HTML]{DAE8FC}20  & \cellcolor[HTML]{DAE8FC}10.005 & \cellcolor[HTML]{DAE8FC}0.00024 \\ \cline{2-5} 
                       & PDTT 8x16x16                        & 12                          & 8.723                          & 0.00033                         \\ \cline{2-5} 
\multirow{-3}{*}{2048} & \cellcolor[HTML]{DAE8FC}\nameshort{} LP SYM & \cellcolor[HTML]{DAE8FC}8   & \cellcolor[HTML]{DAE8FC}5.355  & \cellcolor[HTML]{DAE8FC}0.00056 \\ \hline
                       & PT 12x16x16                         & 22                          & 11.004                         & 0.00016                         \\ \cline{2-5} 
                       & \cellcolor[HTML]{DAE8FC}PT 4x4x192  & \cellcolor[HTML]{DAE8FC}100 & \cellcolor[HTML]{DAE8FC}50.016 & \cellcolor[HTML]{DAE8FC}0.00001 \\ \cline{2-5} 
\multirow{-3}{*}{3072} & \nameshort{} LP SYM                         & 9                           & 5.641                          & 0.00034                         \\ \hline
                       & \cellcolor[HTML]{DAE8FC}PT 16x16x16 & \cellcolor[HTML]{DAE8FC}24  & \cellcolor[HTML]{DAE8FC}12.003 & \cellcolor[HTML]{DAE8FC}0.00012 \\ \cline{2-5} 
\multirow{-2}{*}{4096} & \nameshort{} LP SYM                         & 9                           & 5.877                          & 0.00025                         \\ \hline
                       & \cellcolor[HTML]{DAE8FC}PT 16x16x24 & \cellcolor[HTML]{DAE8FC}28  & \cellcolor[HTML]{DAE8FC}14.002 & \cellcolor[HTML]{DAE8FC}0.00005 \\ \cline{2-5} 
\multirow{-2}{*}{6144} & \nameshort{} LP SYM                         & 10                          & 6.201                          & 0.00015                         \\ \hline
                       & \cellcolor[HTML]{DAE8FC}PT 16x16x32 & \cellcolor[HTML]{DAE8FC}32  & \cellcolor[HTML]{DAE8FC}16.002 & \cellcolor[HTML]{DAE8FC}0.00003 \\ \cline{2-5} 
                       & PDTT 16x16x32                       & 24                          & 13.986                         & 0.00005                         \\ \cline{2-5} 
\multirow{-3}{*}{8192} & \cellcolor[HTML]{DAE8FC}\nameshort{} LP SYM & \cellcolor[HTML]{DAE8FC}10  & \cellcolor[HTML]{DAE8FC}6.464  & \cellcolor[HTML]{DAE8FC}0.00012 \\ \hline
\end{tabular}
\end{table}

\section{Example Topology Visualization}
\label{ap:viz}

\begin{figure}[t]
    \centering
    \begin{subfigure}[b]{0.13\textwidth}
        \centering
        \includegraphics[width=\textwidth]{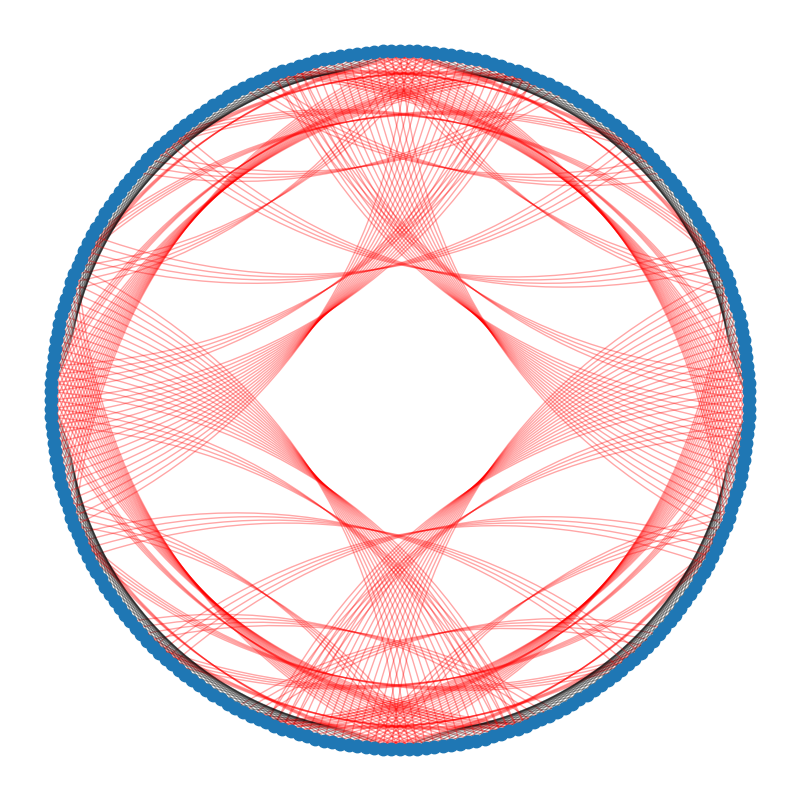}
        \caption{PT}
        \label{fig:topologies_pt}
    \end{subfigure}
    \hspace{1em}
    \begin{subfigure}[b]{0.13\textwidth}
        \centering
        \includegraphics[width=\textwidth]{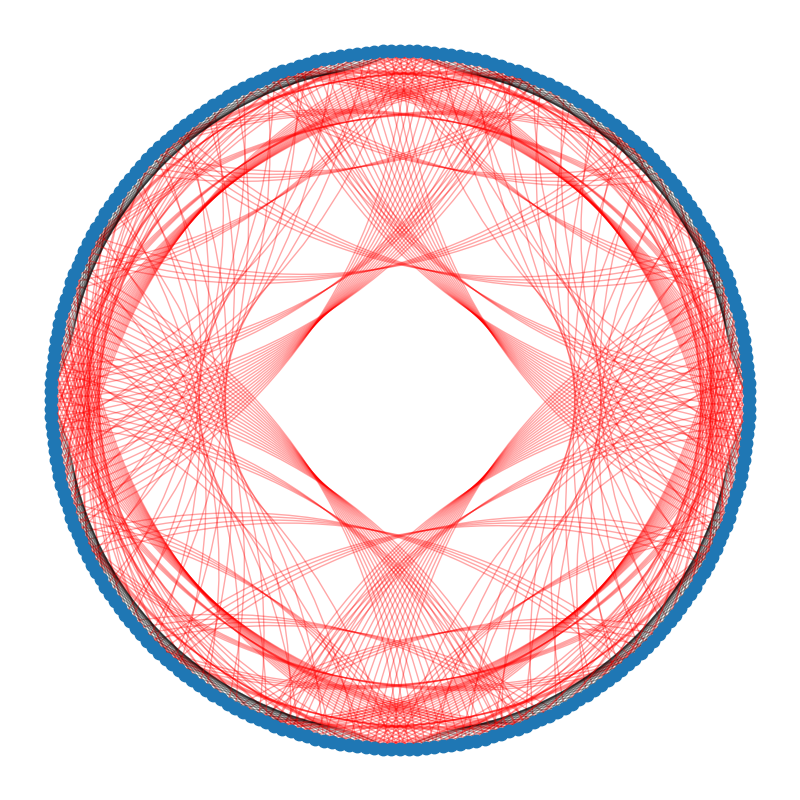}
        \caption{PDTT}
        \label{fig:topologies_pdtt}
    \end{subfigure}
    \hspace{1em}
    \begin{subfigure}[b]{0.13\textwidth}
        \centering
        \includegraphics[width=\textwidth]{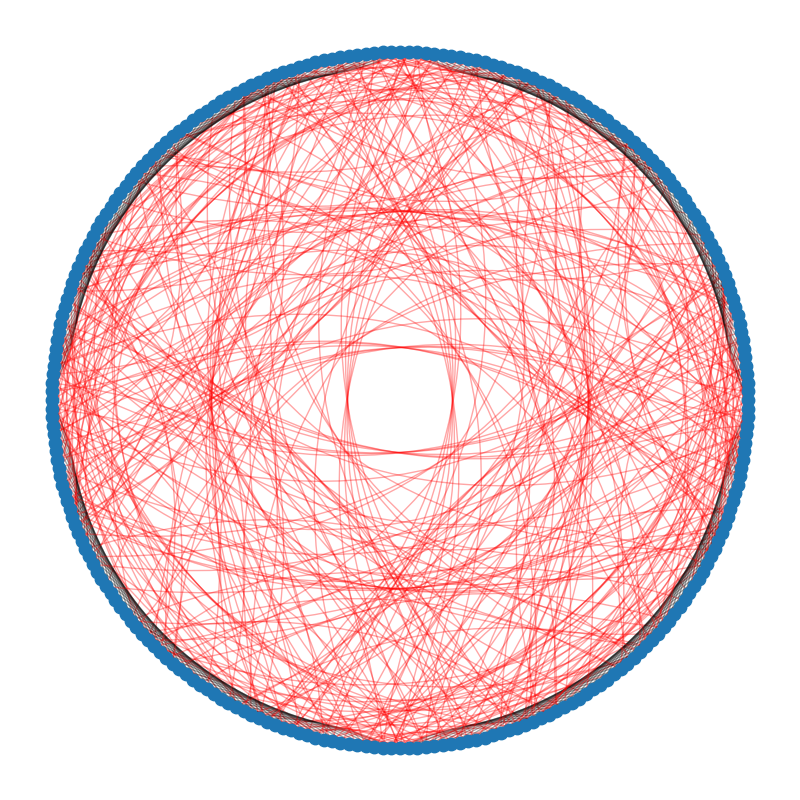}
        \caption{\nameshort{}}
        \label{fig:topologies_asc}
    \end{subfigure}
    
    \captionsetup{skip=2pt}
    \caption{Visualization of the 256 node topologies between nodes (blue) with electrical (black) and optical (red) links.}
    \label{fig:topologies}
    \vspace{-1.2em}
\end{figure}

To provide intuition what causes these higher MCF values, we visualize one size (256 nodes) in~\autoref{fig:topologies}. 
The prismatic torus baselines concentrate optical connectivity into regular patterns.
Visually, the topologies in~\autoref{fig:topologies} suggest that edge selection differs significantly across designs even when degree is identical.
For reference, the bisection improved by PDTT over PT by design~\cite{camara2010twisted} is  across the tightest two bisection cuts, 45$\degree$ slanted diameters in~\autoref{fig:topologies_pdtt}.

\section{Feasibility of OCS Fault-Tolerance}
\label{ap:ocs_fault}

We study when a TPU-style pod network admits $t\ge 2$ \emph{OCS-disjoint spanning trees}. Such a collection implies connectivity under up to $f\le t-1$ OCS-domain (color) faults, because at least one tree remains intact when at most $f$ colors fail. We proceed by relating a
throughput proxy---the maximum concurrent flow (MCF) value---to the Nash--Williams
criterion for the existence of $t$ edge-disjoint spanning trees~\cite{nash1961edge}.
Our key step is a conservative lower bound: if the network's MCF is large enough,
then every partition required by Nash--Williams has enough \emph{distinct OCS colors}
crossing it, and $t$ OCS-disjoint trees exist.

\subsection{Setup and Definitions}
\label{subsec:ocs_setup}

Let $G=(V,E)$ be a valid TPU pod graph with $n \coloneqq |V|$ routers arranged into
$\frac{n}{n_c}$ electrical cubes of size $n_c=64$. We form the \emph{cube graph}
$G'=(V',E')$ by contracting each cube into a supernode, so $|V'| = n/n_c$.
Edges $E'$ correspond to inter-cube optical connections (OCS links); electrical links
remain inside cubes and are not represented in $G'$.

\paragraph{Cuts and MCF.}
For any subset $S\subseteq V$, let $\delta_G(S)$ be the set of edges with exactly one
endpoint in $S$ and let $|\delta_G(S)|$ be its cardinality. Define the (uniform-demand)
\emph{edge expansion} (a.k.a.\ sparsity)
\begin{equation}
\label{eq:phi_def}
\Phi_G(S) \;\coloneqq\; \frac{|\delta_G(S)|}{|S|\,(n-|S|)}.
\end{equation}
Let $\lambda$ denote the maximum concurrent flow value of $G$ under uniform all-pairs
demands and the given edge capacities. A standard cut argument implies that for every
nontrivial cut $S$,
\begin{equation}
\label{eq:mcf_cut_bound}
\lambda \;\le\; \Phi_G(S).
\end{equation}
Equivalently, if we have a certified lower bound $\underline{\lambda}$ on the MCF,
then every cut must have at least
\begin{equation}
\label{eq:cut_edges_from_lambda}
|\delta_G(S)| \;\ge\; \underline{\lambda}\,|S|\,(n-|S|).
\end{equation}
In what follows, we apply \eqref{eq:cut_edges_from_lambda} to cuts induced by unions
of cubes, and we interpret the resulting edge-count lower bounds as lower bounds on
the number of OCS links leaving a set of cubes in $G'$.

\paragraph{OCS colors.}
Each OCS link in $E'$ is labeled by a \emph{color} (an OCS domain ID) from a set of
$48$ colors. By construction, across the entire pod each color appears exactly twice
(i.e., there are two links of each color), so
\begin{equation}
\label{eq:two_per_color}
|E'| = 96 \quad\text{and}\quad
\forall\text{ colors }c,\; |\{e\in E' : \mathrm{col}(e)=c\}| = 2.
\end{equation}
Consequently, any set of $m$ OCS links must contain at least $\lceil m/2\rceil$
distinct colors (since one color can contribute at most two links).

\subsection{From Nash--Williams to a Throughput and Color Budget Condition}
\label{subsec:ocs_lambda_bound}

We work with partitions that respect cube boundaries: each part is a union of whole cubes.
Let $\mathcal{P}=\{P_0,\dots,P_{k-1}\}$ be such a partition of $V$ into $k\ge 2$ parts and
write $p_i\coloneqq |P_i|$ (routers). Let $C(P_0,\dots,P_{k-1})$ denote the number of
\emph{directed} edges of $G$ that cross between distinct parts (equivalently, the sum of
directed cut sizes, divided by two to correct double counting).

\paragraph{Nash--Williams.}
For an undirected graph, the Nash--Williams / Tutte theorem states that $G'$ contains
$t$ edge-disjoint spanning trees iff every partition of $V'$ into $k$ parts has at least
$t(k-1)$ inter-part edges~\cite{nash1961edge}. In our setting we strengthen the requirement
to \emph{OCS-disjoint} spanning trees, meaning the trees are edge-disjoint and use disjoint
sets of OCS colors.

\paragraph{Color-disjoint sufficient condition.}
A conservative sufficient condition for the existence of $t$ OCS-disjoint spanning trees is:
for every cube-respecting partition $\mathcal{P}$ into $k$ parts, there are at least $t(k-1)$
\emph{distinct OCS colors} on inter-part edges. We lower bound this quantity using the cut
guarantee induced by a throughput certificate.

\paragraph{From throughput to distinct inter-part colors.}
Assume a throughput certificate $\underline{\lambda}$ such that for every part $P_i$,
the directed cut size satisfies
\begin{equation}
\label{eq:cut_edges_from_lambda}
|\delta_G(P_i)| \;\ge\; \underline{\lambda}\,p_i\,(n-p_i).
\end{equation}
Summing over parts and correcting double counting yields
\begin{equation}
\label{eq:cross_edges_from_sumcuts}
C(P_0,\dots,P_{k-1})
\;\ge\;
\frac{1}{2}\sum_{i=0}^{k-1} |\delta_G(P_i)|
\;\ge\;
\frac{\underline{\lambda}}{2}\sum_{i=0}^{k-1} p_i (n-p_i).
\end{equation}

To relate crossing \emph{edges} to crossing \emph{colors}, we use the per-cube wiring
constraint \eqref{eq:two_per_color}. In the most adversarial arrangement, each distinct
color contributes two directed arcs at a cube, so exposing $m$ directed arcs across a
cube-level cut reveals at least $\lceil m/2\rceil$ distinct colors, but never more than $48$.
Aggregating this argument across a partition, we obtain the
following throughput-driven sufficient condition.

\begin{lemma}[Throughput-driven sufficient condition]
\label{lem:lambda_to_t}
If
\begin{equation}
\label{eq:lambda_to_t}
\underline{\lambda} \;\ge\; \frac{t}{32n},
\end{equation}
then every cube-respecting partition $\mathcal{P}$ satisfies the strengthened
Nash--Williams requirement of at least $t(k-1)$ distinct inter-part OCS colors.
Consequently, $G'$ admits $t$ OCS-disjoint spanning trees.
\end{lemma}

\paragraph{Color budget saturation and the effective upper bound on $t$.}
The condition \eqref{eq:lambda_to_t} captures a \emph{throughput} limitation. Independently,
OCS-disjoint trees cannot share colors and there are only $48$ colors, so
\begin{equation}
\label{eq:t_color_budget}
t \;\le\; 48.
\end{equation}
Combining \eqref{eq:lambda_to_t} and \eqref{eq:t_color_budget}, the maximum number of
OCS-disjoint spanning trees that can be certified from $\underline{\lambda}$ is
\begin{equation}
\label{eq:t_max_min}
t_{\max} \;\le\; \min\big\{\lfloor 32n\underline{\lambda}\rfloor,\;48\big\}.
\end{equation}
This explains why a topology may satisfy $32n\underline{\lambda}>48$ (suggesting more than
$48$ trees from the throughput bound alone) while still being limited to at most $48$
OCS-disjoint trees by the finite color alphabet.

\paragraph{Connectivity under $f$ color faults.}
Let a \emph{color fault} remove all edges of a failed color from $G'$. Because the $t$
spanning trees are OCS-disjoint, any single color fault can destroy edges from at most one
tree; therefore, under $f$ color faults at most $f$ trees are destroyed and at least
$t-f$ trees remain intact. If $f\le t-1$, then at least one spanning tree survives, which
implies cube-to-cube connectivity. Since routers within a cube remain electrically connected,
this implies that every router can still reach every other router in the pod.

Thus, a sufficient condition for tolerating up to $f$ color faults is the existence of
$t\ge f+1$ OCS-disjoint spanning trees. Using \eqref{eq:t_max_min}, a conservative
throughput-and-budget condition is
\begin{equation}
\label{eq:lambda_faults}
\underline{\lambda} \;\ge\; \frac{f+1}{32n}
\quad\text{and}\quad
f \le 47.
\end{equation}

\paragraph{Empirical check.}
In our experiments, we report both terms in \eqref{eq:t_max_min}: the throughput-implied
count $\lfloor 32n\underline{\lambda}\rfloor$ and the hard cap of $48$ colors. This makes
clear when fault tolerance is limited by global throughput versus by the finite OCS color
budget.

\bibliographystyle{ACM-Reference-Format}
\bibliography{reference}

@ARTICLE{duato,
  author={Duato, J.},
  journal={IEEE Transactions on Parallel and Distributed Systems}, 
  title={A necessary and sufficient condition for deadlock-free adaptive routing in wormhole networks}, 
  year={1995},
  volume={6},
  number={10},
  pages={1055-1067},
  doi={10.1109/71.473515}}

@ARTICLE{dally1987deadlock,
  author={Dally and Seitz},
  journal={IEEE Transactions on Computers}, 
  title={Deadlock-Free Message Routing in Multiprocessor Interconnection Networks}, 
  year={1987},
  volume={C-36},
  number={5},
  pages={547-553},
  doi={10.1109/TC.1987.1676939}}

@inproceedings{glass1992turnmodel,
author = {Glass, Christopher J. and Ni, Lionel M.},
title = {The Turn Model for Adaptive Routing},
year = {1992},
isbn = {0897915097},
publisher = {Association for Computing Machinery},
address = {New York, NY, USA},
url = {https://doi.org/10.1145/139669.140384},
doi = {10.1145/139669.140384},
booktitle = {Proceedings of the 19th Annual International Symposium on Computer Architecture},
pages = {278–287},
numpages = {10},
location = {Queensland, Australia},
series = {ISCA '92}
}

@inproceedings{domke2016nue,
author = {Domke, Jens and Hoefler, Torsten and Matsuoka, Satoshi},
title = {Routing on the Dependency Graph: A New Approach to Deadlock-Free High-Performance Routing},
year = {2016},
isbn = {9781450343145},
publisher = {Association for Computing Machinery},
address = {New York, NY, USA},
url = {https://doi.org/10.1145/2907294.2907313},
doi = {10.1145/2907294.2907313},
booktitle = {Proceedings of the 25th ACM International Symposium on High-Performance Parallel and Distributed Computing},
pages = {3–14},
numpages = {12},
location = {Kyoto, Japan},
series = {HPDC '16}
}

@INPROCEEDINGS{jerger2014nocarchitectures,
author={Natalie Enright Jerger et. al.},
booktitle={2014 47th Annual IEEE/ACM International Symposium on Microarchitecture}, 
title={NoC Architectures for Silicon Interposer Systems: Why Pay for more Wires when you Can Get them (from your interposer) for Free?}, 
year={2014},
volume={},
number={},
pages={458-470},
doi={10.1109/MICRO.2014.61}}

@article{hartmanis1982miphard,
author = {Hartmanis, Juris},
title = {Computers and Intractability: A Guide to the Theory of NP-Completeness (Michael R. Garey and David S. Johnson)},
journal = {SIAM Review},
volume = {24},
number = {1},
pages = {90-91},
year = {1982},
doi = {10.1137/1024022},
URL = { https://doi.org/10.1137/1024022
},
}

@article{stewart2006kautz,
  title={A new generation of cluster interconnect},
  author={Stewart, Lawrence C and Gingold, David},
  journal={White Paper, SiCortex Inc},
  year={2006}
}

@article{kautz1968kautz,
  title={Bounds on directed (d, k) graphs},
  author={Kautz, W},
  journal={Theory of cellular logic networks and machines, Final Report},
  pages={20--28},
  year={1968}
}

@inproceedings{greenberg2009vl2,
author = {Greenberg, Albert and Hamilton, James R. and Jain, Navendu and Kandula, Srikanth and Kim, Changhoon and Lahiri, Parantap and Maltz, David A. and Patel, Parveen and Sengupta, Sudipta},
title = {VL2: A Scalable and Flexible Data Center Network},
year = {2009},
isbn = {9781605585949},
publisher = {Association for Computing Machinery},
address = {New York, NY, USA},
url = {https://doi.org/10.1145/1592568.1592576},
doi = {10.1145/1592568.1592576},
booktitle = {Proceedings of the ACM SIGCOMM 2009 Conference on Data Communication},
pages = {51–62},
numpages = {12},
location = {Barcelona, Spain},
series = {SIGCOMM '09}
}

@inproceedings{gangwar2020automatedsynthesis,
author = {Anup Gangwar et. al.},
title = {Automated Synthesis of Custom Networks-on-Chip for Real World Applications},
year = {2020},
isbn = {9781450380263},
publisher = {Association for Computing Machinery},
address = {New York, NY, USA},
url = {https://doi.org/10.1145/3400302.3415656},
doi = {10.1145/3400302.3415656},
booktitle = {Proceedings of the 39th International Conference on Computer-Aided Design},
articleno = {41},
numpages = {9},
location = {Virtual Event, USA},
series = {ICCAD '20}
}

@INPROCEEDINGS{jyoth2016measuringthroughput,
  author={Sangeetha Abdu Jyothi et. al.},
  booktitle={SC '16: Proceedings of the International Conference for High Performance Computing, Networking, Storage and Analysis}, 
  title={Measuring and Understanding Throughput of Network Topologies}, 
  year={2016},
  volume={},
  number={},
  pages={761-772},
  doi={10.1109/SC.2016.64}}

@INPROCEEDINGS{curtis2012rewire,
  author={Curtis, Andrew R. and Carpenter, Tommy and Elsheikh, Mustafa and López-Ortiz, Alejandro and Keshav, S.},
  booktitle={2012 Proceedings IEEE INFOCOM}, 
  title={REWIRE: An optimization-based framework for unstructured data center network design}, 
  year={2012},
  volume={},
  number={},
  pages={1116-1124},
  doi={10.1109/INFCOM.2012.6195470}}

@ARTICLE{lpbt,
  author={Srinivasan, K. and Chatha, K.S. and Konjevod, G.},
  journal={IEEE Transactions on Very Large Scale Integration (VLSI) Systems}, 
  title={Linear-programming-based techniques for synthesis of network-on-chip architectures}, 
  year={2006},
  volume={14},
  number={4},
  pages={407-420},
  doi={10.1109/TVLSI.2006.871762}}

@article{gem5,
  author       = {Jason Lowe{-}Power et. al.},
  title        = {The gem5 Simulator: Version 20.0+},
  journal      = {CoRR},
  volume       = {abs/2007.03152},
  year         = {2020},
  url          = {https://arxiv.org/abs/2007.03152},
  eprinttype    = {arXiv},
  eprint       = {2007.03152},
  bibsource    = {dblp computer science bibliography, https://dblp.org}
}

@misc{gurobi,
  author = {{Gurobi Optimization, LLC}},
  title = {{Gurobi Optimizer Reference Manual}},
  year = 2023,
  url = "https://www.gurobi.com"
}

@inproceedings{towles2002worst,
  title={Worst-case traffic for oblivious routing functions},
  author={Towles, Brian and Dally, William J},
  booktitle={Proceedings of the fourteenth annual ACM symposium on Parallel algorithms and architectures},
  pages={1--8},
  year={2002}
}

@inproceedings{netsmith,
author = {Green, Conor James and Thottethodi, Mithuna},
title = {NetSmith: An Optimization Framework for Machine-Discovered Network Topologies},
year = {2024},
isbn = {9798400717932},
publisher = {Association for Computing Machinery},
address = {New York, NY, USA},
url = {https://doi.org/10.1145/3673038.3673060},
doi = {10.1145/3673038.3673060},
booktitle = {Proceedings of the 53rd International Conference on Parallel Processing},
pages = {421–432},
numpages = {12},
location = {Gotland, Sweden},
series = {ICPP '24}
}

@article{leighton1999lrapproxsc,
author = {Leighton, Tom and Rao, Satish},
title = {Multicommodity max-flow min-cut theorems and their use in designing approximation algorithms},
year = {1999},
issue_date = {Nov. 1999},
publisher = {Association for Computing Machinery},
address = {New York, NY, USA},
volume = {46},
number = {6},
issn = {0004-5411},
url = {https://doi.org/10.1145/331524.331526},
doi = {10.1145/331524.331526},
journal = {J. ACM},
month = nov,
pages = {787–832},
numpages = {46}
}

@article{shahrokhi1990mcf,
author = {Shahrokhi, Farhad and Matula, D. W.},
title = {The maximum concurrent flow problem},
year = {1990},
issue_date = {April 1990},
publisher = {Association for Computing Machinery},
address = {New York, NY, USA},
volume = {37},
number = {2},
issn = {0004-5411},
url = {https://doi.org/10.1145/77600.77620},
doi = {10.1145/77600.77620},
journal = {J. ACM},
month = apr,
pages = {318–334},
numpages = {17}
}

@INPROCEEDINGS{domke2011dfsssp,
  author={Domke, Jens and Hoefler, Torsten and Nagel, Wolfgang E.},
  booktitle={2011 IEEE International Parallel and Distributed Processing Symposium}, 
  title={Deadlock-Free Oblivious Routing for Arbitrary Topologies}, 
  year={2011},
  volume={},
  number={},
  pages={616-627},
  doi={10.1109/IPDPS.2011.65}}

@ARTICLE{lysne2006lash,
  author={Lysne, O. and Skeie, T. and Reinemo, S.-A. and Theiss, I.},
  journal={IEEE Transactions on Parallel and Distributed Systems}, 
  title={Layered routing in irregular networks}, 
  year={2006},
  volume={17},
  number={1},
  pages={51-65},
  doi={10.1109/TPDS.2006.12}}

@ARTICLE{Schroeder1991updown,
  author={Schroeder, M.D. and Birrell, A.D. and Burrows, M. and Murray, H. and Needham, R.M. and Rodeheffer, T.L. and Satterthwaite, E.H. and Thacker, C.P.},
  journal={IEEE Journal on Selected Areas in Communications}, 
  title={Autonet: a high-speed, self-configuring local area network using point-to-point links}, 
  year={1991},
  volume={9},
  number={8},
  pages={1318-1335},
  doi={10.1109/49.105178}}

@INPROCEEDINGS{dally1990vcs,
  author={Dally, W.J.},
  booktitle={[1990] Proceedings. The 17th Annual International Symposium on Computer Architecture}, 
  title={Virtual-channel flow control}, 
  year={1990},
  volume={},
  number={},
  pages={60-68},
  doi={10.1109/ISCA.1990.134508}}

@inproceedings{towles2003thruputrouting,
  title={Throughput-centric routing algorithm design},
  author={Towles, Brian and Dally, William J and Boyd, Stephen},
  booktitle={Proceedings of the fifteenth annual ACM symposium on Parallel algorithms and architectures},
  pages={200--209},
  year={2003}
}

@inproceedings {zu2024resiliency,
author = {Yazhou Zu and Alireza Ghaffarkhah and Hoang-Vu Dang and Brian Towles and Steven Hand and Safeen Huda and Adekunle Bello and Alexander Kolbasov and Arash Rezaei and Dayou Du and Steve Lacy and Hang Wang and Aaron Wisner and Chris Lewis and Henri Bahini},
title = {Resiliency at Scale: Managing {Google{\textquoteright}s} {TPUv4} Machine Learning Supercomputer},
booktitle = {21st USENIX Symposium on Networked Systems Design and Implementation (NSDI 24)},
year = {2024},
isbn = {978-1-939133-39-7},
address = {Santa Clara, CA},
pages = {761--774},
url = {https://www.usenix.org/conference/nsdi24/presentation/zu},
publisher = {USENIX Association},
month = apr
}

@INPROCEEDINGS{huang2021multitree,
  author={Huang, Jiayi and Majumder, Pritam and Kim, Sungkeun and Muzahid, Abdullah and Yum, Ki Hwan and Kim, Eun Jung},
  booktitle={2021 ACM/IEEE 48th Annual International Symposium on Computer Architecture (ISCA)}, 
  title={Communication Algorithm-Architecture Co-Design for Distributed Deep Learning}, 
  year={2021},
  volume={},
  number={},
  pages={181-194},
  doi={10.1109/ISCA52012.2021.00023}}

@inproceedings{ahn2009hyperx,
author = {Ahn, Jung Ho and Binkert, Nathan and Davis, Al and McLaren, Moray and Schreiber, Robert S.},
title = {HyperX: topology, routing, and packaging of efficient large-scale networks},
year = {2009},
isbn = {9781605587448},
publisher = {Association for Computing Machinery},
address = {New York, NY, USA},
url = {https://doi.org/10.1145/1654059.1654101},
doi = {10.1145/1654059.1654101},
booktitle = {Proceedings of the Conference on High Performance Computing Networking, Storage and Analysis},
articleno = {41},
numpages = {11},
location = {Portland, Oregon},
series = {SC '09}
}

@inproceedings{besta2014slimfly,
  title={Slim fly: A cost effective low-diameter network topology},
  author={Besta, Maciej and Hoefler, Torsten},
  booktitle={SC'14: proceedings of the international conference for high performance computing, networking, storage and analysis},
  pages={348--359},
  year={2014},
  organization={IEEE}
}

@inproceedings{shah2023taccl,
  title={ TACCL : Guiding Collective Algorithm Synthesis using Communication Sketches},
  author={Shah, Aashaka and Chidambaram, Vijay and Cowan, Meghan and Maleki, Saeed and Musuvathi, Madan and Mytkowicz, Todd and Nelson, Jacob and Saarikivi, Olli and Singh, Rachee},
  booktitle={20th USENIX Symposium on Networked Systems Design and Implementation (NSDI 23)},
  pages={593--612},
  year={2023}
}

@INPROCEEDINGS{jiang2013booksim,
  author={Nan Jiang and Becker, Daniel U. and Michelogiannakis, George and Balfour, James and Towles, Brian and Shaw, D. E. and Kim, John and Dally, William J.},
  booktitle={2013 IEEE International Symposium on Performance Analysis of Systems and Software (ISPASS)}, 
  title={A detailed and flexible cycle-accurate Network-on-Chip simulator}, 
  year={2013},
  volume={},
  number={},
  pages={86-96},
  doi={10.1109/ISPASS.2013.6557149}}

@INPROCEEDINGS{won2023astrasim2,
  author={Won, William and Heo, Taekyung and Rashidi, Saeed and Sridharan, Srinivas and Srinivasan, Sudarshan and Krishna, Tushar},
  booktitle={2023 IEEE International Symposium on Performance Analysis of Systems and Software (ISPASS)}, 
  title={ASTRA-sim2.0: Modeling Hierarchical Networks and Disaggregated Systems for Large-model Training at Scale}, 
  year={2023},
  volume={},
  number={},
  pages={283-294},
  doi={10.1109/ISPASS57527.2023.00035}}

@article{singh2015jupiter,
  title={Jupiter rising: A decade of clos topologies and centralized control in google's datacenter network},
  author={Singh, Arjun and Ong, Joon and Agarwal, Amit and Anderson, Glen and Armistead, Ashby and Bannon, Roy and Boving, Seb and Desai, Gaurav and Felderman, Bob and Germano, Paulie and others},
  journal={ACM SIGCOMM computer communication review},
  volume={45},
  number={4},
  pages={183--197},
  year={2015},
  publisher={ACM New York, NY, USA}
}

@inproceedings {singla2012jellyfish,
author = {Ankit Singla and Chi-Yao Hong and Lucian Popa and P. Brighten Godfrey},
title = {Jellyfish: Networking Data Centers Randomly},
booktitle = {9th USENIX Symposium on Networked Systems Design and Implementation (NSDI 12)},
year = {2012},
isbn = {978-931971-92-8},
address = {San Jose, CA},
pages = {225--238},
url = {https://www.usenix.org/conference/nsdi12/technical-sessions/presentation/singla},
publisher = {USENIX Association},
month = {apr}
}

@article{gyarmati2010scafida,
  title={Scafida: A scale-free network inspired data center architecture},
  author={Gyarmati, L{\'a}szl{\'o} and Trinh, Tuan Anh},
  journal={ACM SIGCOMM Computer Communication Review},
  volume={40},
  number={5},
  pages={4--12},
  year={2010},
  publisher={ACM New York, NY, USA}
}

@inproceedings{guo2009bcube,
  title={BCube: a high performance, server-centric network architecture for modular data centers},
  author={Guo, Chuanxiong and Lu, Guohan and Li, Dan and Wu, Haitao and Zhang, Xuan and Shi, Yunfeng and Tian, Chen and Zhang, Yongguang and Lu, Songwu},
  booktitle={Proceedings of the ACM SIGCOMM 2009 conference on Data communication},
  pages={63--74},
  year={2009}
}

@inproceedings{guo2008dcell,
  title={Dcell: a scalable and fault-tolerant network structure for data centers},
  author={Guo, Chuanxiong and Wu, Haitao and Tan, Kun and Shi, Lei and Zhang, Yongguang and Lu, Songwu},
  booktitle={Proceedings of the ACM SIGCOMM 2008 conference on Data communication},
  pages={75--86},
  year={2008}
}

@book{dally2004textbook,
  title={Principles and practices of interconnection networks},
  author={Dally, William James and Towles, Brian Patrick},
  year={2004},
  publisher={Elsevier}
}

@inproceedings{narayanan2019pipedream,
  title={PipeDream: Generalized pipeline parallelism for DNN training},
  author={Narayanan, Deepak and Harlap, Aaron and Phanishayee, Amar and Seshadri, Vivek and Devanur, Nikhil R and Ganger, Gregory R and Gibbons, Phillip B and Zaharia, Matei},
  booktitle={Proceedings of the 27th ACM symposium on operating systems principles},
  pages={1--15},
  year={2019}
}

@article{kaplan2020scaling,
  title={Scaling laws for neural language models},
  author={Kaplan, Jared and McCandlish, Sam and Henighan, Tom and Brown, Tom B and Chess, Benjamin and Child, Rewon and Gray, Scott and Radford, Alec and Wu, Jeffrey and Amodei, Dario},
  journal={arXiv preprint arXiv:2001.08361},
  year={2020}
}

@misc{openai2024gpt4technicalreport,
      title={GPT-4 Technical Report}, 
      author={OpenAI and Josh Achiam and Steven Adler and Sandhini Agarwal and Lama Ahmad and Ilge Akkaya and Florencia Leoni Aleman and Diogo Almeida and Janko Altenschmidt and Sam Altman and Shyamal Anadkat and Red Avila and Igor Babuschkin and Suchir Balaji and Valerie Balcom and Paul Baltescu and Haiming Bao and Mohammad Bavarian and Jeff Belgum and Irwan Bello and Jake Berdine and Gabriel Bernadett-Shapiro and Christopher Berner and Lenny Bogdonoff and Oleg Boiko and Madelaine Boyd and Anna-Luisa Brakman and Greg Brockman and Tim Brooks and Miles Brundage and Kevin Button and Trevor Cai and Rosie Campbell and Andrew Cann and Brittany Carey and Chelsea Carlson and Rory Carmichael and Brooke Chan and Che Chang and Fotis Chantzis and Derek Chen and Sully Chen and Ruby Chen and Jason Chen and Mark Chen and Ben Chess and Chester Cho and Casey Chu and Hyung Won Chung and Dave Cummings and Jeremiah Currier and Yunxing Dai and Cory Decareaux and Thomas Degry and Noah Deutsch and Damien Deville and Arka Dhar and David Dohan and Steve Dowling and Sheila Dunning and Adrien Ecoffet and Atty Eleti and Tyna Eloundou and David Farhi and Liam Fedus and Niko Felix and Simón Posada Fishman and Juston Forte and Isabella Fulford and Leo Gao and Elie Georges and Christian Gibson and Vik Goel and Tarun Gogineni and Gabriel Goh and Rapha Gontijo-Lopes and Jonathan Gordon and Morgan Grafstein and Scott Gray and Ryan Greene and Joshua Gross and Shixiang Shane Gu and Yufei Guo and Chris Hallacy and Jesse Han and Jeff Harris and Yuchen He and Mike Heaton and Johannes Heidecke and Chris Hesse and Alan Hickey and Wade Hickey and Peter Hoeschele and Brandon Houghton and Kenny Hsu and Shengli Hu and Xin Hu and Joost Huizinga and Shantanu Jain and Shawn Jain and Joanne Jang and Angela Jiang and Roger Jiang and Haozhun Jin and Denny Jin and Shino Jomoto and Billie Jonn and Heewoo Jun and Tomer Kaftan and Łukasz Kaiser and Ali Kamali and Ingmar Kanitscheider and Nitish Shirish Keskar and Tabarak Khan and Logan Kilpatrick and Jong Wook Kim and Christina Kim and Yongjik Kim and Jan Hendrik Kirchner and Jamie Kiros and Matt Knight and Daniel Kokotajlo and Łukasz Kondraciuk and Andrew Kondrich and Aris Konstantinidis and Kyle Kosic and Gretchen Krueger and Vishal Kuo and Michael Lampe and Ikai Lan and Teddy Lee and Jan Leike and Jade Leung and Daniel Levy and Chak Ming Li and Rachel Lim and Molly Lin and Stephanie Lin and Mateusz Litwin and Theresa Lopez and Ryan Lowe and Patricia Lue and Anna Makanju and Kim Malfacini and Sam Manning and Todor Markov and Yaniv Markovski and Bianca Martin and Katie Mayer and Andrew Mayne and Bob McGrew and Scott Mayer McKinney and Christine McLeavey and Paul McMillan and Jake McNeil and David Medina and Aalok Mehta and Jacob Menick and Luke Metz and Andrey Mishchenko and Pamela Mishkin and Vinnie Monaco and Evan Morikawa and Daniel Mossing and Tong Mu and Mira Murati and Oleg Murk and David Mély and Ashvin Nair and Reiichiro Nakano and Rajeev Nayak and Arvind Neelakantan and Richard Ngo and Hyeonwoo Noh and Long Ouyang and Cullen O'Keefe and Jakub Pachocki and Alex Paino and Joe Palermo and Ashley Pantuliano and Giambattista Parascandolo and Joel Parish and Emy Parparita and Alex Passos and Mikhail Pavlov and Andrew Peng and Adam Perelman and Filipe de Avila Belbute Peres and Michael Petrov and Henrique Ponde de Oliveira Pinto and Michael and Pokorny and Michelle Pokrass and Vitchyr H. Pong and Tolly Powell and Alethea Power and Boris Power and Elizabeth Proehl and Raul Puri and Alec Radford and Jack Rae and Aditya Ramesh and Cameron Raymond and Francis Real and Kendra Rimbach and Carl Ross and Bob Rotsted and Henri Roussez and Nick Ryder and Mario Saltarelli and Ted Sanders and Shibani Santurkar and Girish Sastry and Heather Schmidt and David Schnurr and John Schulman and Daniel Selsam and Kyla Sheppard and Toki Sherbakov and Jessica Shieh and Sarah Shoker and Pranav Shyam and Szymon Sidor and Eric Sigler and Maddie Simens and Jordan Sitkin and Katarina Slama and Ian Sohl and Benjamin Sokolowsky and Yang Song and Natalie Staudacher and Felipe Petroski Such and Natalie Summers and Ilya Sutskever and Jie Tang and Nikolas Tezak and Madeleine B. Thompson and Phil Tillet and Amin Tootoonchian and Elizabeth Tseng and Preston Tuggle and Nick Turley and Jerry Tworek and Juan Felipe Cerón Uribe and Andrea Vallone and Arun Vijayvergiya and Chelsea Voss and Carroll Wainwright and Justin Jay Wang and Alvin Wang and Ben Wang and Jonathan Ward and Jason Wei and CJ Weinmann and Akila Welihinda and Peter Welinder and Jiayi Weng and Lilian Weng and Matt Wiethoff and Dave Willner and Clemens Winter and Samuel Wolrich and Hannah Wong and Lauren Workman and Sherwin Wu and Jeff Wu and Michael Wu and Kai Xiao and Tao Xu and Sarah Yoo and Kevin Yu and Qiming Yuan and Wojciech Zaremba and Rowan Zellers and Chong Zhang and Marvin Zhang and Shengjia Zhao and Tianhao Zheng and Juntang Zhuang and William Zhuk and Barret Zoph},
      year={2024},
      eprint={2303.08774},
      archivePrefix={arXiv},
      primaryClass={cs.CL},
      url={https://arxiv.org/abs/2303.08774}, 
}

@article{meta2025llama,
  title={The llama 4 herd: The beginning of a new era of natively multimodal ai innovation},
  author={Meta, AI},
  journal={https://ai. meta. com/blog/llama-4-multimodal-intelligence/, checked on},
  volume={4},
  number={7},
  pages={2025},
  year={2025}
}

@article{mann2020language,
  title={Language models are few-shot learners},
  author={Mann, Ben and Ryder, Nick and Subbiah, Melanie and Kaplan, J and Dhariwal, P and Neelakantan, A and Shyam, P and Sastry, G and Askell, A and Agarwal, S and others},
  journal={arXiv preprint arXiv:2005.14165},
  volume={1},
  number={3},
  pages={3},
  year={2020}
}

@article{thoppilan2022lamda,
  title={Lamda: Language models for dialog applications},
  author={Thoppilan, Romal and De Freitas, Daniel and Hall, Jamie and Shazeer, Noam and Kulshreshtha, Apoorv and Cheng, Heng-Tze and Jin, Alicia and Bos, Taylor and Baker, Leslie and Du, Yu and others},
  journal={arXiv preprint arXiv:2201.08239},
  year={2022}
}

@misc{tpuv8announcement,
	author = {Vahdat, Amin},
	title = {Our eighth generation TPUs: two chips for the agentic era},
	howpublished = {\url{https://blog.google/innovation-and-ai/infrastructure-and-cloud/google-cloud/eighth-generation-tpu-agentic-era/}},
	year = {2026},
	note = {[Accessed 09-13-2026]},
}

@misc{tpuv5announcement,
	author = {Vahdat, Amin and Lohmeyer, Mark},
	title = {Enabling next-generation AI workloads: Announcing TPU v5p and AI Hypercomputer},
	howpublished = {\url{https://cloud.google.com/blog/products/ai-machine-learning/introducing-cloud-tpu-v5p-and-ai-hypercomputer}},
	year = {2023},
	note = {[Accessed 11-09-2025]},
}

@misc{tpuv5specs,
	title = {TPU v5p},
	howpublished = {\url{https://docs.cloud.google.com/tpu/docs/v5p}},
	year = {2026},
	note = {[Accessed 11-09-2025]},
}

@inproceedings{tpuv4,
  title={Tpu v4: An optically reconfigurable supercomputer for machine learning with hardware support for embeddings},
  author={Jouppi, Norm and Kurian, George and Li, Sheng and Ma, Peter and Nagarajan, Rahul and Nai, Lifeng and Patil, Nishant and Subramanian, Suvinay and Swing, Andy and Towles, Brian and others},
  booktitle={Proceedings of the 50th annual international symposium on computer architecture},
  pages={1--14},
  year={2023}
}

@inproceedings{poutievski2022jupiterevolving,
author = {Poutievski, Leon and Mashayekhi, Omid and Ong, Joon and Singh, Arjun and Tariq, Mukarram and Wang, Rui and Zhang, Jianan and Beauregard, Virginia and Conner, Patrick and Gribble, Steve and Kapoor, Rishi and Kratzer, Stephen and Li, Nanfang and Liu, Hong and Nagaraj, Karthik and Ornstein, Jason and Sawhney, Samir and Urata, Ryohei and Vicisano, Lorenzo and Yasumura, Kevin and Zhang, Shidong and Zhou, Junlan and Vahdat, Amin},
title = {Jupiter evolving: transforming google's datacenter network via optical circuit switches and software-defined networking},
year = {2022},
isbn = {9781450394208},
publisher = {Association for Computing Machinery},
address = {New York, NY, USA},
url = {https://doi.org/10.1145/3544216.3544265},
doi = {10.1145/3544216.3544265},
booktitle = {Proceedings of the ACM SIGCOMM 2022 Conference},
pages = {66–85},
numpages = {20},
location = {Amsterdam, Netherlands},
series = {SIGCOMM '22}
}

@article{urata2022missionapollo,
  title={Mission Apollo: Landing optical circuit switching at datacenter scale},
  author={Urata, Ryohei and Liu, Hong and Yasumura, Kevin and Mao, Erji and Berger, Jill and Zhou, Xiang and Lam, Cedric and Bannon, Roy and Hutchinson, Darren and Nelson, Daniel and others},
  journal={arXiv preprint arXiv:2208.10041},
  year={2022}
}

@inproceedings{liu2023lightwave,
  title={Lightwave fabrics: at-scale optical circuit switching for datacenter and machine learning systems},
  author={Liu, Hong and Urata, Ryohei and Yasumura, Kevin and Zhou, Xiang and Bannon, Roy and Berger, Jill and Dashti, Pedram and Jouppi, Norm and Lam, Cedric and Li, Sheng and others},
  booktitle={Proceedings of the ACM SIGCOMM 2023 Conference},
  pages={499--515},
  year={2023}
}

@inproceedings{liu2024reconfigurable,
  title={Reconfigurable Lightwave Fabrics for ML Supercomputers},
  author={Liu, Hong and Urata, Ryohei and Yasumura, Kevin and Zhou, Xiang and Bannon, Roy and Berger, Jill and Dashti, Pedram and Jouppi, Norm and Lam, Cedric and Li, Sheng and others},
  booktitle={2024 Optical Fiber Communications Conference and Exhibition (OFC)},
  pages={1--3},
  year={2024},
  organization={IEEE}
}

@article{camara2010twisted,
  title={Twisted torus topologies for enhanced interconnection networks},
  author={Camara, Jose M and Moreto, Miquel and Vallejo, Enrique and Beivide, Ramon and Miguel-Alonso, Jose and Mart{\'\i}nez, Carmen and Navaridas, Javier},
  journal={IEEE Transactions on Parallel and Distributed Systems},
  volume={21},
  number={12},
  pages={1765--1778},
  year={2010},
  publisher={IEEE}
}

@misc{tpuv4documentation,
	title = {TPU v4},
	howpublished = {\url{https://cloud.google.com/tpu/docs/v4}},
	note = {[Accessed 11-09-2025]},
}

@misc{teh2020couder,
      title={COUDER: Robust Topology Engineering for Optical Circuit Switched Data Center Networks}, 
      author={Min Yee Teh and Shizhen Zhao and Peirui Cao and Keren Bergman},
      year={2020},
      eprint={2010.00090},
      archivePrefix={arXiv},
      primaryClass={cs.NI},
      url={https://arxiv.org/abs/2010.00090}, 
}

@inproceedings{shin2011swdc,
author = {Shin, Ji-Yong and Wong, Bernard and Sirer, Emin G\"{u}n},
title = {Small-world datacenters},
year = {2011},
isbn = {9781450309769},
publisher = {Association for Computing Machinery},
address = {New York, NY, USA},
url = {https://doi.org/10.1145/2038916.2038918},
doi = {10.1145/2038916.2038918},
booktitle = {Proceedings of the 2nd ACM Symposium on Cloud Computing},
articleno = {2},
numpages = {13},
location = {Cascais, Portugal},
series = {SOCC '11}
}

@inproceedings{valadarsky2016xpander,
  title={Xpander: Towards optimal-performance datacenters},
  author={Valadarsky, Asaf and Shahaf, Gal and Dinitz, Michael and Schapira, Michael},
  booktitle={Proceedings of the 12th International on Conference on emerging Networking EXperiments and Technologies},
  pages={205--219},
  year={2016}
}

@inproceedings{abu2010camcube,
  title={Symbiotic routing in future data centers},
  author={Abu-Libdeh, Hussam and Costa, Paolo and Rowstron, Antony and O'Shea, Greg and Donnelly, Austin},
  booktitle={Proceedings of the ACM SIGCOMM 2010 conference},
  pages={51--62},
  year={2010}
}

@inproceedings{mudigonda2011taming,
  title={Taming the flying cable monster: A topology design and optimization framework for $\{$Data-Center$\}$ networks},
  author={Mudigonda, Jayaram and Yalagandula, Praveen and Mogul, Jeffrey C},
  booktitle={2011 USENIX Annual Technical Conference (USENIX ATC 11)},
  year={2011}
}

@article{valiant1982scheme,
  title={A scheme for fast parallel communication},
  author={Valiant, Leslie G.},
  journal={SIAM journal on computing},
  volume={11},
  number={2},
  pages={350--361},
  year={1982},
  publisher={SIAM}
}

@inproceedings{zhang2008faultvlb,
  title={Designing a fault-tolerant network using valiant load-balancing},
  author={Zhang-Shen, Rui and McKeown, Nick},
  booktitle={IEEE INFOCOM 2008-The 27th Conference on Computer Communications},
  pages={2360--2368},
  year={2008},
  organization={IEEE}
}

@misc{chowdhery2022palm,
      title={PaLM: Scaling Language Modeling with Pathways}, 
      author={Aakanksha Chowdhery and Sharan Narang and Jacob Devlin and Maarten Bosma and Gaurav Mishra and Adam Roberts and Paul Barham and Hyung Won Chung and Charles Sutton and Sebastian Gehrmann and Parker Schuh and Kensen Shi and Sasha Tsvyashchenko and Joshua Maynez and Abhishek Rao and Parker Barnes and Yi Tay and Noam Shazeer and Vinodkumar Prabhakaran and Emily Reif and Nan Du and Ben Hutchinson and Reiner Pope and James Bradbury and Jacob Austin and Michael Isard and Guy Gur-Ari and Pengcheng Yin and Toju Duke and Anselm Levskaya and Sanjay Ghemawat and Sunipa Dev and Henryk Michalewski and Xavier Garcia and Vedant Misra and Kevin Robinson and Liam Fedus and Denny Zhou and Daphne Ippolito and David Luan and Hyeontaek Lim and Barret Zoph and Alexander Spiridonov and Ryan Sepassi and David Dohan and Shivani Agrawal and Mark Omernick and Andrew M. Dai and Thanumalayan Sankaranarayana Pillai and Marie Pellat and Aitor Lewkowycz and Erica Moreira and Rewon Child and Oleksandr Polozov and Katherine Lee and Zongwei Zhou and Xuezhi Wang and Brennan Saeta and Mark Diaz and Orhan Firat and Michele Catasta and Jason Wei and Kathy Meier-Hellstern and Douglas Eck and Jeff Dean and Slav Petrov and Noah Fiedel},
      year={2022},
      eprint={2204.02311},
      archivePrefix={arXiv},
      primaryClass={cs.CL},
      url={https://arxiv.org/abs/2204.02311}, 
}

@misc{brown2020gpt3,
      title={Language Models are Few-Shot Learners}, 
      author={Tom B. Brown and Benjamin Mann and Nick Ryder and Melanie Subbiah and Jared Kaplan and Prafulla Dhariwal and Arvind Neelakantan and Pranav Shyam and Girish Sastry and Amanda Askell and Sandhini Agarwal and Ariel Herbert-Voss and Gretchen Krueger and Tom Henighan and Rewon Child and Aditya Ramesh and Daniel M. Ziegler and Jeffrey Wu and Clemens Winter and Christopher Hesse and Mark Chen and Eric Sigler and Mateusz Litwin and Scott Gray and Benjamin Chess and Jack Clark and Christopher Berner and Sam McCandlish and Alec Radford and Ilya Sutskever and Dario Amodei},
      year={2020},
      eprint={2005.14165},
      archivePrefix={arXiv},
      primaryClass={cs.CL},
      url={https://arxiv.org/abs/2005.14165}, 
}

@inproceedings{narayanan2021megatron,
  title={Efficient large-scale language model training on gpu clusters using megatron-lm},
  author={Narayanan, Deepak and Shoeybi, Mohammad and Casper, Jared and LeGresley, Patrick and Patwary, Mostofa and Korthikanti, Vijay and Vainbrand, Dmitri and Kashinkunti, Prethvi and Bernauer, Julie and Catanzaro, Bryan and others},
  booktitle={Proceedings of the international conference for high performance computing, networking, storage and analysis},
  pages={1--15},
  year={2021}
}

@inproceedings{jiang2024megascale,
  title={$\{$MegaScale$\}$: Scaling large language model training to more than 10,000 $\{$GPUs$\}$},
  author={Jiang, Ziheng and Lin, Haibin and Zhong, Yinmin and Huang, Qi and Chen, Yangrui and Zhang, Zhi and Peng, Yanghua and Li, Xiang and Xie, Cong and Nong, Shibiao and others},
  booktitle={21st USENIX Symposium on Networked Systems Design and Implementation (NSDI 24)},
  pages={745--760},
  year={2024}
}

@misc{rajbhandari2020zero,
      title={ZeRO: Memory Optimizations Toward Training Trillion Parameter Models}, 
      author={Samyam Rajbhandari and Jeff Rasley and Olatunji Ruwase and Yuxiong He},
      year={2020},
      eprint={1910.02054},
      archivePrefix={arXiv},
      primaryClass={cs.LG},
      url={https://arxiv.org/abs/1910.02054}, 
}

@misc{xu2021gspmd,
      title={GSPMD: General and Scalable Parallelization for ML Computation Graphs}, 
      author={Yuanzhong Xu and HyoukJoong Lee and Dehao Chen and Blake Hechtman and Yanping Huang and Rahul Joshi and Maxim Krikun and Dmitry Lepikhin and Andy Ly and Marcello Maggioni and Ruoming Pang and Noam Shazeer and Shibo Wang and Tao Wang and Yonghui Wu and Zhifeng Chen},
      year={2021},
      eprint={2105.04663},
      archivePrefix={arXiv},
      primaryClass={cs.DC},
      url={https://arxiv.org/abs/2105.04663}, 
}

@inproceedings{zheng2022alpa,
  title={Alpa: Automating inter-and $\{$Intra-Operator$\}$ parallelism for distributed deep learning},
  author={Zheng, Lianmin and Li, Zhuohan and Zhang, Hao and Zhuang, Yonghao and Chen, Zhifeng and Huang, Yanping and Wang, Yida and Xu, Yuanzhong and Zhuo, Danyang and Xing, Eric P and others},
  booktitle={16th USENIX Symposium on Operating Systems Design and Implementation (OSDI 22)},
  pages={559--578},
  year={2022}
}

@article{barham2022pathways,
  title={Pathways: Asynchronous distributed dataflow for ml},
  author={Barham, Paul and Chowdhery, Aakanksha and Dean, Jeff and Ghemawat, Sanjay and Hand, Steven and Hurt, Daniel and Isard, Michael and Lim, Hyeontaek and Pang, Ruoming and Roy, Sudip and others},
  journal={Proceedings of Machine Learning and Systems},
  volume={4},
  pages={430--449},
  year={2022}
}

@misc{sergeev2018horovod,
      title={Horovod: fast and easy distributed deep learning in TensorFlow}, 
      author={Alexander Sergeev and Mike Del Balso},
      year={2018},
      eprint={1802.05799},
      archivePrefix={arXiv},
      primaryClass={cs.LG},
      url={https://arxiv.org/abs/1802.05799}, 
}

@inproceedings {jiang2020byteps,
    author = {Yimin Jiang and Yibo Zhu and Chang Lan and Bairen Yi and Yong Cui and Chuanxiong Guo},
    title = {A Unified Architecture for Accelerating Distributed {DNN} Training in Heterogeneous {GPU/CPU} Clusters},
    booktitle = {14th USENIX Symposium on Operating Systems Design and Implementation (OSDI 20)},
    year = {2020},
    isbn = {978-1-939133-19-9},
    pages = {463--479},
    url = {https://www.usenix.org/conference/osdi20/presentation/jiang},
    publisher = {USENIX Association},
    month = nov
}

@misc{shoeybi2020megatron,
      title={Megatron-LM: Training Multi-Billion Parameter Language Models Using Model Parallelism}, 
      author={Mohammad Shoeybi and Mostofa Patwary and Raul Puri and Patrick LeGresley and Jared Casper and Bryan Catanzaro},
      year={2020},
      eprint={1909.08053},
      archivePrefix={arXiv},
      primaryClass={cs.CL},
      url={https://arxiv.org/abs/1909.08053}, 
}

@article{fernandez2025diminishing,
  title={Efficient Hardware Scaling and Diminishing Returns in Large-Scale Training of Language Models},
  author={Fernandez, Jared and Wehrstedt, Luca and Shamis, Leonid and Elhoushi, Mostafa and Saladi, Kalyan and Bisk, Yonatan and Strubell, Emma and Kahn, Jacob},
  journal={Transactions on Machine Learning Research},
    year={2025}
}

@article{huang2019gpipe,
  title={Gpipe: Efficient training of giant neural networks using pipeline parallelism},
  author={Huang, Yanping and Cheng, Youlong and Bapna, Ankur and Firat, Orhan and Chen, Dehao and Chen, Mia and Lee, HyoukJoong and Ngiam, Jiquan and Le, Quoc V and Wu, Yonghui and others},
  journal={Advances in neural information processing systems},
  volume={32},
  year={2019}
}

@article{jiang2024lancet,
  title={Lancet: Accelerating mixture-of-experts training via whole graph computation-communication overlapping},
  author={Jiang, Chenyu and Tian, Ye and Jia, Zhen and Zheng, Shuai and Wu, Chuan and Wang, Yida},
  journal={Proceedings of Machine Learning and Systems},
  volume={6},
  pages={74--86},
  year={2024}
}

@inproceedings{cowan2023mscclang,
  title={Mscclang: Microsoft collective communication language},
  author={Cowan, Meghan and Maleki, Saeed and Musuvathi, Madanlal and Saarikivi, Olli and Xiong, Yifan},
  booktitle={Proceedings of the 28th ACM International Conference on Architectural Support for Programming Languages and Operating Systems, Volume 2},
  pages={502--514},
  year={2023}
}

@inproceedings{wu2024mccs,
  title={MCCS: A Service-based Approach to Collective Communication for Multi-Tenant Cloud},
  author={Wu, Yongji and Xu, Yechen and Chen, Jingrong and Wang, Zhaodong and Zhang, Ying and Lentz, Matthew and Zhuo, Danyang},
  booktitle={Proceedings of the ACM SIGCOMM 2024 Conference},
  pages={679--690},
  year={2024}
}

@inproceedings{feng2024cnsim,
  title={Evaluating Chiplet-based Large-Scale Interconnection Networks via Cycle-Accurate Packet-Parallel Simulation},
  author={Feng, Yinxiao and Wei, Yuchen and Xiang, Dong and Ma, Kaisheng},
  booktitle={2024 USENIX Annual Technical Conference (USENIX ATC 24)},
  pages={731--747},
  year={2024}
}

@article{hestness2011netrace,
  title={Netrace: Dependency-tracking traces for efficient network-on-chip experimentation},
  author={Hestness, Joel and Keckler, Stephen W},
  journal={The University of Texas at Austin, Dept. of Computer Science, Tech. Rep},
  year={2011}
}

@article{nash1961edge,
  title={Edge-disjoint spanning trees of finite graphs},
  author={Nash-Williams, C St JA},
  journal={Journal of the London Mathematical Society},
  volume={1},
  number={1},
  pages={445--450},
  year={1961},
  publisher={Oxford University Press}
}

@article{edmonds1965maximum,
  title={Maximum matching and a polyhedron with 0, 1-vertices},
  author={Edmonds, Jack},
  journal={Journal of research of the National Bureau of Standards B},
  volume={69},
  number={125-130},
  pages={55--56},
  year={1965}
}

@article{imase1983genkautz,
  title={A design for directed graphs with minimum diameter},
  author={Imase and Itoh},
  journal={IEEE Transactions on Computers},
  volume={100},
  number={8},
  pages={782--784},
  year={1983},
  publisher={IEEE}
}

@inproceedings{basu2024efficientalltoall,
  title={Efficient all-to-all collective communication schedules for direct-connect topologies},
  author={Basu, Prithwish and Zhao, Liangyu and Fantl, Jason and Pal, Siddharth and Krishnamurthy, Arvind and Khoury, Joud},
  booktitle={Proceedings of the 33rd International Symposium on High-Performance Parallel and Distributed Computing},
  pages={28--41},
  year={2024}
}

@inproceedings{zhao2025efficientdirect,
  title={Efficient $\{$Direct-Connect$\}$ Topologies for Collective Communications},
  author={Zhao, Liangyu and Pal, Siddharth and Chugh, Tapan and Wang, Weiyang and Fantl, Jason and Basu, Prithwish and Khoury, Joud and Krishnamurthy, Arvind},
  booktitle={22nd USENIX Symposium on Networked Systems Design and Implementation (NSDI 25)},
  pages={705--737},
  year={2025}
}

@inproceedings{hu2006communication,
  title={Communication latency aware low power NoC synthesis},
  author={Hu, Yuanfang and Zhu, Yi and Chen, Hongyu and Graham, Ronald and Cheng, Chung-Kuan},
  booktitle={Proceedings of the 43rd annual Design Automation Conference},
  pages={574--579},
  year={2006}
}

@article{griner2021cerberus,
  title={Cerberus: The power of choices in datacenter topology design-a throughput perspective},
  author={Griner, Chen and Zerwas, Johannes and Blenk, Andreas and Ghobadi, Manya and Schmid, Stefan and Avin, Chen},
  journal={Proceedings of the ACM on Measurement and Analysis of Computing Systems},
  volume={5},
  number={3},
  pages={1--33},
  year={2021},
  publisher={ACM New York, NY, USA}
}

@inproceedings {singla2014highthroughput,
author = {Ankit Singla and P. Brighten Godfrey and Alexandra Kolla},
title = {High Throughput Data Center Topology Design},
booktitle = {11th USENIX Symposium on Networked Systems Design and Implementation (NSDI 14)},
year = {2014},
isbn = {978-1-931971-09-6},
address = {Seattle, WA},
pages = {29--41},
url = {https://www.usenix.org/conference/nsdi14/technical-sessions/presentation/singla},
publisher = {USENIX Association},
month = apr
}

@misc{shukla2025tamingtail,
      title={Taming the Tail: NoI Topology Synthesis for Mixed DL Workloads on Chiplet-Based Accelerators}, 
      author={Arnav Shukla and Harsh Sharma and Srikant Bharadwaj and Vinayak Abrol and Sujay Deb},
      year={2025},
      eprint={2510.24113},
      archivePrefix={arXiv},
      primaryClass={cs.AR},
      url={https://arxiv.org/abs/2510.24113}, 
}

@inproceedings{schlinker2015condor,
  title={Condor: Better topologies through declarative design},
  author={Schlinker, Brandon and Mysore, Radhika Niranjan and Smith, Sean and Mogul, Jeffrey C and Vahdat, Amin and Yu, Minlan and Katz-Bassett, Ethan and Rubin, Michael},
  booktitle={Proceedings of the 2015 ACM Conference on Special Interest Group on Data Communication},
  pages={449--463},
  year={2015}
}

@inproceedings{mellette2017rotornet,
  title={Rotornet: A scalable, low-complexity, optical datacenter network},
  author={Mellette, William M and McGuinness, Rob and Roy, Arjun and Forencich, Alex and Papen, George and Snoeren, Alex C and Porter, George},
  booktitle={Proceedings of the Conference of the ACM Special Interest Group on Data Communication},
  pages={267--280},
  year={2017}
}

@inproceedings{de2024swing,
  title={Swing: Short-cutting rings for higher bandwidth allreduce},
  author={De Sensi, Daniele and Bonato, Tommaso and Saam, David and Hoefler, Torsten},
  booktitle={21st USENIX Symposium on Networked Systems Design and Implementation (NSDI 24)},
  pages={1445--1462},
  year={2024}
}

@article{nguyen2021linear,
  title={Linear size MIP formulation of Max-Cut: new properties, links with cycle inequalities and computational results},
  author={Nguyen, Viet Hung and Minoux, Michel},
  journal={Optimization Letters},
  volume={15},
  number={4},
  pages={1041--1060},
  year={2021},
  publisher={Springer}
}

@book{schrijver1998duality,
  title={Theory of linear and integer programming},
  author={Schrijver, Alexander},
  year={1998},
  publisher={John Wiley \& Sons}
}

@inproceedings{man2026stage,
  author    = {Changhai Man and
               Joongun Park and
               Hanjiang Wu and
               Huan Xu and
               Srinivas Sridharan and
               Tushar Krishna},
  title     = {Scalable Synthesis of Distributed LLM Workloads through Symbolic Tensor Graphs},
  booktitle = {Proceedings of the 53rd Annual IEEE/ACM International Symposium on Computer Architecture (ISCA ’26)},
  year      = {2026}
}

@article{sridharan2023chakra,
  title={Chakra: Advancing performance benchmarking and co-design using standardized execution traces},
  author={Sridharan, Srinivas and Heo, Taekyung and Feng, Louis and Wang, Zhaodong and Bergeron, Matt and Fu, Wenyin and Zheng, Shengbao and Coutinho, Brian and Rashidi, Saeed and Man, Changhai and others},
  journal={arXiv preprint arXiv:2305.14516},
  year={2023}
}

@article{hawks2025mlcommons,
  title={An MLCommons Scientific Benchmarks Ontology},
  author={Hawks, Ben and von Laszewski, Gregor and Sinclair, Matthew D and Colombo, Marco and Venkataraman, Shivaram and Jain, Rutwik and Jiang, Yiwei and Tran, Nhan and Fox, Geoffrey},
  journal={arXiv preprint arXiv:2511.05614},
  year={2025}
}

@inproceedings{shen2025atlahs,
    author = {Shen, Siyuan and Bonato, Tommaso and Hu, Zhiyi and Jordan, Pasquale and Chen, Tiancheng and Hoefler, Torsten},
    title = {ATLAHS: An Application-centric Network Simulator Toolchain for AI, HPC, and Distributed Storage},
    year = {2025},
    isbn = {9798400714665},
    publisher = {Association for Computing Machinery},
    address = {New York, NY, USA},
    url = {https://doi.org/10.1145/3712285.3759838},
    doi = {10.1145/3712285.3759838},
    booktitle = {Proceedings of the International Conference for High Performance Computing, Networking, Storage and Analysis},
    pages = {349–367},
    numpages = {19},
    location = {
    },
    series = {SC '25}
}

@misc{spcl_atlahs,
	author = {},
	title = {{S}{P}{C}{L} - {A}{T}{L}{A}{H}{S} --- spcl.inf.ethz.ch},
	howpublished = {\url{https://spcl.inf.ethz.ch/Research/Scalable\_Networking/ATLAHS/}},
	year = {},
	note = {[Accessed 10-06-2026]},
}

@inproceedings{mudigere2022dlrm,
author = {Mudigere, Dheevatsa and Hao, Yuchen and Huang, Jianyu and Jia, Zhihao and Tulloch, Andrew and Sridharan, Srinivas and Liu, Xing and Ozdal, Mustafa and Nie, Jade and Park, Jongsoo and Luo, Liang and Yang, Jie (Amy) and Gao, Leon and Ivchenko, Dmytro and Basant, Aarti and Hu, Yuxi and Yang, Jiyan and Ardestani, Ehsan K. and Wang, Xiaodong and Komuravelli, Rakesh and Chu, Ching-Hsiang and Yilmaz, Serhat and Li, Huayu and Qian, Jiyuan and Feng, Zhuobo and Ma, Yinbin and Yang, Junjie and Wen, Ellie and Li, Hong and Yang, Lin and Sun, Chonglin and Zhao, Whitney and Melts, Dimitry and Dhulipala, Krishna and Kishore, KR and Graf, Tyler and Eisenman, Assaf and Matam, Kiran Kumar and Gangidi, Adi and Chen, Guoqiang Jerry and Krishnan, Manoj and Nayak, Avinash and Nair, Krishnakumar and Muthiah, Bharath and khorashadi, Mahmoud and Bhattacharya, Pallab and Lapukhov, Petr and Naumov, Maxim and Mathews, Ajit and Qiao, Lin and Smelyanskiy, Mikhail and Jia, Bill and Rao, Vijay},
title = {Software-hardware co-design for fast and scalable training of deep learning recommendation models},
year = {2022},
isbn = {9781450386104},
publisher = {Association for Computing Machinery},
address = {New York, NY, USA},
url = {https://doi.org/10.1145/3470496.3533727},
doi = {10.1145/3470496.3533727},
booktitle = {Proceedings of the 49th Annual International Symposium on Computer Architecture},
pages = {993–1011},
numpages = {19},
location = {New York, New York},
series = {ISCA '22}
}

@article{grattafiori2025llama3,
  title={The Llama 3 herd of models},
  author={Grattafiori, Aaron and Dubey, Abhimanyu and Jauhri, Abhinav and Pandey, Abhinav and Kadian, Abhishek and Al-Dahle, Ahmad and Letman, Aiesha and Mathur, Akhil and Schelten, Alan and Vaughan, Alex and others},
  year={2025}
}

@misc{qwen3technicalreport,
      title={Qwen3 Technical Report}, 
      author={Qwen Team},
      year={2025},
      eprint={2505.09388},
      archivePrefix={arXiv},
      primaryClass={cs.CL},
      url={https://arxiv.org/abs/2505.09388}, 
}

@article{qwen2.5-1m,
      title={Qwen2.5-1M Technical Report}, 
      author={An Yang and Bowen Yu and Chengyuan Li and Dayiheng Liu and Fei Huang and Haoyan Huang and Jiandong Jiang and Jianhong Tu and Jianwei Zhang and Jingren Zhou and Junyang Lin and Kai Dang and Kexin Yang and Le Yu and Mei Li and Minmin Sun and Qin Zhu and Rui Men and Tao He and Weijia Xu and Wenbiao Yin and Wenyuan Yu and Xiafei Qiu and Xingzhang Ren and Xinlong Yang and Yong Li and Zhiying Xu and Zipeng Zhang},
      journal={arXiv preprint arXiv:2501.15383},
      year={2025}
}

@ARTICLE{yoo2025collectiveapi, author={Yoo, Jinsun and Won, William and Cowan, Meghan and Jiang, Nan and Klenk, Benjamin and Sridharan, Srinivas and Krishna, Tushar}, journal={IEEE Micro}, title={Toward a Standardized Representation for Deep Learning Collective Algorithms}, year={2025}, volume={45}, number={2}, pages={46-55}, doi={10.1109/MM.2025.3547363}}

@article{zhao2023fsdp,
  title={Pytorch fsdp: experiences on scaling fully sharded data parallel},
  author={Zhao, Yanli and Gu, Andrew and Varma, Rohan and Luo, Liang and Huang, Chien-Chin and Xu, Min and Wright, Less and Shojanazeri, Hamid and Ott, Myle and Shleifer, Sam and others},
  journal={arXiv preprint arXiv:2304.11277},
  year={2023}
}

@inproceedings{wang2021dcnv2,
author = {Wang, Ruoxi and Shivanna, Rakesh and Cheng, Derek and Jain, Sagar and Lin, Dong and Hong, Lichan and Chi, Ed},
title = {DCN V2: Improved Deep \& Cross Network and Practical Lessons for Web-scale Learning to Rank Systems},
year = {2021},
isbn = {9781450383127},
publisher = {Association for Computing Machinery},
address = {New York, NY, USA},
url = {https://doi.org/10.1145/3442381.3450078},
doi = {10.1145/3442381.3450078},
booktitle = {Proceedings of the Web Conference 2021},
pages = {1785–1797},
numpages = {13},
location = {Ljubljana, Slovenia},
series = {WWW '21}
}

@article{wong1974circulant,
  title={A combinatorial problem related to multimodule memory organizations},
  author={Wong, CK and Coppersmith, Don},
  journal={Journal of the ACM (JACM)},
  volume={21},
  number={3},
  pages={392--402},
  year={1974},
  publisher={ACM New York, NY, USA}
}

@article{hoffman1960moore,
  title={On Moore graphs with diameters 2 and 3},
  author={Hoffman, Alan J and Singleton, Robert R},
  journal={IBM Journal of Research and Development},
  volume={4},
  number={5},
  pages={497--504},
  year={1960},
  publisher={IBM}
}

@misc{openxla2024xla,
  author       = {{The OpenXLA Project}},
  title        = {{XLA: Accelerated Linear Algebra}},
  howpublished = {\url{https://github.com/openxla/xla}},
  year         = {2024},
  note         = {Accessed: 2026-09-16}
}

@inproceedings{juerss2026trivance,
  title={Trivance: Latency-optimal allreduce by shortcutting multiport networks},
  author={Juerss, Anton and Addanki, Vamsi and Schmid, Stefan},
  booktitle={Proceedings of the ACM SIGCOMM 2026 Conference},
  pages={160--176},
  year={2026}
}

@inproceedings{isaev2023calculon,
  title={Calculon: a methodology and tool for high-level co-design of systems and large language models},
  author={Isaev, Mikhail and McDonald, Nic and Dennison, Larry and Vuduc, Richard},
  booktitle={Proceedings of the International Conference for High Performance Computing, Networking, Storage and Analysis},
  pages={1--14},
  year={2023}
}

\end{document}